\documentclass[preprint2]{emulateapj-rtx4} 
\pdfoutput=1
\usepackage{amsmath,natbib,graphicx}
\usepackage{epsf}
\input{epsf}
\usepackage{epstopdf}
\usepackage{multirow}
\usepackage{epsfig}
\usepackage{color}
\usepackage{lscape}

\usepackage{ifthen}
\newcommand{\ispreprint}{true}
\DeclareGraphicsExtensions{.jpg,.pdf,.png,.eps,.ps}
\graphicspath{{figures/}}

\DeclareGraphicsExtensions{.jpg,.pdf,.png,.eps,.ps}

\newcommand{\spitzer}{{\it Spitzer}}

\newcommand{\Tcmb}{\mbox{$T_{\mbox{\tiny CMB}}$}}

\newcommand{\LCDM}{\mbox{$\Lambda$CDM}}

\newcommand{\ltsima}{$\; \buildrel < \over \sim \;$}
\newcommand{\ltsim}{\lower.5ex\hbox{\ltsima}}

\newcommand{\be}{\begin{equation}}
\newcommand{\ee}{\end{equation}}
\newcommand{\bea}{\begin{eqnarray}}
\newcommand{\eea}{\end{eqnarray}}

\newcommand{\spt}{{ South Pole Telescope}}
\newcommand{\degs}{deg$^2$}
\newcommand{\msun}{\mbox{$M_\odot$}}
\newcommand{\nSN}{\ensuremath{\zeta}}

\newcommand{\uks}{$\mu$K-arcmin}

\newcommand{\fsz}{f_\mathrm{\mbox{\tiny{SZ}}}}
\newcommand{\ysz}{y_\mathrm{\mbox{\tiny{SZ}}}}
\newcommand{\fnl}{\ensuremath{f_{NL}}}
\newcommand{\chandra}{{\it Chandra }}

\def\KICPChicago{1}
\def\AAUChicago{2}
\def\EFIChicago{3}
\def\McGill{4}
\def\Cardiff{5}
\def\UChicago{6}
\def\Munich{7}
\def\MIT{8}
\def\NCSA{9}
\def\CfA{10}
\def\ExcellenceCluster{11}
\def\IAPFrance{12}
\def\PhysicsUChicago{13}
\def\NASA{14}
\def\ANL{15}
\def\IoA{16}
\def\PUC{17}
\def\Illinois{18}
\def\PennState{19}
\def\Berkeley{20}
\def\UFlorida{21}
\def\Colorado{22}
\def\Davis{23}
\def\LBNL{24}
\def\Michigan{25}
\def\MPE{26}
\def\CaseWestern{27}
\def\Caltech{28}
\def\Harvard{29}
\def\STScI{30}
\def\SAIC{31}
\def\Yale{32}

\begin{document}

\title{An SZ-selected sample of the most massive galaxy clusters in the
2500-square-degree South Pole Telescope survey}

\slugcomment{Accepted by \apj}

\author{
R.~Williamson\altaffilmark{\KICPChicago,\AAUChicago}, 
B.~A.~Benson\altaffilmark{\KICPChicago,\EFIChicago},
F.~W.~High\altaffilmark{\KICPChicago,\AAUChicago}, 
K.~Vanderlinde\altaffilmark{\McGill},
P.~A.~R.~Ade\altaffilmark{\Cardiff},
K.~A.~Aird\altaffilmark{\UChicago},
K.~Andersson\altaffilmark{\Munich,\MIT},
R.~Armstrong\altaffilmark{\NCSA},
M.~L.~N.~Ashby\altaffilmark{\CfA},
M.~Bautz\altaffilmark{\MIT},
G.~Bazin\altaffilmark{\Munich,\ExcellenceCluster},
E.~Bertin\altaffilmark{\IAPFrance},
L.~E.~Bleem\altaffilmark{\KICPChicago,\PhysicsUChicago},
M.~Bonamente\altaffilmark{\NASA},
M.~Brodwin\altaffilmark{\CfA},
J.~E.~Carlstrom\altaffilmark{\KICPChicago,\AAUChicago,\EFIChicago,\PhysicsUChicago,\ANL}, 
C.~L.~Chang\altaffilmark{\KICPChicago,\EFIChicago,\ANL}, 
S.~C.~Chapman\altaffilmark{\IoA}, 
A.~Clocchiatti\altaffilmark{\PUC},
T.~M.~Crawford\altaffilmark{\KICPChicago,\AAUChicago},
A.~T.~Crites\altaffilmark{\KICPChicago,\AAUChicago},
T.~de~Haan\altaffilmark{\McGill},
S.~Desai\altaffilmark{\NCSA,\Illinois},
M.~A.~Dobbs\altaffilmark{\McGill},
J.~P.~Dudley\altaffilmark{\McGill},
G.~G.~Fazio\altaffilmark{\CfA},
R.~J.~Foley\altaffilmark{\CfA}, 
W.~R.~Forman\altaffilmark{\CfA},
G.~Garmire\altaffilmark{\PennState},
E.~M.~George\altaffilmark{\Berkeley},
M.~D.~Gladders\altaffilmark{\KICPChicago,\AAUChicago},
A.~H.~Gonzalez\altaffilmark{\UFlorida},
N.~W.~Halverson\altaffilmark{\Colorado},
G.~P.~Holder\altaffilmark{\McGill},
W.~L.~Holzapfel\altaffilmark{\Berkeley},
S.~Hoover\altaffilmark{\KICPChicago,\EFIChicago},
J.~D.~Hrubes\altaffilmark{\UChicago},
C.~Jones\altaffilmark{\CfA},
M.~Joy\altaffilmark{\NASA},
R.~Keisler\altaffilmark{\KICPChicago,\PhysicsUChicago},
L.~Knox\altaffilmark{\Davis},
A.~T.~Lee\altaffilmark{\Berkeley,\LBNL},
E.~M.~Leitch\altaffilmark{\KICPChicago,\AAUChicago},
M.~Lueker\altaffilmark{\Berkeley},
D.~Luong-Van\altaffilmark{\UChicago},
D.~P.~Marrone\altaffilmark{\KICPChicago,\UChicago},
J.~J.~McMahon\altaffilmark{\KICPChicago,\EFIChicago,\Michigan},
J.~Mehl\altaffilmark{\KICPChicago,\AAUChicago},
S.~S.~Meyer\altaffilmark{\KICPChicago,\AAUChicago,\EFIChicago,\PhysicsUChicago},
J.~J.~Mohr\altaffilmark{\Munich,\ExcellenceCluster,\MPE},
T.~E.~Montroy\altaffilmark{\CaseWestern},
S.~S.~Murray\altaffilmark{\CfA},
S.~Padin\altaffilmark{\KICPChicago,\AAUChicago,\Caltech},
T.~Plagge\altaffilmark{\KICPChicago,\AAUChicago},
C.~Pryke\altaffilmark{\KICPChicago,\AAUChicago,\EFIChicago}, 
C.~L.~Reichardt\altaffilmark{\Berkeley},
A.~Rest\altaffilmark{\Harvard,\STScI},
J.~Ruel\altaffilmark{\Harvard},
J.~E.~Ruhl\altaffilmark{\CaseWestern}, 
B.~R.~Saliwanchik\altaffilmark{\CaseWestern}, 
A.~Saro\altaffilmark{\Munich},
K.~K.~Schaffer\altaffilmark{\KICPChicago,\EFIChicago,\SAIC}, 
L.~Shaw\altaffilmark{\McGill,\Yale},
E.~Shirokoff\altaffilmark{\Berkeley}, 
J.~Song\altaffilmark{\Michigan},
H.~G.~Spieler\altaffilmark{\LBNL},
B.~Stalder\altaffilmark{\CfA},
S.~A.~Stanford\altaffilmark{\Davis},
Z.~Staniszewski\altaffilmark{\CaseWestern},
A.~A.~Stark\altaffilmark{\CfA}, 
K.~Story\altaffilmark{\KICPChicago,\PhysicsUChicago},
C.~W.~Stubbs\altaffilmark{\Harvard,\CfA}, 
J.~D.~Vieira\altaffilmark{\KICPChicago,\PhysicsUChicago,\Caltech},
A. Vikhlinin\altaffilmark{\CfA},
and
A.~Zenteno\altaffilmark{\Munich,\ExcellenceCluster}
}

\altaffiltext{\KICPChicago}{Kavli Institute for Cosmological Physics, University of Chicago, 5640 South Ellis Avenue, Chicago, IL 60637}
\altaffiltext{\AAUChicago}{Department of Astronomy and Astrophysics, University of Chicago, 5640 South Ellis Avenue, Chicago, IL 60637}
\altaffiltext{\EFIChicago}{Enrico Fermi Institute, University of Chicago, 5640 South Ellis Avenue, Chicago, IL 60637}
\altaffiltext{\McGill}{Department of Physics, McGill University, 3600 Rue University, Montreal, Quebec H3A 2T8, Canada}
\altaffiltext{\Cardiff}{Department of Physics and Astronomy, Cardiff University, CF24 3YB, UK}
\altaffiltext{\UChicago}{University of Chicago, 5640 South Ellis Avenue, Chicago, IL 60637}
\altaffiltext{\Munich}{Department of Physics,
Ludwig-Maximilians-Universit\"{a}t, Scheinerstr.\ 1, 81679 M\"{u}nchen,
Germany}
\altaffiltext{\MIT}{MIT Kavli Institute for Astrophysics and Space
Research, Massachusetts Institute of Technology, 77 Massachusetts Avenue,
Cambridge, MA 02139}
\altaffiltext{\NCSA}{National Center for Supercomputing Applications,
University of Illinois, 1205 West Clark Street, Urbanan, IL 61801}
\altaffiltext{\CfA}{Harvard-Smithsonian Center for Astrophysics, 60 Garden Street, Cambridge, MA 02138}
\altaffiltext{\ExcellenceCluster}{Excellence Cluster Universe,
Boltzmannstr.\ 2, 85748 Garching, Germany}
\altaffiltext{\IAPFrance}{Institut d'Astrophysique de Paris, UMR 7095
CNRS, Universit\'e Pierre et Marie Curie, 98 bis boulevard Arago, F-75014
Paris, France}
\altaffiltext{\PhysicsUChicago}{Department of Physics, University of Chicago, 5640 South Ellis Avenue, Chicago, IL 60637}
\altaffiltext{\NASA}{Department of Space Science, VP62, NASA Marshall Space Flight Center, Huntsville, AL 35812}
\altaffiltext{\ANL}{Argonne National Laboratory, 9700 S. Cass Avenue, Argonne, IL, USA 60439}
\altaffiltext{\IoA}{Institute of Astronomy, Madingley Road, Cambridge, CB3 0HA, U.K.}
\altaffiltext{\PUC}{Departamento de Astronom'a y Astrof'sica, PUC Casilla 306, Santiago 22, Chile}
\altaffiltext{\Illinois}{Department of Astronomy, University of Illinois,
1002 West Green Street, Urbana, IL 61801}
\altaffiltext{\PennState}{Department of Astronomy and Astrophysics, Pennsylvania State University, 525 Davey Lab, University Park, PA 16802}
\altaffiltext{\Berkeley}{Department of Physics, University of California, Berkeley, CA 94720}
\altaffiltext{\UFlorida}{Department of Astronomy, University of Florida, Gainesville, FL 32611}
\altaffiltext{\Colorado}{Department of Astrophysical and Planetary Sciences and Department of Physics, University of Colorado, Boulder, CO 80309}
\altaffiltext{\Davis}{Department of Physics, University of California, One Shields Avenue, Davis, CA 95616}
\altaffiltext{\LBNL}{Physics Division, Lawrence Berkeley National Laboratory, Berkeley, CA 94720}
\altaffiltext{\Michigan}{Department of Physics, University of Michigan,
450 Church Street, Ann Arbor, MI, 48109}
\altaffiltext{\MPE}{Max-Planck-Institut f\"{u}r extraterrestrische Physik,
Giessenbachstr.\ 85748 Garching, Germany}
\altaffiltext{\CaseWestern}{Physics Department and CERCA, Case Western Reserve University, 10900 Euclid Ave., Cleveland, OH 44106}
\altaffiltext{\Caltech}{California Institute of Technology, 1200 E. California Blvd., Pasadena, CA 91125}
\altaffiltext{\Harvard}{Department of Physics, Harvard University, 17 Oxford Street, Cambridge, MA 02138}
\altaffiltext{\STScI}{Space Telescope Science Institute, 3700 San Martin
Dr., Baltimore, MD 21218}
\altaffiltext{\SAIC}{Liberal Arts Department, 
School of the Art Institute of Chicago, 
112 S Michigan Ave, Chicago, IL 60603}
\altaffiltext{\Yale}{Department of Physics, Yale University, P.O. Box 208210, New Haven, CT 06520-8120}

\shorttitle{Massive clusters in the SPT survey}

\shortauthors{Williamson et al.}

\email{rw247@kicp.uchicago.edu}

\begin{abstract}
The \spt\ (SPT) is currently surveying 2500 \degs\ of the southern
sky 
to detect massive galaxy clusters out to the epoch 
of their formation using the Sunyaev-Zel'dovich (SZ) effect.
This paper presents a catalog of the 26 most 
significant SZ cluster detections in the full survey region.
The catalog includes 14 clusters which have been previously identified and
12 that are new discoveries.
These clusters were
identified in fields observed to two differing noise depths:
1500~\degs\ at the final SPT survey depth
of $18\,$\uks\ at $150\,$GHz, and 1000 \degs\ at a depth of $54\,$\uks. 
Clusters were selected on the basis 
of their SZ signal-to-noise ratio (S/N) in SPT maps, a quantity which has been demonstrated to correlate tightly with cluster mass.  
The S/N thresholds were chosen to achieve a comparable mass selection across 
survey fields of both depths.
Cluster redshifts were obtained with 
optical and infrared imaging and spectroscopy from a variety of 
ground- and space-based
facilities. The redshifts range from $0.098 \le z \le 1.132$ with a median of $z_\mathrm{med}=0.40$. 
The measured SZ S/N and redshifts lead to 
unbiased mass estimates ranging from $9.8 \times 10^{14}  \,M_\odot\, h_{70}^{-1} \le M_{200}(\rho_{\mathrm{mean}}) \le 3.1 \times 10^{15}\,M_\odot\, h_{70}^{-1}$. 
Based on the SZ mass estimates, we find that none of the clusters are individually in significant tension with the $\Lambda$CDM cosmological model.
We also 
test for evidence of non-Gaussianity based on the cluster sample and find the data show 
no preference for non-Gaussian perturbations.
\end{abstract}

\keywords{galaxies: clusters: individual, cosmology: observations}

\bigskip\bigskip

\defcitealias{andersson10}{A10}
\defcitealias{vanderlinde10}{V10}
\defcitealias{high10}{H10}
\defcitealias{mortonson11}{M11}

\section{Introduction}
\label{sec:intro}

Galaxy clusters are the most massive 
collapsed objects in the Universe, with masses that
range from $10^{14}$ \msun\ to over $10^{15}$ \msun.  
Their abundance as a function of mass and redshift
can be used to constrain cosmological parameters
\citep{wang98,haiman01,holder01b,battye03,molnar04,wang04,lima07},
and this constraining power has now been demonstrated with real cluster 
samples identified in optical \citep[e.g.,][]{rozo10}, 
X-ray \citep[e.g.,][]{vikhlinin09}, and, most recently, 
millimeter (mm) \citep{vanderlinde10,sehgal10b} data.
The most massive
clusters are of particular interest, especially at high 
redshifts.  
As tracers of the most extreme
tails of the cosmological density field,
these clusters can be used to place limits on 
the Gaussianity of the initial density perturbations of the Universe
\citep[e.g.,][]{matarrese00}.
Furthermore, massive, high-redshift clusters provide laboratories for the study of astrophysics
(particularly galaxy formation and evolution) in dense environments
in the early Universe. 

Although the largest existing catalogs of galaxy clusters
are derived from optical and X-ray observations, 
clusters can also be identified by their interaction
with cosmic microwave background (CMB) photons.
The thermal Sunyaev-Zel'dovich (SZ) effect is a spectral 
distortion of the CMB caused by 
inverse-Compton scattering with hot cluster gas
\citep{sunyaev72}.
The surface brightness of the effect is independent of redshift, 
and the integrated thermal SZ effect from a cluster is
expected to trace cluster mass
with low scatter \citep{barbosa96, holder01a, motl05, nagai07, stanek09}, 
implying that SZ cluster surveys should deliver nearly mass-limited 
catalogs of clusters 
to arbitrarily high redshift.
With the recent development 
of bolometric receivers with hundreds or
thousands of pixels, dedicated mm-wave SZ
surveys over large areas of the sky are now being carried out by the \spt\
\citep[SPT,][]{carlstrom11} and the Atacama Cosmology Telescope
\citep[ACT,][]{fowler07}.  Such surveys promise to be 
powerful tools for cluster cosmology.  

The SPT is currently surveying 2500 \degs\ of the southern 
sky at $95$, $150$, and $220\,$GHz.  
To date, 
roughly 1500~\degs\ have been observed to a depth of 
$18\,$\uks\ at $150\,$GHz.\footnote{In this work, ``\uks'' refers
to the rms noise in equivalent CMB fluctuation temperature in a map with
square $1^\prime \times 1^\prime$ pixels.}  
Motivated by preliminary evidence of surprisingly massive, 
high-redshift clusters in these first 1500~\degs,
we conducted
``preview" observations of the remaining
$\sim 1000$~deg$^2$ of the SPT survey field
during a three week period of the 2010 Austral winter, 
mapping this region to a noise level three times higher
than the full survey depth 
(54\,\uks\, at $150\,$GHz).  In this paper, we present a catalog of
the 26 most significant galaxy clusters in the full 2500 \degs\ SPT
survey field
and test whether the cluster masses and redshift distribution are
consistent with those expected in a \LCDM\ cosmology.

We complement our SZ cluster catalog with data from 
observations at other wavelengths.
Spectroscopic or photometric redshifts were obtained 
for each cluster as a part of a dedicated optical and infrared (IR)
follow-up campaign.  X-ray luminosities were also determined for 
each cluster using a combination of pointed observations with
the \chandra satellite 
and measurements from the {\it Roentgensatellit} (ROSAT) mission \citep{voges99}.  


This paper is presented as follows:  Section \ref{sec:obsredux} describes 
the SPT observations, data reduction pipeline, and cluster-finding methodology.
The catalog is presented in Section \ref{sec:catalog}.  Section \ref{sec:sims} 
introduces the simulations carried out to test purity and completeness,
and to determine the scaling between the observable quantity (signal-to-noise ratio, S/N)
and cluster mass.  Section \ref{sec:optical} describes the optical and IR 
follow-up measurements used to determine the cluster redshifts, and
Section \ref{sec:xray} presents X-ray luminosities for these clusters. 
Cosmological implications of this cluster catalog are discussed
in Section \ref{sec:discussion}, and our conclusions are presented in
Section \ref{sec:conclusions}.

Unless otherwise noted, we have assumed a {\sl WMAP}7+BAO+{\sl $H_0$}
$\Lambda$CDM cosmology \citep{komatsu11} with $\Omega_M = 0.272$,
$\Omega_\Lambda = 0.728$ and $H_0 = 70.2~$km$~$s$^{-1}~$Mpc$^{-1}$ with
distance measurements from Baryon Acoustic Oscillations (BAO) in the
distribution of galaxies \citep{percival10} and the Hubble constant ($H_0$)
measurement from \citet{riess09}.  Cluster mass estimates are reported in 
terms of $M_{200}(\rho_{\mathrm{mean}})$, the mass enclosed within
a radius corresponding to an average density of $200$ times the mean density
of the Universe.  For the purposes of comparison with certain scaling relations
in the literature, we also convert these masses into 
$M_{500}(\rho_{\mathrm{crit}})$, or the mass enclosed within
a radius corresponding to an average density of $500$ times the critical density.
The conversion factor for each cluster is calculated assuming a 
Navarro-Frenk-White density profile and the mass-concentration relation of \citet{duffy08}.

\section{Instrument, Observations, and Data Reduction}
\label{sec:obsredux}

\subsection{The South Pole Telescope}

The SPT, a 10-meter off-axis Gregorian design with a 1 \degs\ field of view, has been searching for galaxy clusters in the mm-wave sky since its commissioning in 2007.
The SPT is located within 1 km of the geographical South Pole.
At an altitude of 2800 meters 
above sea level, the South Pole is one of the premier locations for mm-wave astronomy. The 
high altitude and low temperatures ensure an atmosphere with low water-vapor content and excellent transparency.  Meanwhile, the location near the Earth's rotational axis allows 24-hour access to the target fields.  

The SZ receiver currently mounted on the telescope consists of 960 transition-edge-sensor bolometers \citep{lee98}, cooled to a temperature of 280\,mK. 
These bolometers are split into six  wedges each containing 160 detectors. 
The sensitivity and configuration of these wedges have changed over the four years of scientific operation. 
In 2007, we fielded a preliminary array with wedges at all three frequencies but with limited sensitivity.  
In 2008, the array contained a single 95\,GHz wedge, three 150\,GHz wedges and two 220\,GHz wedges.  The 95\,GHz wedge did not produce science-quality data, but the 150 and 220~GHz wedges performed to specification.  
In 2009, the 95\,GHz wedge was replaced with a wedge with much higher sensitivity, and one of the 220\,GHz wedges was replaced by a 150\,GHz wedge, resulting in an array
with one wedge at 95 GHz, four wedges at 150 GHz, and one wedge at 220 GHz.
The focal plane configuration has remained the same since 2009.
The 10-meter primary is conservatively illuminated, resulting in beam sizes (FWHM) of approximately 
$1.6^\prime$, $1.1^\prime$, and $1.0^\prime$ at $95$, $150$, and $220$~GHz.

The SPT team has previously published two SZ-selected cluster samples: 
\citet[][hereafter S09]{staniszewski09} presented four clusters
(including three newly discovered clusters) selected from $\sim 40$~\degs\ of 
2007 and 2008 $150$~GHz data, while \citet[][hereafter V10]{vanderlinde10}
presented 21 clusters (including 12 new discoveries beyond S09)
selected from $150$~GHz data in the full 
$200$~\degs\ of 2008 observations.

\subsection{Observations}
\label{sec:obs}

\begin{figure}[]
\includegraphics[scale=0.4]{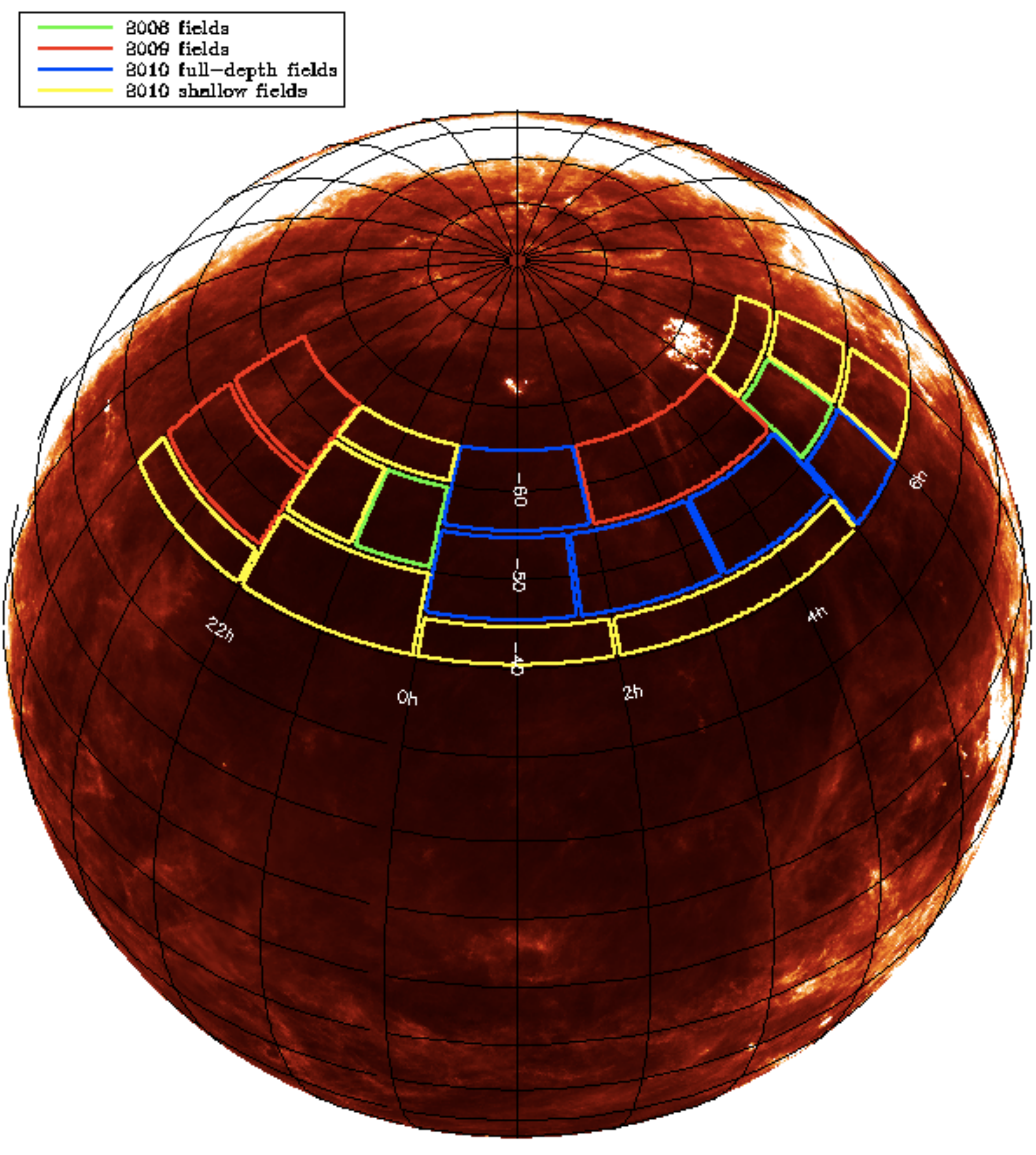}
  \caption[]{Outlines of the SPT-SZ survey fields overlaid on an orthographic projection of the IRAS $100 \, \mu$m dust map from \citet{schlegel98}.  The 
  sky is rotated such that the South Celestial Pole is at the top of the globe, and $\mathrm{R.A.}=1\mathrm{h}$ faces the viewer.  Green lines indicate fields observed in 2008, red lines indicate fields 
  observed in 2009, blue lines indicate fields observed to full depth in 2010, and yellow lines indicate fields 
  observed to preview depth in 2010, which will be completed to full depth in 2011.}
\label{fig:iras_100}
\end{figure}

The SPT-SZ survey area is a contiguous 2500 \degs\, region defined by the boundaries 
$20\mathrm{h} \le \mathrm{R.A.} \le 24\mathrm{h}; 0\mathrm{h} \le \mathrm{R.A.} \le 7\mathrm{h}$ and
$-65^\circ \le \delta \le -40^\circ$.  This region comprises the majority of the low-dust-emission southern sky below $\delta=-40^\circ$.  Observing fields north of this declination becomes difficult with the SPT because of the increased atmospheric loading and attenuation at low elevation.
We split the survey region into 19 fields, ranging in size from $\sim 70$~\degs\ to 
$\sim 230$~\degs.  Single observations of fields of this size can be completed in an 
hour or two, allowing a regular schedule of interleaved calibrations 
(see S09 and \citet{carlstrom11} for details).
The exact size, shape, location, and order of observation of these fields are determined 
by a combination of factors including availability of data at other wavelengths, sun avoidance 
(some of our observations take place during the Austral spring and summer), and the desire
to have a final survey area that is easy to define. 
The location, year of observation and size
for each field are shown in Table \ref{tab:spt_fields}.
Figure \ref{fig:iras_100} shows the full survey region and the individual field borders 
overlaid on the $100 \, \mu$m dust map from \citet{schlegel98}.

The standard operating mode of the SPT is to observe a target field by scanning back and forth in azimuth across the field followed by a step in elevation.  
These steps are large compared to the beam size, so subsequent observations of the field have a small offset in elevation applied in order to oversample the sky.  
Certain fields were observed in what is called a ``lead-trail'' mode.  
In this observing mode, the lead half of a field was scanned followed by the trail half, 
as opposed to scanning the entire azimuth range of the field in a single scan. 
This strategy was employed to safeguard against possible ground contamination, but 
we see no evidence of such contamination on the angular scales of interest to this work.
Approximately two-thirds of the ra21hdec-50 observations\footnote{Coordinates in the field 
names refer to the R.A.\ and $\delta$\ of the center of the field.}
were taken using an elevation scan mode rather than scanning in azimuth.
In this mode, the telescope was parked at a fixed azimuth and scanned up and down in elevation, allowing the sky to drift through the field of view.
We include data from both azimuth and elevation scans on this field.  We have investigated
the effects of these different scan strategies on noise properties and cluster finding
and found these effects to be negligible.

As mentioned in the introduction, observations have been completed to full survey depth
(a noise level of 18\,\uks\, at $150$~GHz) for roughly $1500$~\degs\, of the SPT survey region.  
Initial preview observations of the remaining $\sim 1000$~\degs\, were performed in late 2010, 
to a noise level of 54\,\uks\, at $150$~GHz, or three times the full-depth noise level.
The results presented in this work are based on data from both the full-depth and 
preview-depth fields.

One field, the ra23h30dec-55 field, was observed in both 2008 and 2010. 
Given the higher quality of the 95\,GHz wedge in 2010, that year's data is used in preference to the 2008 data, which was used 
in \citetalias{vanderlinde10}.  
For this reason, the properties of SPT-CL~J2337-5942 are not identical 
to those reported in \citetalias{vanderlinde10}. 
We choose not to combine the 2008 and 2010 data in order to reduce the number of different 
map depths considered.

\begin{center}
\begin{deluxetable}{lcc}
\tablecaption{2008-2011 SPT Fields
\label{tab:spt_fields}}
\tablehead{
\colhead{Field Name} & 
\colhead{Obs Year} & 
\colhead{Area [\degs]} 
}
\startdata
ra5h30dec-55 & 2008 & 90 \\ 
ra3h30dec-60 & 2009 & 230 \\ 
ra21hdec-50 & 2009 & 200 \\ 
ra21hdec-60 & 2009 & 150 \\ 
ra0h50dec-50 & 2010 & 160 \\ 
ra1hdec-60 & 2010 & 150 \\ 
ra2h30dec-50 & 2010 & 160 \\ 
ra4h10dec-50 & 2010 & 160 \\ 
ra5h30dec-45 & 2010 & 110 \\ 
ra23h30dec-55 & 2010* & 100 \\ 
ra1hdec-42.5 & 2010S & 110 \\ 
ra3h30dec-42.5 & 2010S & 170 \\ 
ra6hdec-62.5 & 2010S & 70 \\ 
ra6h30dec-45 & 2010S & 110 \\ 
ra6h30dec-55 & 2010S & 90 \\ 
ra21hdec-42.5 & 2010S & 110 \\ 
ra22h30dec-55 & 2010S & 80 \\ 
ra23hdec-45 & 2010S & 210 \\ 
ra23hdec-62.5 & 2010S & 70
\enddata
\tablecomments{The field centers, year of observation and area of the
  SPT fields. The 19 SPT fields cover a total of roughly 2500
  \degs. The nominal noise level of an SPT field is 18
  \uks. ``2010S" refers to those fields observed to three times the survey
  noise level (54 \uks). The ra23h30dec-55 field was also observed in 2008, 
  but only the 2010 data is used in this work. }
\end{deluxetable}
\end{center}

\subsection{Data processing, calibration, and map-making}
\label{sec:data_proc}

The data reduction pipeline applied to the SPT data is very similar to that described in 
previous SPT papers such as S09, 
\citetalias{vanderlinde10}, and \citet{shirokoff11}.
An overview of the processing is presented here, highlighting differences with earlier SPT releases.
The same data-processing and map-making procedure is used 
for each field, with minor adaptations in filtering to produce uniform map  
properties regardless of scan strategy.

The first steps in processing are to flag regions of compromised data (for instance, time samples with cosmic ray events) and to reconstruct the pointing for each detector.
We then calibrate the time-ordered data (TOD) to CMB temperature units. 
As in S09, this calibration is based on observations of a galactic HII region (RCW38).
The TOD is filtered and co-added into the final single-frequency map with inverse-noise weighting.

The filtering consists of bandpass filtering the TOD and removing correlated noise between detectors.
The high-pass filter is implemented by removing a ninth-order Legendre polynomial and a set of Fourier modes from each scan.  The highest-frequency Fourier modes removed correspond to an 
angular frequency of $k = 400$ in the scan direction.\footnote{We use the flat-sky approximation
throughout this work, so $| {\bf k} | \equiv \ell$.}
Depending on the scan strategy used for the observation, this filter acts as a high-pass filter in the R.A.\ or decl.\ direction.
This differs slightly from \citetalias{vanderlinde10} where only a first-order polynomial was removed, and the set of Fourier modes removed was defined by temporal frequency ($f<.25\,$Hz) rather than angular frequency ($k<400$).
The cutoff definition was altered to handle variable scan speeds. 
For the 2008 scan speeds, 0.25 Hz corresponds to $k\simeq360$, which means that
the $k$-space high-pass cutoff is slightly higher in this work than in \citetalias{vanderlinde10}.
A low-pass filter was also applied (with a cutoff at $k \sim 30000$) to avoid aliasing of
high-frequency TOD noise when the data is binned into a map.

Atmospheric noise is correlated across the entire focal plane. 
\citetalias{vanderlinde10} removed the mean and slope across all detectors in a frequency band at each 
time sample.
However, the number of detectors at the two frequencies used in this work (95 and $150\,$GHz) differ by a factor of four, so this scheme would filter different spatial modes on the sky at each frequency.
Instead, as was done in \citet{shirokoff11}, the mean of the TOD across a geometrically compact set of one quarter of the $150\,$GHz detectors or all the $95\,$GHz detectors (i.e., across one detector wedge) is subtracted at each time sample.  This acts as an isotropic high-pass filter with a cutoff at roughly $k=500$.  

\subsection{Cluster Finding}
\label{sec:clusterfind}

As discussed in Section \ref{sec:obs}, most of the SPT fields have been observed in three 
frequency bands, centered at $95$, $150$, and $220$~GHz.  
Multiple sky signals and sources of noise contribute to each single-frequency co-added
map of a field, and each of these contributions has unique spatial and
spectral properties.  Primary CMB fluctuations, emissive point sources, and
noise (both atmospheric and instrumental) contribute to the maps at all
three frequencies.  A small signal from the kinetic SZ (kSZ) effect 
(due to the interaction between CMB photons and free electrons with a bulk velocity)
also contributes to all three frequencies.
Most importantly for this work, the $95$~GHz and $150$~GHz maps contain 
an additional signal due to the thermal SZ
(tSZ) effect from clusters.
Because we can 
predict the spectral signature of the tSZ effect (up to a small relativistic
correction), we can combine the maps from the 
three bands to maximize sensitivity to tSZ and minimize noise and other contaminants.
Furthermore, we can use the fact that the galaxy clusters we expect to find have a different
spatial profile than other signals and noise to construct a spatial filter that maximizes 
sensitivity to cluster-shaped signals.  

As shown by \citet{melin06} and others, the 
optimal\footnote{This method is in fact optimal only under certain assumptions, the 
most important of which are that all sources of noise and unwanted astrophysical 
signals are random and translationally invariant, and that the exact spectral and 
spatial behavior of every component of signal and noise are known perfectly.}
way to extract a cluster-shaped tSZ signal from our data is to construct a simultaneous
spatial-spectral filter.  
We begin
by assuming that the maps are fully described by
\begin{align}
T(\mathbf{x},\nu_i) =&  \\
\nonumber & B(\mathbf{x},\nu_i) * [ \fsz(\nu_i) \Tcmb \ysz(\mathbf{x}) + 
n_\mathrm{astro}(\mathbf{x},\nu_i)] \\
\nonumber & + n_\mathrm{noise}(\mathbf{x},\nu_i), 
\end{align}
where $\ysz$ is the true tSZ sky signal in units of the Compton $y$ parameter, 
$\Tcmb$ is the mean temperature of the CMB, 
$\fsz$ encodes the frequency scaling of the tSZ effect relative to primary 
CMB fluctuations \citep[e.g.,][]{carlstrom02}, $n_\mathrm{astro}$ and
$n_\mathrm{noise}$ are the astrophysical signals and instrument/atmospheric noise
we wish to de-weight, $B(\mathbf{x},\nu_i)$ encodes the instrument beam and any filtering
applied in the analysis, and ``$*$" denotes convolution.
Given this assumption, the matched spatial-spectral filter is given by
\begin{equation}
\boldsymbol{\psi}(k_x,k_y,\nu_i) = \sigma_\psi^{-2} \ 
\sum_j \mathbf{N}_{ij}^{-1}(k_x,k_y) \fsz(\nu_j) S_\mathrm{filt}(k_x,k_y,\nu_j).
\label{eqn:optfilt}
\end{equation}
Here, $\sigma_\psi^{-2}$ is the predicted variance in the filtered map
%
\begin{align}
\sigma_\psi^{-2} = 
\sum_{i,j}& \fsz(\nu_i) S_\mathrm{filt}(k_x,k_y,\nu_i) \ \mathbf{N}_{ij}^{-1}(k_x,k_y) \ \times \\
\nonumber & \fsz(\nu_j) S_\mathrm{filt}(k_x,k_y,\nu_j),
\end{align}
%
and $S_\mathrm{filt}$ is the assumed cluster profile convolved with $B(\mathbf{x},\nu_i)$.  
The $k_x$ and $k_y$ arguments 
are included explicitly in $S_\mathrm{filt}$ because, while the underlying
cluster profile is assumed to be azimuthally symmetric, the filtering described in 
Section \ref{sec:data_proc} is anisotropic.
The $\nu$ argument is included explicitly to account for the fact that the 
filtering and instrument
beam can be different in the different SPT bands.  $\mathbf{N}$ is the band-band, pixel-pixel
covariance matrix describing the noise and non-tSZ signal
%
\begin{align}
\mathbf{N}_{abij} &= \\
\nonumber & \langle \left [ B(\mathbf{x}_a,\nu_i) * n_\mathrm{astro}(\mathbf{x}_a,\nu_i) + n_\mathrm{noise}(\mathbf{x}_a,\nu_i) \right ] \\
\nonumber & \left [ B(\mathbf{x}_b,\nu_j) * n_\mathrm{astro}(\mathbf{x}_b,\nu_j) + n_\mathrm{noise}(\mathbf{x}_b,\nu_j) \right ] \rangle.
\end{align}
%


Under the assumption that 
these components are translationally invariant, the pixel-pixel part of this matrix will be 
diagonal in the Fourier domain, which is why we only include the band indices in 
Equation \ref{eqn:optfilt}.  This also
means that $\boldsymbol{\psi}$ can be evaluated separately at each value of $\{ k_x, k_y \}$, 
and the largest matrix that needs to be inverted is $N_{\mathrm{bands}}$-by-$N_{\mathrm{bands}}$.

There should be no correlation between the astrophysical signals and instrumental/atmospheric
noise, in which case $\mathbf{N}$ can be separated into $\mathbf{N}_\mathrm{astro}$ and 
$\mathbf{N}_\mathrm{noise}$.  Furthermore, the instrumental noise should be uncorrelated
between bands, although the atmospheric noise may have correlations.  We have performed correlation
analyses on SPT maps similar to those used in this work 
and found little, if any, noise correlation between bands, which is expected 
because the correlated part of the atmospheric emission is largely removed in the filtering
described in Section \ref{sec:data_proc}.  In this case, we can estimate $\mathbf{N}_\mathrm{noise}$
individually in each band.  We do so using the jackknife procedure described in \citetalias{vanderlinde10} and S09.

Our model for $\mathbf{N}_\mathrm{astro}$
is a combination of primary and lensed CMB fluctuations, point sources below the SPT
detection threshold, kSZ, and tSZ from clusters below the SPT detection threshold. 
The power-spectrum shapes and $150$~GHz amplitudes for these components are 
identical to those used in \citetalias{vanderlinde10}.  The spectral behavior of the primary CMB, kSZ, and
tSZ components are known (up to the relativistic correction for tSZ, which we 
ignore).  The spectral behavior of the point sources is assumed to be such that the flux density
of a given source follows a power law in frequency $(\nu/\nu_0)^\alpha$, with 
$\alpha=3.6$.  This is consistent with the behavior of dusty, star-forming galaxies (DSFGs) below the 
SPT detection threshold.  Radio sources below the SPT detection
threshold are expected to contribute negligibly to the map rms compared to noise
\citep{hall10,shirokoff11}.

As in \citetalias{vanderlinde10} and S09, the source template $S$ is described by a projected spherical $\beta$-model, with $\beta$ fixed to 1,
\begin{equation}
\Delta T = \Delta T_0 (1+\theta^2/\theta_c^2)^{-1},
\end{equation}
where the normalization $\Delta T_0$ and the core radius $\theta_c$ are free parameters.
As in \citetalias{vanderlinde10} and S09, twelve different matched filters were constructed and applied to the data, 
each with a different core radius, spaced evenly between $0.25^\prime$ and $3.0^\prime$.
As in \citetalias{vanderlinde10}, point sources detected above $5 \sigma$ were masked out to a radius of $4^\prime$,  with the value inside that radius set to the average of the surrounding pixels. For extended sources, a custom mask was applied, covering the shape of the emission.  In both cases, cluster detections inside the masked region or within
$4^\prime$ of the outer edge of the masked region were rejected, meaning that any 
detection within $8^\prime$ of a masked point source was rejected.

The method of extracting clusters from the filtered maps in this work (including S/N
estimation and peak detection) is identical to that used in \citetalias{vanderlinde10}.
As in V10, we refer to the detection significance maximized across all twelve 
matched filters as $\xi$, and we use $\xi$ as the primary SZ observable.

We have measured the performance of the multi-band cluster detection algorithm 
relative to single-band detection and found significant improvement when we add
the $95$~GHz data---which has roughly a factor of two higher noise in CMB fluctuation 
temperature than the $150$~GHz data---over $150$~GHz data alone.  
We find very little further improvement
when we add the $220$~GHz data, which has a factor of five higher noise than 
the $150\,$GHz data.
As the $220\,$GHz SPT maps are not currently deep enough to add significantly to the cluster 
detection efficiency, we use only $95$ and $150$~GHz data for all the results 
presented in this work.  

For one field, ra5h30dec-55, only $150$ and $220$~GHz data exist, 
and we only perform single-frequency $150$~GHz cluster finding on this field's map
(with results indistinguishable from those reported in \citetalias{vanderlinde10}).  
Using the measured
improvement in cluster S/N in multi-band vs.\ single-band data (roughly 
$20\%$ averaged across all redshifts and cluster sizes), we determined that the most significant 
cluster in this field would not have made it into the catalog in this work even if we had
$95$~GHz data on this field (see Section \ref{sec:catalog} for details of the selection of 
clusters for this work).


\section{Catalog}
\label{sec:catalog}

The cluster-finding pipeline described above returns hundreds or even thousands of candidates (depending on the threshold value of $\xi$) 
within the 2500 \degs\ survey area.  Some of these candidates have already been 
reported in S09 and V10, 
and upcoming publications will continue to expand the SPT catalog, including 
candidates down to S/N values as low as $\xi=4.5$, where the purity of the sample 
is still estimated to be well over $50\%$. 
The aim for this paper, however, is to search the full survey area for the clusters that
have the greatest potential to test the current cosmological model---in other words, the 
most massive clusters.  At any given redshift, the most massive clusters will correspond
to the clusters with the highest SZ significance, so we present here a catalog 
of the 26 most significant detections in the 2500 \degs\ survey area.

This catalog is constructed by setting a high significance threshold of $\xi \ge 7$ for all shallow fields. 
This threshold was chosen to ensure zero false detections at high confidence (see Section \ref{sec:sims}
for details on the false detection rate estimate) and to limit the scattering 
of low-mass systems into the sample. 
To choose a deep-field threshold to match this shallow-field threshold, 
we use simulated observations 
(described in Section~\ref{sec:sims}) 
at both field depths using the same underlying simulated SZ skies.  For each cluster which met the $\xi \ge 7$ threshold in the shallow-field simulations, a deep-field $\xi$ was calculated. The shallow-to-deep $\xi$ ratio was then calculated for each cluster and found to vary from 0.4-0.6, depending on cluster characteristics.  These ratios were averaged to yield an approximately equivalent threshold value of $\xi \ge 13$ for the full-depth fields. Note, however, that significance values do not scale simply with field depth. The matched-filter noise is a combination of the CMB, point sources and observational noise, and the relative contribution of each is $k$-dependent, varying significantly between the different depths. For example, in shallow fields the relative contribution of the CMB is less, tilting the matched filter to prefer larger scales; correspondingly, it prefers smaller scales in deep fields. A simple, direct translation of ``preview-depth $\xi$'' to ``full-depth $\xi$'' is therefore not possible on a cluster-by-cluster basis; the matching of thresholds is approximate and designed only to match the average scaling.

\begin{figure}[]
\includegraphics[scale=0.60]{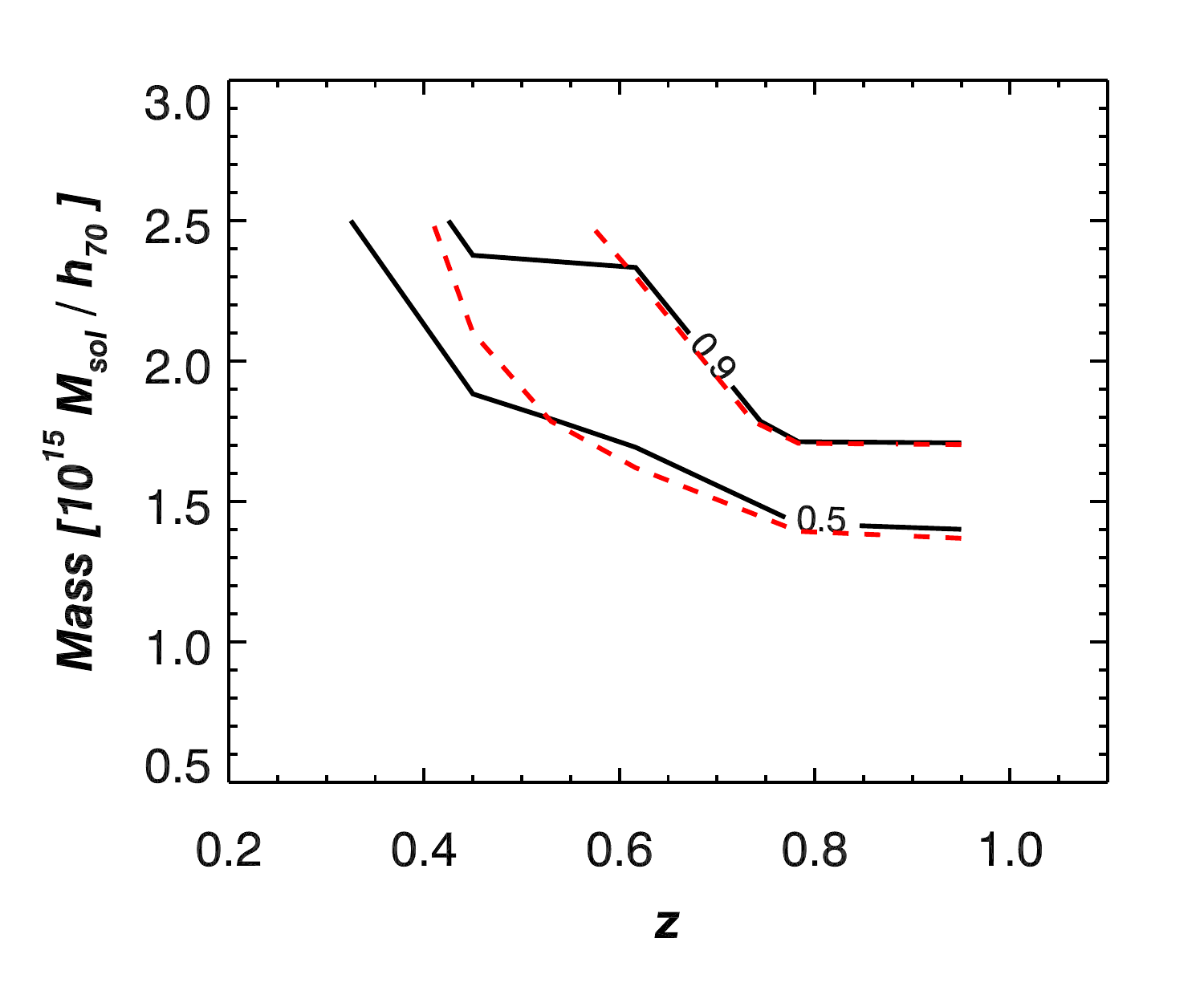}
  \caption[]{Completeness contours as a function of mass ($M_{200}(\rho_{\mathrm{mean}})$)  and redshift for both sets of map 
  depths and $\xi$ thresholds, estimated from simulated observations.  $50\%$ and $90\%$
  contours are shown for shallow fields with $\xi \ge 7$ (black, solid lines) and for deep fields
  with $\xi \ge 13$ (red, dashed lines).}
\label{fig:selfun}
\end{figure}

Nevertheless, as shown in Figure \ref{fig:selfun}, the rough $50\%$ and $90\%$
completeness contours using the two depths are fairly well matched, particularly
at high redshift.  These contours are calculated
as in V10, by matching detections in simulated observations with halos identified in the underlying
dark matter simulations in bins of mass and redshift.  For the high $\xi$ thresholds in this work---which
correspond to very massive dark matter halos---the number of matching halos is small in most bins, 
so the completeness contours are poorly sampled and noisy, but the general agreement between the 
two sets of contours is clear.
Note that the uncertainty in these curves does not directly affect the remainder of this work,
as the selection function is always considered as a strict threshold in $\xi$; 
the selection as a function of mass and redshift is presented in Figure \ref{fig:selfun} purely 
for illustrative purposes.

For each cluster, both the redshift, and 
the X-ray luminosity were determined (as outlined in Section \ref{sec:optical} and Section \ref{sec:xray}).  The  catalog is presented in Table \ref{tab:clusters} and thumbnail images of each cluster in the SZ and optical/IR are shown in Appendix A.




\subsection{Notable Clusters}
\label{sec:notable}

Of the 26 clusters reported in this paper, 12 are 
new discoveries, one was previously reported 
in V10, and 13 others have been identified in other 
optical, X-ray, and SZ cluster catalogs, with 7 having multiple 
identifications.  The previously identified clusters include seven clusters in the optical 
Abell catalog \citep{abell89}, nine clusters in the X-ray 
ROSAT-ESO Flux Limited X-ray Galaxy cluster 
survey catalog \citep[REFLEX,][]{bohringer04}, and six clusters in the mm-wave 
Atacama Cosmology Telescope catalog \citep[ACT,][]{marriage11}.
These cross-associations and alternative identifications are noted in Table \ref{tab:clusters}.   
In this section, we discuss particularly notable clusters in the 
SPT catalog.  

\paragraph{SPT-CL~J0102-4915}
This cluster was first reported in \citet{marriage11}.  It is the most significant 
detection to date in the full SPT survey by nearly a factor of two. 
It has a comparable X-ray luminosity and beam-averaged SZ decrement to the Bullet cluster 
and AS1063, whose SZ significances should be similar when the SPT survey is 
completed to full depth.   
Given the redshift of this cluster ($z=0.78$), it is expected to be one of 
the rarest objects in the SPT survey (see Figure \ref{fig:mortplot}).



\paragraph{SPT-CL~J0615-5746}
This cluster has the 
second highest redshift of any cluster in this paper, with a redshift of $z=0.972$.  
Based on its ROSAT faint source catalog counterpart, it is measured to be the fourth most 
X-ray luminous cluster in this catalog (see Section \ref{sec:xray}).  




\paragraph{SPT-CL~J0658-5556}
This cluster is the well-known Bullet cluster, otherwise known as 1ES 0657-558.  
It has been extensively studied in multiple wavelengths \citep[e.g.,][]{clowe06} 
and is known to be one of the most massive and X-ray-luminous clusters 
in the Universe.  
It is expected to be the most massive cluster in the final SPT catalog.  



\paragraph{SPT-CL~J2106-5844}
Multi-wavelength observations of this SPT-discovered cluster are 
discussed in detail in \citet{foley11}. This is 
the highest-redshift cluster ($z=1.132$) spectroscopically confirmed in the SPT survey.  
X-ray observations from \chandra measure an X-ray luminosity of 
$L_X$[0.5-2.0 keV]$=13.9\times10^{44}$ erg s$^{-1}$ \citep{foley11}, 
comparable to the X-ray luminosity of the Bullet cluster.  The mass
we report in Table \ref{tab:clusters} for SPT-CL~J2106-5844 is slightly ($5$-$10\%$)
discrepant with the SZ mass reported in \citet{foley11}, although the difference
is much smaller than either value's $1\sigma$ uncertainty.
The difference arises because 
\citet{foley11} use the single-band $150$~GHz $\xi$ and the exact
V10 scaling relation to derive the mass, whereas this work uses the
multi-band $\xi$ and the multi-band scaling relation developed specifically
for this work.  (See Section \ref{sec:sim_scaling} and V10 for details on the mass
estimation and scaling relations.)


\paragraph{SPT-CL~J2248-4431}
This cluster is also known as AS1063.  It is the second most X-ray luminous 
cluster in the REFLEX X-ray survey \citep{bohringer04}, even more luminous 
than the Bullet cluster.  It has the second highest estimated mass for 
any cluster in this paper.


\paragraph{SPT-CL~J2344-4243}
 From its 
redshift and its ROSAT bright source catalog counterpart (see Section \ref{sec:xray}), 
this cluster is measured to have the largest X-ray luminosity of any cluster
in this paper.
A bright Type 2 Seyfert galaxy at redshift $0.5975$, 2MASX
J23444387-4243124, is located $19\arcsec$ from the SZ cluster
centroid.  This redshift is consistent with our photometric
red-sequence redshift estimate of $0.62$ for the cluster (see Section
\ref{sec:optical}), suggesting this galaxy may be in or near this
cluster.

\subsection{Point Source Veto}
\label{sec:psource}

As discussed in Section \ref{sec:clusterfind}, emissive sources above $5 \sigma$ are 
masked in the cluster finding procedure, and any cluster detections within $8'$ of 
a masked point source are rejected, because residual source flux or artifacts due
to the masking can cause spurious decrements when the maps are filtered.
However, this rather conservative procedure can result in rejecting a cluster
detection that was only marginally affected by the nearby emissive
source.  

To test this scenario, we re-ran the cluster finding algorithm on all the fields
used in this work with only the very brightest ($S_{150 \mathrm{GHz}}>50$~mJy) 
sources masked (as compared to the $5 \sigma$ thresholds of roughly 
$6.4$ and $17$~mJy in the full-depth and shallow fields).  Each detection above
the $\xi$ threshold for this paper was visually inspected, and the vast majority 
were rejected as obvious point-source-related artifacts.  However, two objects 
were clearly real detections.  One of these detections, with $\xi=19.3$ (in a full-depth field),
is a known cluster (Abell S0295), and we include it in our catalog as 
SPT-CL~J0245-5302.  There are two $>5 \sigma$ sources within $8$~arcmin 
of this cluster, but neither is strong enough to affect the $\xi$ measurement 
by more than $1$ or $2 \sigma$.  Furthermore, the sources lie to the north and 
south of the cluster, and wings from the matched filter are predominantly along the 
scan direction (east-west in all SPT fields but one).  However, because this cluster
was not found by the original version of the cluster finding algorithm, we do not 
include it any cosmological analysis.

The other detection, 
$\xi=13.3$ (full-depth), has an $8 \sigma$ source almost directly to the east.  
Because this cluster 
(SPT-CL~J2142-6419, which will likely appear in a future catalog) 
could have been bumped over our $\xi=13$ full-depth threshold by the filtering wing
of the point source, we choose not to include it in this work.

\section{Simulations}
\label{sec:sims}

Simulated observations were used to characterize the catalog presented in this work. These simulations were modeled after those used in \citetalias{vanderlinde10}, where full details can be found; a brief summary is provided here.

We create random Gaussian realizations of the power spectra of primary CMB 
anisotropy, kSZ, and point sources below the SPT detection threshold.
The primary CMB power spectrum is chosen to match the WMAP7 best-fit $\Lambda$CDM model \citep{larson10},
and we add the \citet{sehgal10} predicted kSZ power spectrum to the expected CMB anisotropy.
The point-source power spectrum includes
terms corresponding to synchrotron-dominated sources and DSFGs with amplitudes based on the results of \citet{shirokoff11}.
We include Gaussian realizations of the power from Poisson distributions of both sources with amplitudes at $150\,$GHz and $k = 3000$ of $D_{3000}^\mathrm{r} = 1.3\,\mu{\rm K^2}$ for the synchrotron sources and $D_{3000}^\mathrm{dsfg} = 7.7\,\mu{\rm K^2}$ for the DSFGs.
We assume a spectral index of $\alpha=-0.6$ for the synchrotron sources and $\alpha=3.6$ for DSFGs.
In addition to the Poisson power, we model a clustered DSFG component with an angular multipole dependence of $D_k \propto k$ and amplitude at $150\,$GHz and $k = 3000$ of $D_{3000}^c = 5.9\,\mu{\rm K^2}$.
SZ skies were simulated as in \citetalias{vanderlinde10}
using the methodology of \citet{shaw09}; the fiducial simulations from that work are used here. Briefly, they consist of semi-analytic gas models pasted over halos identified in N-body dark matter simulations.
At each frequency ($95$ and $150$~GHz), forty 100 \degs\  sky maps were generated.

These simulated skies were processed with an analytic approximation to the SPT transfer function, consisting of a Gaussian beam, an isotropic high-pass filter and a high-pass filter along the R.A.\  direction. The filters were arranged to model the effect of the SPT data processing described
in Section \ref{sec:data_proc}, with the R.A.\ high-pass set to match the $k_{\rm R.A.}=400$ cutoff of the 
Fourier mode removal
and the isotropic high-pass set at $k=500$ to approximate the spatial template removal.

Noise realizations were generated at each frequency for both the full survey and preview depths.
The noise power was measured in differenced (jackknife) maps for each field, and these powers were averaged across the set of fields at each frequency and depth. Forty Gaussian random realizations of each of these averages were then generated and added to the processed simulated sky maps.

These simulations were then subjected to the same cluster-finding pipeline applied to the real data, and recovered clusters were matched with the underlying catalog of massive halos associated with the SZ simulations.

\subsection{Mass Scaling Relations and Unbiased Mass Estimates}
\label{sec:sim_scaling}
A number of different techniques are available for obtaining cluster mass estimates from SZ measurements.  The integrated SZ flux, $Y$, is expected to be a tight proxy for the cluster mass \citep{barbosa96, holder01a, motl05, nagai07, shaw08, stanek09}. 
Unfortunately, the difficulty in determining the correct filter scale $\theta_c$ from SPT data alone adds significant scatter to the scaling relation of mass with $Y$ (see V10 for details).\footnote{Effort is currently underway to use our multi-frequency data to improve the determination of $\theta_c$.}  
In simulations, the SPT significance $\xi$ has a smaller scatter than our current integrated $Y$ estimates, and, as in V10, we use the significance as a mass proxy.

Due to the significant impact of noise biases, a direct $\xi$-M scaling relation is complex and difficult to characterize. Instead, following the prescription of \citetalias{vanderlinde10}, we introduce an unbiased significance $\zeta$, whose scaling with mass $M_{200}(\rho_{\mathrm{mean}})$ takes the form
\begin{equation}
\label{eq:mass_scaling}
\nSN = A \left(\frac{M}{5\times10^{14}\,M_\odot h^{-1}}\right)^B \left(\frac{1+z}{1.6}\right)^C,
\end{equation}
where $A$ is a normalization, $B$ a mass evolution and $C$ a redshift 
evolution.\footnote{Note that, for consistency with V10, this relation is given in terms of
$h$ (i.e., $h_{100}$), not $h_{70}$}
In simulated
maps of both depths, $\zeta$ was calculated for each cluster as in V10, by determining 
the preferred filter scale and cluster position in the absence of noise then averaging the
detection significance at that filter scale and position over many noise realizations.
Mass scaling relations were fit to the subset of these with $M > 2 \times 10^{14} h^{-1} M_\odot$ and 
$z>0.3$ by minimizing the residual logarithmic scatter in $\zeta$ about the 
relation.\footnote{The redshift cutoff is due to the fact that it was found in V10 that the power-law parametrization of the scaling relation fails to fully capture the behavior of the SPT selection function below $z=0.3$.} These relations are given in Table \ref{tab:sr}. As these are based on the same SZ simulations used in \citetalias{vanderlinde10}, they can be viewed as equivalent to the relations presented in that work.

\begin{deluxetable}{lcccc}
\tablecaption{Mass Scaling Relations\label{tab:sr}}
\tablewidth{0pt}
\tablehead{
\colhead{Depth}		& 
\colhead{A}		& 
\colhead{B}		& 
\colhead{C}		& 
\colhead{Scatter}
}
\startdata
Full survey	& 7.50	& 1.32	& 1.64	& 0.21 \\
Preview		& 3.50	& 1.29	& 0.87	& 0.16
\enddata
\end{deluxetable}

Uncertainties in the SZ modeling lead to significant systematic uncertainties on these scaling relation parameters. Following \citetalias{vanderlinde10}, we apply conservative 30\%, 20\%, 50\%, 20\% Gaussian uncertainties to A, B, C, and scatter, respectively.

Mass estimates are constructed as in \citetalias{vanderlinde10} with slight modifications to account for the different field depths.  Details of the conversion from $\xi$ to mass are given in Appendices
B and D of V10.
Briefly, we calculate the conditional probability of detecting a cluster of mass $M$
at a given value of $\xi$, $P(\xi | M)$, and then apply a mass-function prior to create
the posterior probability $P(M | \xi)$.
This procedure accounts for two types of bias in the mass estimate, the first 
due to the fact that we have maximized $\xi$ over many filter choices and 
positions in the map, and the
second due to the combination of the steepness of the cluster mass 
function and observational noise or scatter in the mass-observable scaling-relation.
The latter effect, which results in more low-mass systems scattering up 
into a given $\xi$ bin than high-mass systems scattering down, is 
related to the phenomenon of Eddington bias\footnote{Strictly speaking, ``Eddington bias" refers to
the bias in number counts caused by this asymmetric scatter.  The bias in 
the measured properties of individual objects is sometimes erroneously referred
to as ``Malmquist bias".}
\citep{eddington13}.
For the very 
high detection significances used in this work, the maximization bias
is completely negligible compared to the bias due to this asymmetric scatter.
As in V10, we use the \citet{tinker08} mass function evaluated at the maximum
likelihood point in the WMAP7+V10 chain as our prior.  This method produces
unbiased posterior estimates for cluster masses, assuming the validity of the 
simulations and of the various priors applied.


In \citet{andersson10}, we compared SZ inferred masses
calculated with this method to X-ray mass estimates  
for the 15 clusters from V10 that had X-ray measurements.
Overall, we found agreement between the 
SZ and X-ray mass estimates near the quoted level of the
systematic uncertainties of the SZ mass estimates.  However, 
there was a significant statistical offset, with the SZ-inferred masses 
lower by a factor of $0.78 \pm 0.06$ averaged over 
the sample, which we do not correct for in the masses in Table \ref{tab:clusters}.
This factor could have a redshift or mass 
dependence and naively applying a correction factor would ignore these effects.  
There was some evidence for this in \citet{andersson10}, where 
the lowest redshift and most massive cluster had the most 
discrepant SZ and X-ray inferred masses.  
We are currently pursuing an analysis
that jointly constrains the SZ and X-ray mass-observable 
relations with cosmology \citep{benson11}, which will 
more accurately quantify any systematic offset between the SZ and 
X-ray mass estimates.  For this work, 
we consider the quoted uncertainty of the SZ mass estimates to 
be a reasonable estimate of the systematic uncertainty, and note 
that there is some evidence that the SZ mass estimates are low 
by $\sim$25\%.  We show in Section \ref{sec:discussion} that none of the
conclusions in our cosmological analyses would change if we naively applied
this scaling factor.

Finally, we note that, although the scaling relation fits were only performed on 
simulated clusters above $z=0.3$, we nevertheless report SZ-derived masses
for several $z<0.3$ clusters in Table \ref{tab:clusters}.  These mass estimates
are extrapolations of the scaling relations to areas of parameter space in which
they have not been tied to simulations and may therefore be subject 
to further systematic uncertainties.  For this reason, we do not use any clusters
at $z<0.3$ in the likelihood calculations described in Section \ref{sec:fnl}.






\subsection{Purity and Completeness}
\label{sec:sim_purity}

To test the likelihood of false detections, simulated observations were generated omitting the SZ signal, and run through the cluster-finding pipeline. The false detection rate was found to be a rapidly falling function of the detection threshold. No false detections were found above a significance of $\xi=6$ in simulations of 4000 deg$^2$ at either depth. Given that the lowest threshold used in generating the catalog presented in this work is $\xi \ge 7$, it is highly improbable that it contains any false detections. This is confirmed by the multi-band followup, which shows counterparts for each cluster in the catalog.




The catalog is complete above threshold $\xi$ values by construction. As discussed in 
Section \ref{sec:catalog}, two factors make it difficult to quantitatively convert this into a mass and redshift completeness. 
The clusters in this catalog lie at the extreme high-mass end, and, as such, are rare in the simulated skies, yielding insufficient statistics to obtain a robust estimate of their detectability. Furthermore, there is large uncertainty associated with modeling the gas attached to halos in the simulation that makes any threshold uncertain at the $\sim30\%$ level.  Figure \ref{fig:selfun}
shows our best estimate of $50\%$ and $90\%$ completeness for the two sets of field depths
and $\xi$ thresholds.
We note that, as the same simulated SZ skies were used for both depths,
the variance due to the limited sample size will appear as a coherent shift,
rather than a scatter, between the two sets of curves


\section{Optical and Infrared Data}
\label{sec:optical}

\begin{figure}[]
\begin{center}
\includegraphics[scale=0.6]{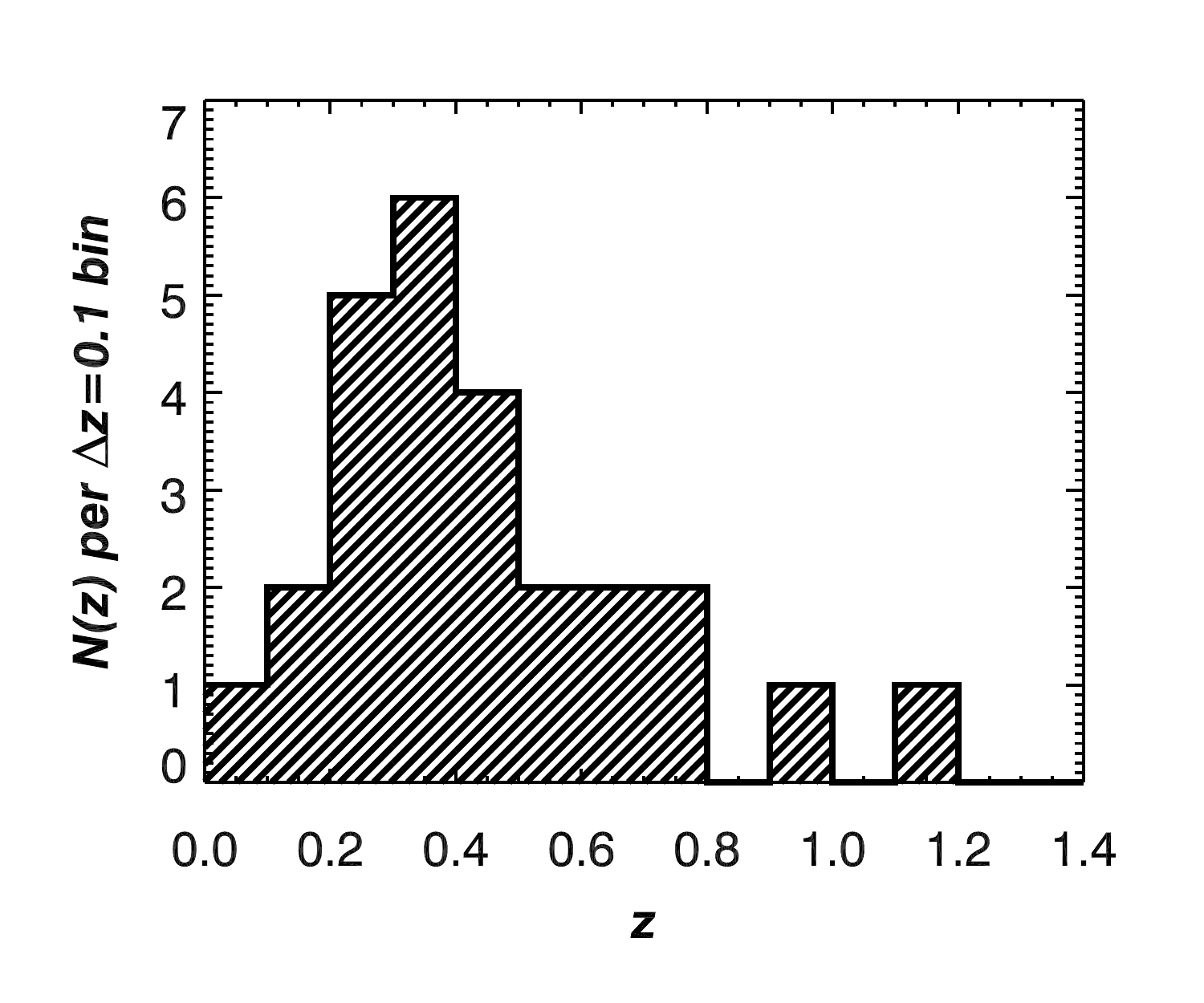}
  \caption[]{Redshift histogram of the sample.}
\label{fig:zhist}
\end{center}
\end{figure}

Multi-band imaging from both ground- and space-based facilities has been obtained
for clusters in the catalog, for the purpose of cluster redshift
estimation where no previous spectroscopic redshift could be
determined from the literature. We have also carried out new
multi-slit spectroscopy on some of the clusters.
A summary of the cluster redshifts is shown in Table
\ref{tab:redshift}, and a description of our methods is outlined
below.
Figure \ref{fig:zhist} shows the redshift histogram of our cluster sample.

\ifthenelse{\equal{\ispreprint}{true}}{ 
\begin{deluxetable*}{llllll}
}{
\begin{deluxetable}{llllll}
}
\tablecaption{Cluster Redshift Data\label{tab:redshift}}
\tablewidth{0pt}
\tablehead{
\colhead{Object Name} & 
\colhead{$z_{\textrm{spec}}$} & 
\colhead{Spectroscopy Ref.\tablenotemark{a}} & 
\colhead{$\#$ members} & 
\colhead{$z_{\textrm{rs}}$} & 
\colhead{Imaging Ref.\tablenotemark{b}} 
}
\startdata
 SPT-CL J0040-4407 &   \nodata &              \nodata & \nodata &      0.40 &                  im1 \\ 
 SPT-CL J0102-4915 &   \nodata &              \nodata & \nodata &      0.78 &                  im1 \\ 
 SPT-CL J0232-4421 &     0.284 &   \citet{degrandi99} & \nodata &   \nodata &              \nodata \\ 
 SPT-CL J0234-5831 &     0.415 &                  sp2 &    21 &      0.44 &             im3, im4 \\ 
 SPT-CL J0243-4833 &   \nodata &              \nodata & \nodata &      0.53 &             im1, im4 \\ 
 SPT-CL J0245-5302 &     0.300 &       \citet{edge94} & \nodata &      0.35 &                  im4 \\ 
 SPT-CL J0254-5856 &     0.438 &                  sp2 &    32 &      0.43 &             im3, im4 \\ 
 SPT-CL J0304-4401 &   \nodata &              \nodata & \nodata &      0.52 &                  im1 \\ 
 SPT-CL J0411-4819 &   \nodata &              \nodata & \nodata &      0.42 &                  im4 \\ 
 SPT-CL J0417-4748 &   \nodata &              \nodata & \nodata &      0.62 &             im1, im7 \\ 
 SPT-CL J0438-5419 &   \nodata &              \nodata & \nodata &      0.45 &             im1, im4 \\ 
 SPT-CL J0549-6204 &   \nodata &              \nodata & \nodata &      0.32 &                  im1 \\ 
 SPT-CL J0555-6405 &   \nodata &              \nodata & \nodata &      0.42 &                  im1 \\ 
 SPT-CL J0615-5746 &     0.972 &                  sp3 &     1 &       1.0 &             im1, im6 \\ 
 SPT-CL J0628-4143 &     0.176 &   \citet{degrandi99} & \nodata &      0.21 &                  im1 \\ 
 SPT-CL J0638-5358 &     0.222 &   \citet{degrandi99} & \nodata &   \nodata &              \nodata \\ 
 SPT-CL J0645-5413 &     0.167 &   \citet{degrandi99} & \nodata &   \nodata &              \nodata \\ 
 SPT-CL J0658-5556 &     0.296 &     \citet{tucker98} & \nodata &   \nodata &              \nodata \\ 
 SPT-CL J2023-5535 &     0.232 & \citet{boehringer04} & \nodata &      0.23 &        im2, im3, im4 \\ 
 SPT-CL J2031-4037 &     0.342 & \citet{boehringer04} & \nodata &   \nodata &              \nodata \\ 
 SPT-CL J2106-5844 &     1.132 & sp4, see \citet{foley11} &    18 &      1.17 &        im5, im6, im7 \\ 
 SPT-CL J2201-5956 &     0.098 &    \citet{struble99} & \nodata &           &             im2, im6 \\ 
 SPT-CL J2248-4431 &     0.348 & \citet{boehringer04} & \nodata &   \nodata &              \nodata \\ 
 SPT-CL J2325-4111 &   \nodata &              \nodata & \nodata &      0.37 &                  im1 \\ 
 SPT-CL J2337-5942 &     0.776 & sp5; also sp1, see \citetalias{high10} &    20 &      0.77 & im2, see \citetalias{high10}; im7 \\ 
 SPT-CL J2344-4243 &   \nodata &              \nodata & \nodata &      0.62 &                  im1  
\enddata
\tablecomments{Spectroscopic and red-sequence redshift information for the cluster sample.}
\tablenotetext{a}{Cross-reference to spectroscopic redshift data from external and internal sources.  Internal references are denoted ``sp\#'' and refer to Table \ref{tab:oirspec}.}
\tablenotetext{b}{Cross-reference to broadband redshift data from external and internal sources.  Internal references are denoted ``im\#'' and refer to Table \ref{tab:oirim}.}
\ifthenelse{\equal{\ispreprint}{true}}{ 
\end{deluxetable*}
}{
\end{deluxetable}
}

\subsection{Ground-based Imaging}

The previously unknown clusters in this catalog were imaged with the
cameras shown in Table \ref{tab:oirim}. 
The Swope $1\,$m at Las Campanas Observatory, equipped with
the SITe3 CCD detector and $BVRI$ filters, provided
sensitivity to clusters at $z\lesssim0.7$. 
We chose at least two passbands that spanned the 4000~\AA\ break,
as determined initially using cluster redshifts estimated by eye from
Digitized Sky Survey (DSS) images.  We required a detection of
$0.4L^*$ early-type 
cluster galaxies at about $5\sigma$.  We then iterated on this
strategy: if the cluster was still not sufficiently detected in the
initial set of exposures, we updated our best redshift estimate using
the new data, if possible, and reobserved.

The Blanco $4\,$m at Cerro Tololo Inter-American Observatory and
the Magellan $6.5\,$m telescopes at Las Campanas Observatory, provided
sensitivity to clusters at $z\lesssim1$.  The Blanco/MOSAIC-II,
Magellan/LDSS3, and  
Magellan/IMACS CCD cameras were used with $griz$ filters.  
The same iterative strategy
was implemented until we reached our required detection.

As detailed in \citet{foley11}, 
50~min of preimaging with the VLT's FORS2 camera in $I$ provided
additional broadband data for our highest-redshift cluster,
SPT-CL~J2106-5844, in our corresponding spectroscopic program.  




All images were reduced in a uniform manner using the same software
and methods described in detail in the previous, closely related work
of \citet[][hereafter H10]{high10}.  Photometry was calibrated using the Stellar Locus
Regression method \citep{high09}.  The red-sequence redshifts reported in Table
\ref{tab:redshift} were derived using the same red-sequence software
described in \citetalias{high10}.  These redshifts are estimated to be accurate
to $\sigma_z/(1+z)\approx 2$--$3\%$ (statistical plus sytematic),
as determined using a larger
subset of clusters with added spectroscopic redshift data.  An
exception to this are the SWOPE derived redshifts (SPT-CL~J0245-5302
and SPT-CL~J0411-4819) which are accurate
to $\sigma_z/(1+z)\approx 4$--$5\%$.  This was due to using Johnson
filters (as opposed to Gunn-Sloan filters), and calibrating the photometry
with a synthetic stellar locus (rather than the median SDSS stellar locus
from \citet{covey07}) using the PHOENIX stellar model atmosphere library \citep{brott05}.

A parallel reduction of the Blanco/MOSAIC-II data was performed using
the Dark Energy Survey data management system \citep{ngeow06,mohr08}.
Redshifts for the 12 clusters with MOSAIC-II imaging were
independently measured from these data using Artificial Neural Network
\citep[ANN$z$][]{collister04} software and another red-sequence method (Song et al, in prep).
This independent cross-checking led to consistent 
redshift estimates in all cases, except for SPT-CL~J0615-5746, where redshift 
estimates ranged from about $0.9$ to $1.1$ (this cluster was later spectroscopically
confirmed at $z=0.972$); and SPT-CL~J0555-6405, where 
different estimates gave redshifts of $0.27$, $0.35$, and $0.42$.  In the latter
case, because the red sequence appears most pronounced and
unambiguous at $0.42$, we report the redshift from the \citetalias{high10}
software 
only.

An additional cross-check for two of the clusters presented here is 
provided by \citet{menanteau10b}, who estimated photometric
redshifts of $0.75\pm 0.04$ for SPT-CL~J0102-4915 (compare our result
of $0.78$), and $0.54\pm 0.05$ for SPT-CL~J0438-5419 (compare $0.45$).

\subsection{Spitzer Space Telescope Imaging}
\label{sec:spitzer}

\spitzer/IRAC imaging is particularly important for the confirmation
and study of high-redshift SPT clusters, such as SPT-CL~J2106-5844 at
$z=1.132$ (Figure \ref{fig:thumb21}), where the optically faint members
are strongly detected in the mid-infrared. Three of our catalog
clusters were observed as part of a larger program to follow up
clusters identified in the SPT survey. The on-target observations
consisted of $8\times100$\,s and $6\times30$\,s dithered exposures at
3.6 and $4.5\,\mu$m, respectively. The deep 3.6\,$\mu$m observations
are sensitive to passively evolving cluster galaxies down to 0.1 $L^*$
at $z = 1.5$.  The data were reduced exactly as in \citet{brodwin10},
following the method of \citet{ashby09}.  Briefly, we correct for
column pulldown and residual image effects, mosaic the individual exposures, resample to
0\farcs86 pixels (half the solid angle of the native IRAC pixels), and
reject cosmic rays.



\ifthenelse{\equal{\ispreprint}{true}}
{
\begin{deluxetable*}{llllllcl}
}{
\begin{deluxetable}{llllllcl}
}
\tabletypesize{\scriptsize}
\tablecaption{Optical and infrared imagers\label{tab:oirim}}
\tablewidth{0pt}
\tablehead{
\colhead{Alias\tablenotemark{a}} &
\colhead{Site} &
\colhead{Telescope} & 
\colhead{Aperture} & 
\colhead{Camera} &
\colhead{Filters\tablenotemark{b}} &
\colhead{Field} &
\colhead{Pixel scale} \\
\colhead{~} &
\colhead{~} &
\colhead{~} &
\colhead{(m)} &
\colhead{~} &
\colhead{~} &
\colhead{~} &
\colhead{($\arcsec$)}
}
\startdata
im1 & Cerro Tololo & Blanco & 4 & MOSAIC-II & $griz$ & $36\arcmin \times 36\arcmin$ & $0.27$ \\   
im2 & Las Campanas & Magellan/Baade & 6.5 & IMACS f/2 & $griz$ & $27\farcm4 \times 27\farcm4$ & $0.200$ \\  
im3 & Las Campanas & Magellan/Clay & 6.5 & LDSS3 & $griz$ & $8\farcm3$ diam.\ circle  & $0.189$ \\ 
im4 & Las Campanas & Swope & 1 & SITe3 & $BVRI$ & $14\farcm8 \times 22\farcm8$ & $0.435$ \\ 
im5 & Paranal & VLT & 8.2 & FORS2 & $I$ & $6\farcm8 \times 6\farcm8$ & $0.25$ \\ 
im6 & Cerro Tololo & Blanco & 4 & NEWFIRM & $JK_s$ & $28\arcmin \times 28\arcmin$ & $0.4$ \\   
im7 & \nodata & Spitzer Space Telescope & 0.85 & IRAC & $[3.6][4.5]$ & $5\farcm2 \times 5\farcm2$ & $1.2$
\enddata
\tablecomments{The optical and infrared cameras used. The choice of facilities and filters for any given cluster was typically optimized according to our best redshift estimate prior to observation.  Not all imagers, nor all the listed filters, were used on each cluster.}
\tablenotetext{a}{Shorthand alias used in Table \ref{tab:redshift}.}
\tablenotetext{b}{The filters we used, which were in general a subset of all of those available. We did not typically use all listed filters on each cluster.}
\ifthenelse{\equal{\ispreprint}{true}}
{
\end{deluxetable*}
}{
\end{deluxetable}
}

\subsection{Spectroscopy}
\label{sec:spec}

Eleven of the clusters in this work have published spectroscopic
redshifts, which we note in Table \ref{tab:redshift}.
Using the instruments listed in Table \ref{tab:oirspec}, 
we present new spectroscopic redshift measurements on five clusters,
four of which have no such previously published data.
The robust biweight location estimator is used to determine the
cluster spectroscopic redshifts from ensembles of member galaxies.

\ifthenelse{\equal{\ispreprint}{true}}
{
\begin{deluxetable*}{llllll}
}{
\begin{deluxetable}{llllll}
}
\tablecaption{Optical and infrared spectrographs\label{tab:oirspec}}
\tablewidth{0pt}
\tablehead{
\colhead{Alias\tablenotemark{a}} &
\colhead{Site} &
\colhead{Telescope} & 
\colhead{Aperture} & 
\colhead{Camera} &
\colhead{Mode} \\
\colhead{~} &
\colhead{~} &
\colhead{~} &
\colhead{(m)} &
\colhead{~} &
\colhead{~}
}
\startdata
sp1 & Las Campanas & Magellan/Clay & 6.5 & LDSS3 & longslit \\ 
sp2 & Las Campanas & Magellan/Baade & 6.5 & IMACS & GISMO \\ 
sp3 & Las Campanas & Magellan/Baade & 6.5 & IMACS & longslit \\ 
sp4 & Paranal & VLT & 8.2 & FORS2 & MOS \\
sp5 & Cerro Pachon & Gemini South & 8.1 & GMOS & MOS
\enddata
\tablecomments{The spectrographs used.}
\tablenotetext{a}{Shorthand alias used in Table \ref{tab:redshift}.}
\ifthenelse{\equal{\ispreprint}{true}}
{
\end{deluxetable*}
}{
\end{deluxetable}
}

\subsubsection{SPT-CL~J0234-5831 and SPT-CL~J0254-5857}

In a procedure similar to \citet{brodwin10}, multislit spectroscopic
observations were acquired on the 6.5-meter Baade Magellan telescope
on UT 2010 October 8 for SPT-CL~J0234-5831 and SPT-CL~J0254-5857.
Measurements were made using the Gladders Image-Slicing Multislit
Option (GISMO, Gladders et al.\ in prep.)  module on the
Magellan/IMACS spectrograph.  GISMO optically remaps the central
region of the IMACS field of view (roughly $3.5' \times 3.2'$) to
sixteen evenly-spaced regions of the focal plane, allowing for a large
density of slitlets in the cluster core while minimizing trace
overlaps on the CCD.

In designing the multislit mask, galaxies were assigned a weight
proportional to their r-band brightness and adjusted for their
position in color space with respect to a manually-selected red
sequence. The f/4 camera, the 300 l/mm grating and the $z1 430-675$
filter were used. Each cluster was observed with three 30-min exposures 
of one mask in good seeing ($\sim 0.6"$).

The COSMOS reduction package was used for standard CCD processing,
resulting in wavelength-calibrated 2D spectra. The 1D spectra were
then extracted from the sum of the reduced data. Secure redshifts were
obtained for 21 member galaxies of SPT-CL~J0234-5831, and 32 member
galaxies of SPT-CL~J0254-5857.

\subsubsection{SPT-CL~J0615-5746}

Longslit spectroscopy of SPT-CL~J0615-5746 was performed on 
UT 2011 March 8, also with the IMACS spectrograph on the Baade 
Magellan telescope. The longslit was aligned across several 
objects and yielded clear redshifts for the BCG and a second 
cluster member. The reported redshift is that of the BCG.

\subsubsection{SPT-CL~J2106-5844}

We refer the reader to \citet{foley11} for a detailed discription of
spectroscopic measurements of SPT-CL~J2106-5844.  In short, the
redshift given in Table \ref{tab:redshift} is derived from 18 member
galaxies using VLT/FORS2 and Magellan/IMACS-GISMO.
 


\subsubsection{SPT-CL~J2337-5942}

The redshift of SPT-CL~J2337-5942 reported in Table
\ref{tab:clusters}, $z_{\textrm{spec}}=0.776$, is from 
combined
measurements of 19 cluster members using GMOS on the $8.1$~m Gemini
South telescope and 2 members using the Magellan/LDSS3 longslit---one 
of which overlaps with a GMOS member---for a total of 20 cluster
members. The LDSS3 data were described in detail in
\citetalias{high10}, where $z_{\textrm{spec}}=0.781$ was reported from
the two members.

For the new GMOS observations we are presenting here, 
galaxies with $r - i$ color consistent with a cluster
red-sequence at $z = 0.77$, and having non-stellar PSFs in the Gemini
$i$-band pre-image, were used to populate two masks. A total of 31
galaxies were observed for three hours with the R150\_G5326 grism and
the GG455-G0329 filter.
The IRAF Gemini reduction package was used for standard CCD
processing. The wavelength-calibrated 2D spectra were sky-subtracted
using an in-house routine, after which the 1D spectra were extracted
from the coadded 2D spectra.  Secure redshifts for the 19 cluster
members were obtained from the GMOS masks.  All the spectra are from
early-type galaxies, often exhibiting a very strong Ca~H\&K absorption
feature.  

\section{X-ray Data}

\label{sec:xray}

For each cluster in Table \ref{tab:clusters}, we searched the 
ROSAT data archive for possible X-ray 
counterparts, including the REFLEX catalog \citep{bohringer04}, the ROSAT All-Sky Survey Bright 
Source Catalog \citep{voges99}, and the ROSAT All-Sky Survey Faint 
Source Catalog\footnote{
http://heasarc.gsfc.nasa.gov/W3Browse/rosat/rassfsc.html}.
We also used data from the ROSAT All Sky Survey and pointed
Position Sensitive Proportional Counter (PSPC)
observations to measure the X-ray flux for the SPT clusters in
the ``hard" 0.6-2 keV energy band, since the signal to noise is
generally better in this band than in the full ROSAT energy range
\citep{vikhlinin98}. To determine the source counts, we used a source radius corresponding
to 2 Mpc, excluded any sources not associated with the cluster emission 
and used a nearby region for measuring the X-ray background. 
Significant detections were found for 25 
of the 26 clusters in the ROSAT observations.  
The highest-redshift cluster, SPT-CL~J2106-5844, was not detected by ROSAT, but has been 
detected in pointed observations with \chandra \citep{foley11}.

To determine the cluster flux and luminosity, we used 
PIMMS\footnote{http://cxc.harvard.edu/toolkit/pimms.jsp} to
determine the cluster unabsorbed flux in the observer's frame from the
ROSAT ``hard" band observations.   We then used the 
XSPEC\footnote{http://heasarc.nasa.gov/xanadu/xspec/xspec11/index.html}
``flux" and ``lumin" functions to determine the cluster luminosity in the
cluster rest frame, which we report in Table \ref{tab:clusters}. 
We report each in the 0.5-2.0
keV band because of this band's relative insensitivity to the assumed
X-ray temperature.  For example, the inferred flux changes by
$\lesssim$2\% when assuming a range of gas temperatures
between 6 to 10 keV.
For two clusters, SPT-CL~J2106-5944 and SPT-CL~J2337-5942, we give the 
flux and luminosity measured by \chandra and reported in \citet{foley11} 
and \citet{andersson10}, respectively.

\begin{figure}[]
\includegraphics[scale=0.60]{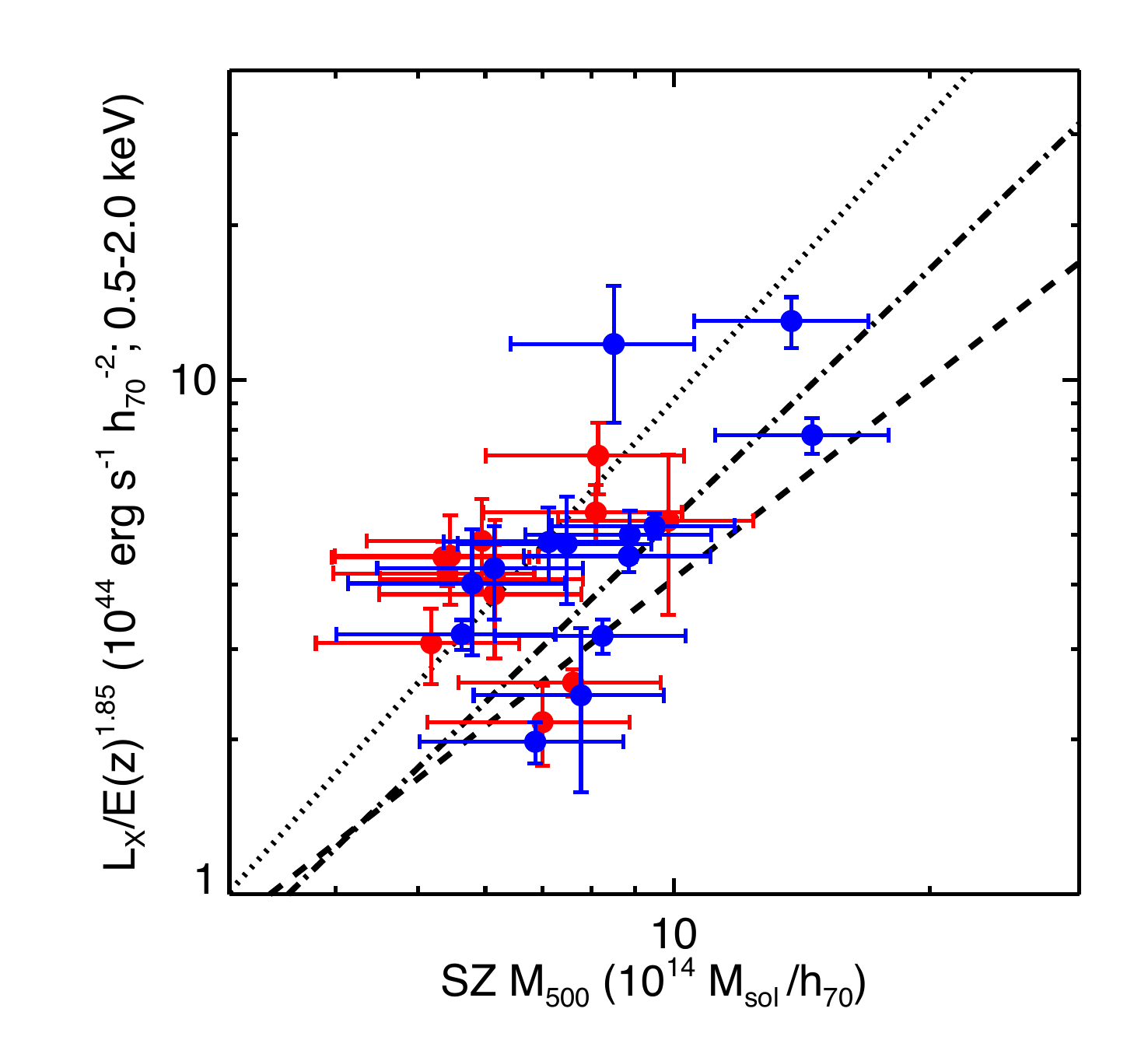}
  \caption[]{The X-ray luminosity and SZ inferred masses 
$M_{500}(\rho_{\mathrm{crit}})$ for our cluster sample.  We 
  plot statistical uncertainties only, and note that 
the statistical uncertainty of the SZ mass estimate is limited 
by the assumed scatter in the SZ significance-mass relation.
Clusters from the shallow fields are in blue, and clusters from the deep fields 
are in red.  We also show the best-fit relations of \citet{pratt09} (dotted), 
\citet{vikhlinin09b} (dash-dot), and \citet{mantz10} (dashed).}
\label{fig:xray}
\end{figure}

In Figure \ref{fig:xray}, we plot the X-ray luminosity, $L_X$, and the SPT measured
mass (converted from $M_{200}(\rho_{\mathrm{mean}})$ to $M_{500}(\rho_{\mathrm{crit}})$) 
from Table \ref{tab:clusters}.  
We plot only statistical uncertainties for both luminosity and mass.  
However, we note that we have ignored
several important systematic uncertainties in the ROSAT
X-ray luminosity measurements, including the effects of
unresolved point sources and cooling cores.  These can add biases and additional
scatter to the X-ray measurements, however cannot observationally
be accounted for because of the relatively large ROSAT beamsize.
Similar phenomena also affect the SZ measurements, however they are
accounted for statistically in the significance estimate and the scatter
in the significance-mass relation, as described 
in Section \ref{sec:sim_scaling}.  The
latter dominates the statistical uncertainty of the SZ mass estimates.
In Figure \ref{fig:xray}, we have assumed the best-fit redshift 
evolution in the $L_X$--$M$ relation measured by \citet{vikhlinin09b}, where 
$L_X \propto M E(z)^{1.85}$.  This is slightly different than the  
self-similar expectation that predicts an evolution of $E(z)^{2}$ 
for the luminosity in the 0.5-2.0 keV band and an evolution of 
$\sim E(z)^{7/3}$ for the bolometric luminosity, which has an 
additional weak dependence of luminosity with temperature.

In Figure \ref{fig:xray}, we also show 
the best-fit $L_X$--$M$ relations from \citet{pratt09}, 
\citet{vikhlinin09b}, and \citet{mantz10}, to compare our 
results with other X-ray cluster studies. 
Both \citet{pratt09} and \citet{mantz10} assume evolution consistent 
with the self-similar relation for bolometric luminosity.  For each 
result, we use their published normalization and slope, but have assumed the 
redshift evolution measured by \citet{vikhlinin09b}.  
Relative to self-similar evolution, this would cause a $<7\%$ 
change in the normalization for the typical redshift range of 
their cluster samples ($z < 0.3$).
\citet{mantz10} also quote their luminosities in the 
rest frame 0.1-2.4 keV band, which we have converted 
to 0.5-2.0 keV by dividing by a factor of 1.61.  This factor is appropriate for a 
cluster with a 8 keV electron temperature, and varies negligibly with 
temperature.  We expect that both approximations 
will not significantly change the normalization of either relation.  
As can be seen in Figure \ref{fig:xray}, there is a fairly large spread in the 
published $L_X$--$M$ relations, although the methods and samples differ 
between each work.  For example, they each use cluster samples with 
somewhat different mass and redshift ranges, and account
for the effects of Eddington bias differently.  
Regardless, we consider the agreement with other cluster 
samples reasonable, and also significant given the unique 
SZ selection of the clusters in this work.     
The X-ray selected cluster samples  
are of generally lower redshift, less massive, and have 
been corrected for Malmquist bias from the X-ray flux selection, a bias
that is completely absent for SZ selected samples.  
This work confirms that the SPT cluster sample consists of very 
massive clusters, which qualitatively follow the $L_X$--$M$  
relation measured from other X-ray selected cluster samples. 

\section{Discussion}
\label{sec:discussion}

The 26 highest-significance SZ-selected clusters from the
2500 \degs\ SPT survey (see Table \ref{tab:clusters}) include all the most
massive galaxy clusters in this region of the sky,
independent of the cluster redshift.  These exceedingly rare
systems populate the high end of the mass function at
each redshift, and predictions for the characteristics of this population
are sensitive to the details of the assumed cosmological model.  An
interesting first step in using the clusters presented in this work to constrain
cosmology is to ask whether their distribution 
in mass and redshift is consistent
with the predictions of the standard \LCDM\ cosmological model.
We investigate this question two ways.  First, we use the framework of 
\citet[][hereafter M11]{mortonson11} and the fitting functions they provide to ask whether the 
existence of any single cluster in our sample is in significant tension with 
\LCDM.  We then fit all available
cosmological data including this new cluster sample to two different cosmological
models---namely, standard \LCDM\ and a single-parameter extension allowing
for non-Gaussian initial conditions---and see if the data prefer the non-standard 
model.

\subsection{Single-cluster Tests}

\begin{figure}[]
\begin{center}
\includegraphics[scale=0.5]{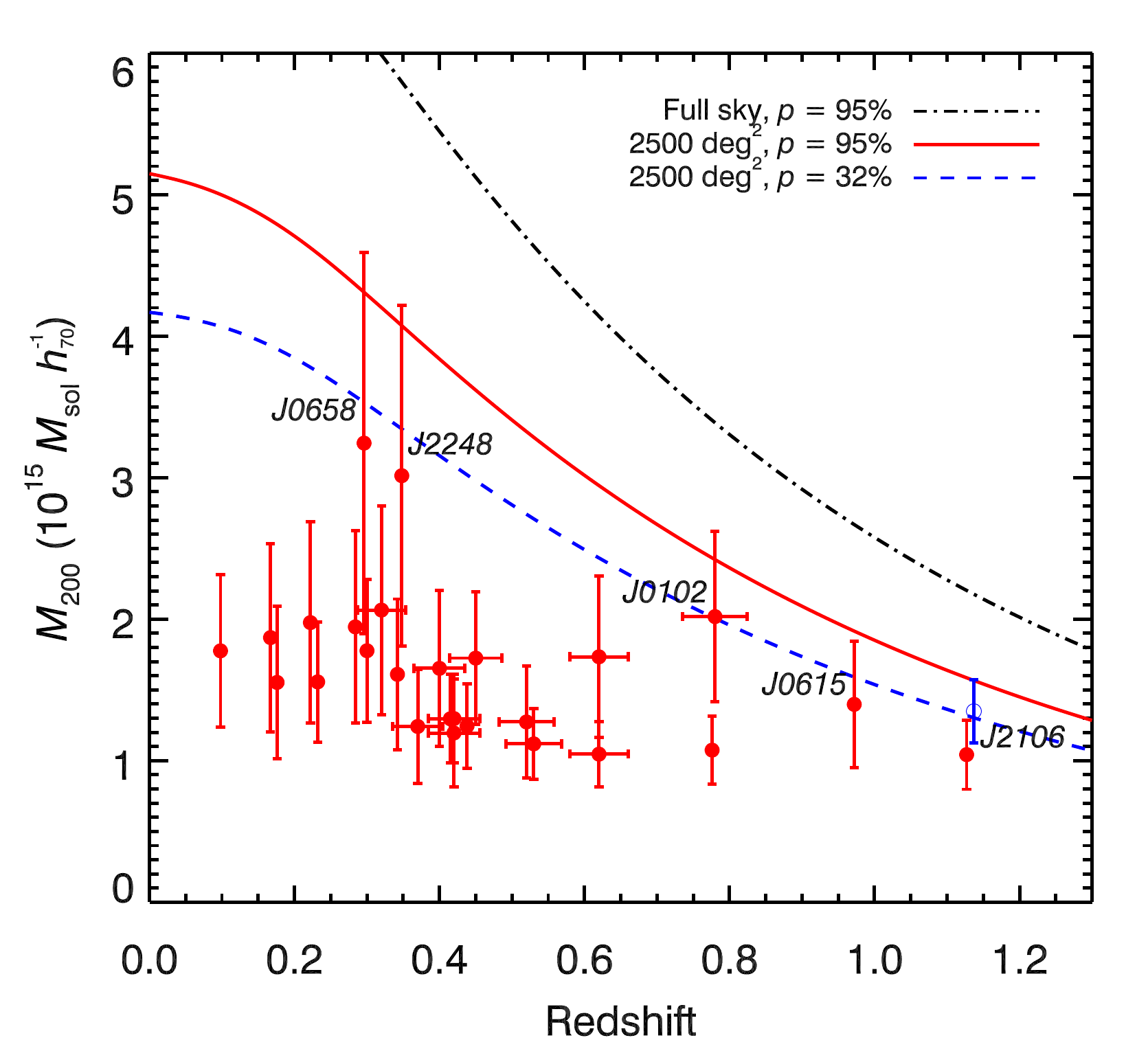}
  \caption[]{An M11-style plot showing the mass 
  $M_{200}(\rho_{\mathrm{mean}})$ and redshift of the clusters
  presented in this paper.  The masses in this plot are slightly 
  ($\sim$5\%) higher than those presented in Table \ref{tab:clusters},
  due to the slightly different treatment of mass bias appropriate to 
  the  M11 calculation (see text and M11
  for details).
  Some of the most extreme objects in the catalog 
  are annotated with the R.A.\ portion of their object name.  The 
  red solid line shows the mass above which 
  a cluster at a given redshift is less than $5\%$ likely to be found in the 
  2500 \degs\ SPT survey region in $95\%$ of the \LCDM\ parameter 
  probability distribution.  The black dot-dashed line shows the analogous limit for the full sky. 
  The blue open data point (redshift slightly offset for clarity)
  denotes the mass estimate for SPT-CL~J2106-5844 from combined X-ray and SZ
  measurements in \citet{foley11}.  That work concludes
  that this cluster is less than $5\%$ likely in $32\%$ of the \LCDM\
  parameter probability distribution, and we
  show the corresponding M11 $p=32\%$ limiting mass
  vs.\ redshift as the dashed blue line.  }
\label{fig:mortplot}
\end{center}
\end{figure}

M11 have published fitting functions that allow us to answer 
the question:  Is this one cluster in significant tension with \LCDM?  In Figure 
\ref{fig:mortplot}, we plot the 
mass vs.\ redshift 
for all 26 clusters and overplot exclusion curves from M11.  As explained
in Appendix C of M11, the mass for a given cluster that is appropriate to 
compare to their exclusion curves is not precisely the best posterior 
estimate of that cluster's mass.  The masses plotted in Figure \ref{fig:mortplot} 
are calculated using Equation C3 from M11, using the conditional 
probability $P(\xi | M)$ to estimate their $M_{\mathrm{obs}}$ and 
$\sigma_{\ln M}$ and using the local slope of the \citet{tinker08} mass
function around the value of $M_{\mathrm{obs}}$ for $\gamma$.  The 
error bars on the masses in Figure \ref{fig:mortplot} are calculated by 
setting the fractional error (i.e., the error in $\ln M$) equal to the fractional
error of the posterior mass estimates in Table \ref{tab:clusters}.
We have confirmed that this is an excellent approximation to the full
probability distribution of the M11-appropriate masses for the cluster 
sample in this work.

The two highest exclusion curves overplotted in Figure \ref{fig:mortplot}
represent the mass and redshift above which 
an individual cluster 
would be less than $5\%$ likely to be found in a given survey region
in $95\%$ of the \LCDM\ parameter probability distribution.
We plot one exclusion curve
for the least likely cluster allowed in a 2500 \degs\ survey and one curve for the least likely 
cluster allowed in the entire sky.  It is clear that, according to the 
formalism of M11, 
no cluster in our sample is individually in strong
tension with \LCDM---a conclusion that would still hold if
we applied the naive scaling factor of $1/0.78$ discussed 
in Section \ref{sec:sim_scaling} to all the cluster masses.  

This result can be compared to the result of \citet{foley11}, in 
which the single cluster SPT-CL~J2106-5844 
is found to be less than $5\%$ likely to exist in the 
2500 \degs\ SPT survey region
in $32\%$ of the \LCDM\ parameter probability distribution.
There are some differences in the two analyses, the most
important of which is that \citet{foley11} use a mass estimate that 
combines SZ and X-ray data, whereas this work only reports
an SZ-derived mass.
The central value of the \citet{foley11} combined SZ/X-ray mass estimate is 
$30\%$ higher than the central value of the SZ-derived mass reported here.
We have included the \citet{foley11} combined mass as a point in
Figure \ref{fig:mortplot}, and we have also plotted the M11
exclusion curve corresponding to $<5\%$ likelihood of finding a cluster
in the SPT survey in $32\%$ of parameter probability.  As expected from the result 
in \citet{foley11}, the $p=32\%$ 
exclusion curve nearly intersects the central SPT-CL~J2106-5844 mass value
from that work (adjusted appropriately for the M11 plot).

\subsection{Extensions to \LCDM}
\label{sec:fnl}

While no individual cluster lies above either $p=95\%$ exclusion line in Figure \ref{fig:mortplot}, 
there are several which come reasonably close.
One might imagine that the
collective ``unlikelihood" of these clusters could indicate the need to go beyond the standard 
\LCDM\ cosmological model.  The most straightforward extension to \LCDM\ that
could explain an excess of massive clusters (including ones at high redshift) is 
the possibility of a non-Gaussian component to the primordial density perturbations.
Different models of inflation predict different levels and types of
non-Gaussianity \citep[e.g.,][]{bartolo04}, with the 
size of the leading-order non-Gaussian term described by the parameter \fnl.
We have included the mass and redshift distribution of the clusters presented 
here in a cosmological likelihood calculation with and without \fnl\ as a free parameter.
The likelihood calculation was implemented as in V10, with the effect of \fnl\ on 
cluster abundance added following the prescription of \citet{dalal08}.  To simplify the selection
function and mass scaling part of the calculation, the preview-depth relation 
(see Table \ref{tab:sr}) was used for all clusters, and preview-depth
values of $\xi$ were estimated for the full-depth clusters by making coadded maps of 
only one ninth of the observations and running the cluster finder on these maps.
As in V10, the scaling relation is not expected to capture the correct behavior at low redshift;
as in V10, we exclude this regime by applying a hard cut $z>0.3$ in this analysis.

The preferred 
value of \fnl\ in the extended model is consistent with zero 
($\fnl = 30 \pm 450$ at $68\%$ confidence).
We note that this is a significantly weaker constraint than that found by 
\citet{komatsu11} using the CMB bispectrum as measured in the WMAP7 data, but 
that the two results are consistent with each other and with $\fnl=0$. This
is in tension with the recent results of \citet{cayon10}, \citet{hoyle10}, 
and \citet{enqvist10}, who found significant evidence for non-zero \fnl\ based on
other high redshift galaxy clusters using a different statistical technique. 
In contrast to those works, this analysis uses 
a likelihood analysis over the full range of mass and redshift space
including the SPT selection function, marginalizing over scaling relation 
uncertainties. This approach naturally incorporates information about both
the mass and redshift distribution and the total number of clusters. 
Our \fnl\ constraints do not change appreciably and are still fully consistent
with $\fnl=0$ if we re-run the analysis using a prior on the $\xi$-$M$ relation
that incorporates the scaling between SZ and X-ray masses discussed in 
Section \ref{sec:sim_scaling}.

%


\section{Conclusions}
\label{sec:conclusions}
We have presented a sample of the most massive galaxy clusters in a 2500 
\degs\ region of the sky, selected via their SZ signature in SPT observations.
These 26 clusters are selected from hundreds of SPT cluster candidates on the 
basis of their SZ detection significance, which has been shown in previous SPT 
analyses to correlate tightly with cluster mass \citep[V10,][]{andersson10}.  
As expected from their high SZ
significance, each one of these objects shows a strong overdensity of similarly 
colored galaxies in optical and/or infrared data, and the X-ray luminosity of 
these systems (as estimated from archival and newly collected data) is consistent
with their SZ-derived masses.

We measure (or collect from the literature) photometric---and, in some cases, 
spectroscopic---redshifts for these 26 clusters.
The cluster sample includes several newly discovered 
high-redshift systems, significantly increasing
the total number of known galaxy clusters with masses of 
$M_{200}(\rho_{\mathrm{mean}}) \gtrsim 10^{15}  \,M_\odot$ and redshifts
$z \gtrsim 0.5$.  In addition to being
interesting targets for studies of cluster physics and 
galaxy formation in the densest environments in the Universe, these massive, 
high-redshift clusters allow us to test the standard \LCDM\ cosmological
model with Gaussian initial conditions.

We test whether the most extreme (in mass or redshift) individual clusters pose a challenge to \LCDM\ by applying the formalism presented by M11. 
No single cluster is in significant tension with the \LCDM\ model, with the caveat that current SZ mass estimates are uncertain at the 30\% level. 
Improved mass estimates would strengthen any test of the \LCDM\ model.
We also examine constraints on primordial non-Gaussianity from the cluster sample. 
The data show no preference for non-Gaussianity as parametrized by $f_{NL}$.
At $68\%$ confidence, we find $\fnl = 20 \pm 450$.

The sample of clusters found with the SPT, the most massive of which are presented in this work, is complementary to the sample expected to be found with  the {\it Planck} satellite.   {\it Planck} will find very massive galaxy clusters over the entire sky---16 times more area than the SPT survey.
However, clusters above $z \sim 0.5$ will have a typical angular size of $1^\prime$, 
meaning that {\it Planck}'s sensitivity to clusters will fall off at high redshift
due to its larger beam ($5^\prime$ FWHM at $150\,$GHz vs.~$1^\prime$ for SPT). 
The clusters found by {\it Planck} will thus be 
distributed towards significantly lower redshift than those found by the SPT.

The SPT observations of the  2500 \degs\ of sky used in this work will be completed to the final survey depth in the Austral winter of 2011.
Extrapolating from current survey yields, the complete SPT SZ survey will 
contain roughly 750 galaxy cluster candidates at a detection
significance of $\xi \ge 4.5$ (and over 400 at $\xi \ge 5$), the vast 
majority of which will correspond to real, massive clusters.
This unique, nearly mass-limited cluster sample will offer an unprecedented opportunity to test the \LCDM\ cosmological model and the properties of dark energy.

\acknowledgments

{\it Facilities:}
\facility{Blanco (NEWFIRM)},
\facility{Blanco (MOSAIC)},
\facility{CXO (ACIS)},
\facility{Gemini-S (GMOS)},
\facility{Magellan:Baade (IMACS)},
\facility{Magellan:Clay (LDSS3)},
\facility{Spitzer (IRAC)},
\facility{South Pole Telescope},
\facility{VLT:Antu (FORS2)}

We thank Wayne Hu, Dragan Huterer, Eduardo Rozo, and an 
anonymous referee for helpful discussions and suggestions, 
and Ryan Chornock and Wen-fai Fong for assistance during 
spectroscopic observations.

The South Pole Telescope program is supported by the National Science
Foundation through grant ANT-0638937.  Partial support is also
provided by the NSF Physics Frontier Center grant PHY-0114422 to the
Kavli Institute of Cosmological Physics at the University of Chicago,
the Kavli Foundation, and the Gordon and Betty Moore Foundation.  This
work is based in part on observations obtained with the Spitzer Space
Telescope (PID 60099), which is operated by the Jet Propulsion
Laboratory, California Institute of Technology under a contract with
NASA.  Support for this work was provided by NASA through an award
issued by JPL/Caltech.  Additional data were obtained with the 6.5~m
Magellan Telescopes located at the Las Campanas Observatory,
Chile. Support for X-ray analysis was provided by NASA through Chandra
Award Numbers 12800071 and 12800088 issued by the Chandra X-ray
Observatory Center, which is operated by the Smithsonian Astrophysical
Observatory for and on behalf of NASA under contract NAS8-03060.
Optical imaging data from the Blanco 4~m at Cerro Tololo Interamerican
Observatories (programs 2005B-0043, 2009B-0400, 2010A-0441,
2010B-0598) and spectroscopic observations from VLT programs
086.A-0741 and 286.A-5021 and Gemini program GS-2009B-Q-16 were
included in this work.
We acknowledge the use of the Legacy Archive for
Microwave Background Data Analysis (LAMBDA).  Support for LAMBDA is
provided by the NASA Office of Space Science.  Galaxy cluster research
at Harvard is supported by NSF grant AST-1009012.  Galaxy cluster
research at SAO is supported in part by NSF grants AST-1009649 and
MRI-0723073.  The McGill group acknowledges funding from the National
Sciences and Engineering Research Council of Canada, Canada Research
Chairs program, and the Canadian Institute for Advanced Research.
X-ray research at the CfA is supported through NASA Contract NAS
8-03060.  The Munich group acknowledges support from the Excellence
Cluster Universe and the DFG research program TR33.  R.J.F.\ is
supported by a Clay Fellowship.  B.A.B\ is supported by a KICP
Fellowship, support for M.Brodwin was provided by the W. M. Keck
Foundation, M.Bautz acknowledges support from contract
2834-MIT-SAO-4018 from the Pennsylvania State University to the
Massachusetts Institute of Technology.  M.D.\ acknowledges support
from an Alfred P.\ Sloan Research Fellowship, W.F.\ and C.J.\
acknowledge support from the Smithsonian Institution, and B.S.\
acknowledges support from the Brinson Foundation.

\bibliography{preview}

\begin{thebibliography}{70}
\expandafter\ifx\csname natexlab\endcsname\relax\def\natexlab#1{#1}\fi

\bibitem[{{Abell} {et~al.}(1989){Abell}, {Corwin}, \& {Olowin}}]{abell89}
{Abell}, G.~O., {Corwin}, Jr., H.~G., \& {Olowin}, R.~P. 1989, \apjs, 70, 1

\bibitem[{{Andersson} {et~al.}(2010){Andersson}, {Benson}, {Ade}, {Aird},
  {Armstrong}, {Bautz}, {Bleem}, {Brodwin}, {Carlstrom}, {Chang}, {Crawford},
  {Crites}, {de Haan}, {Desai}, {Dobbs}, {Dudley}, {Foley}, {Forman},
  {Garmire}, {George}, {Gladders}, {Halverson}, {High}, {Holder}, {Holzapfel},
  {Hrubes}, {Jones}, {Joy}, {Keisler}, {Knox}, {Lee}, {Leitch}, {Lueker},
  {Marrone}, {McMahon}, {Mehl}, {Meyer}, {Mohr}, {Montroy}, {Murray}, {Padin},
  {Plagge}, {Pryke}, {Reichardt}, {Rest}, {Ruel}, {Ruhl}, {Schaffer}, {Shaw},
  {Shirokoff}, {Song}, {Spieler}, {Stalder}, {Staniszewski}, {Stark}, {Stubbs},
  {Vanderlinde}, {Vieira}, {Vikhlinin}, {Williamson}, {Yang}, \&
  {Zahn}}]{andersson10}
{Andersson}, K., {et~al.} 2010, submitted to \apj, arXiv:1006.3068

\bibitem[{{Ashby} {et~al.}(2009){Ashby}, {Stern}, {Brodwin}, {Griffith},
  {Eisenhardt}, {Koz{\l}owski}, {Kochanek}, {Bock}, {Borys}, {Brand}, {Brown},
  {Cool}, {Cooray}, {Croft}, {Dey}, {Eisenstein}, {Gonzalez}, {Gorjian},
  {Grogin}, {Ivison}, {Jacob}, {Jannuzi}, {Mainzer}, {Moustakas},
  {R{\"o}ttgering}, {Seymour}, {Smith}, {Stanford}, {Stauffer}, {Sullivan},
  {van Breugel}, {Willner}, \& {Wright}}]{ashby09}
{Ashby}, M.~L.~N., {et~al.} 2009, \apj,, 701, 428

\bibitem[{Barbosa {et~al.}(1996)Barbosa, Bartlett, Blanchard, \&
  Oukbir}]{barbosa96}
Barbosa, D., Bartlett, J., Blanchard, A., \& Oukbir, J. 1996, \aap, 314, 13

\bibitem[{{Bartolo} {et~al.}(2004){Bartolo}, {Komatsu}, {Matarrese}, \&
  {Riotto}}]{bartolo04}
{Bartolo}, N., {Komatsu}, E., {Matarrese}, S., \& {Riotto}, A. 2004, \physrep,
  402, 103

\bibitem[{{Battye} \& {Weller}(2003)}]{battye03}
{Battye}, R.~A., \& {Weller}, J. 2003, \prd, 68, 083506

\bibitem[{{Benson} {et~al.}(2011){Benson}, {Ade}, {Aird}, {Andersson},
  {Armstrong}, {Bautz}, {Bleem}, {Brodwin}, {Carlstrom}, {Chang}, {Crawford},
  {Crites}, {de Haan}, {Desai}, {Dobbs}, {Dudley}, {Foley}, {Forman},
  {Garmire}, {George}, {Gladders}, {Halverson}, {High}, {Holder}, {Holzapfel},
  {Hrubes}, {Jones}, {Joy}, {Keisler}, {Knox}, {Lee}, {Leitch}, {Lueker},
  {Marrone}, {McMahon}, {Mehl}, {Meyer}, {Mohr}, {Montroy}, {Murray}, {Padin},
  {Plagge}, {Pryke}, {Reichardt}, {Rest}, {Ruel}, {Ruhl}, {Schaffer}, {Shaw},
  {Shirokoff}, {Song}, {Spieler}, {Stalder}, {Staniszewski}, {Stark}, {Stubbs},
  {Vanderlinde}, {Vieira}, {Vikhlinin}, {Williamson}, \& {Zahn}}]{benson11}
{Benson}, B.~A., {et~al.} 2011, In prep.

\bibitem[{{B{\"o}hringer} {et~al.}(2004{\natexlab{a}}){B{\"o}hringer},
  {Schuecker}, {Guzzo}, {Collins}, {Voges}, {Cruddace}, {Ortiz-Gil},
  {Chincarini}, {De Grandi}, {Edge}, {MacGillivray}, {Neumann}, {Schindler}, \&
  {Shaver}}]{bohringer04}
{B{\"o}hringer}, H., {et~al.} 2004{\natexlab{a}}, \aap, 425, 367

\bibitem[{{B{\"o}hringer} {et~al.}(2004{\natexlab{b}}){B{\"o}hringer},
  {Schuecker}, {Guzzo}, {Collins}, {Voges}, {Cruddace}, {Ortiz-Gil},
  {Chincarini}, {De Grandi}, {Edge}, {MacGillivray}, {Neumann}, {Schindler}, \&
  {Shaver}}]{boehringer04}
------. 2004{\natexlab{b}}, \aap, 425, 367

\bibitem[{{Brodwin} {et~al.}(2010){Brodwin}, {Ruel}, {Ade}, {Aird},
  {Andersson}, {Ashby}, {Bautz}, {Bazin}, {Benson}, {Bleem}, {Carlstrom},
  {Chang}, {Crawford}, {Crites}, {de Haan}, {Desai}, {Dobbs}, {Dudley},
  {Fazio}, {Foley}, {Forman}, {Garmire}, {George}, {Gladders}, {Gonzalez},
  {Halverson}, {High}, {Holder}, {Holzapfel}, {Hrubes}, {Jones}, {Joy},
  {Keisler}, {Knox}, {Lee}, {Leitch}, {Lueker}, {Marrone}, {McMahon}, {Mehl},
  {Meyer}, {Mohr}, {Montroy}, {Murray}, {Padin}, {Plagge}, {Pryke},
  {Reichardt}, {Rest}, {Ruhl}, {Schaffer}, {Shaw}, {Shirokoff}, {Song},
  {Spieler}, {Stalder}, {Stanford}, {Staniszewski}, {Stark}, {Stubbs},
  {Vanderlinde}, {Vieira}, {Vikhlinin}, {Williamson}, {Yang}, {Zahn}, \&
  {Zenteno}}]{brodwin10}
{Brodwin}, M., {et~al.} 2010, \apj, 721, 90

\bibitem[{{Brott} \& {Hauschildt}(2005)}]{brott05}
{Brott}, I., \& {Hauschildt}, P.~H. 2005, in ESA Special Publication, Vol. 576,
  The Three-Dimensional Universe with Gaia, ed. {C.~Turon, K.~S.~O'Flaherty, \&
  M.~A.~C.~Perryman}, 565--+

\bibitem[{{Carlstrom} {et~al.}(2011){Carlstrom}, {Ade}, {Aird}, {Benson},
  {Bleem}, {Busetti}, {Chang}, {Chauvin}, {Cho}, {Crawford}, {Crites}, {Dobbs},
  {Halverson}, {Heimsath}, {Holzapfel}, {Hrubes}, {Joy}, {Keisler}, {Lanting},
  {Lee}, {Leitch}, {Leong}, {Lu}, {Lueker}, {Luong-van}, {McMahon}, {Mehl},
  {Meyer}, {Mohr}, {Montroy}, {Padin}, {Plagge}, {Pryke}, {Ruhl}, {Schaffer},
  {Schwan}, {Shirokoff}, {Spieler}, {Staniszewski}, {Stark}, {Tucker},
  {Vanderlinde}, {Vieira}, \& {Williamson}}]{carlstrom11}
{Carlstrom}, J.~E., {et~al.} 2011, \pasp, 123, 568

\bibitem[{{Carlstrom} {et~al.}(2002){Carlstrom}, {Holder}, \&
  {Reese}}]{carlstrom02}
{Carlstrom}, J.~E., {Holder}, G.~P., \& {Reese}, E.~D. 2002, \araa, 40, 643

\bibitem[{{Cay{\'o}n} {et~al.}(2010){Cay{\'o}n}, {Gordon}, \& {Silk}}]{cayon10}
{Cay{\'o}n}, L., {Gordon}, C., \& {Silk}, J. 2010, ArXiv e-prints, 1006.1950

\bibitem[{{Clowe} {et~al.}(2006){Clowe}, {Brada{\v c}}, {Gonzalez},
  {Markevitch}, {Randall}, {Jones}, \& {Zaritsky}}]{clowe06}
{Clowe}, D., {Brada{\v c}}, M., {Gonzalez}, A.~H., {Markevitch}, M., {Randall},
  S.~W., {Jones}, C., \& {Zaritsky}, D. 2006, \apjl, 648, L109

\bibitem[{{Collister} \& {Lahav}(2004)}]{collister04}
{Collister}, A.~A., \& {Lahav}, O. 2004, \pasp, 116, 345

\bibitem[{{Covey} {et~al.}(2007){Covey}, {Ivezi{\'c}}, {Schlegel},
  {Finkbeiner}, {Padmanabhan}, {Lupton}, {Ag{\"u}eros}, {Bochanski}, {Hawley},
  {West}, {Seth}, {Kimball}, {Gogarten}, {Claire}, {Haggard}, {Kaib},
  {Schneider}, \& {Sesar}}]{covey07}
{Covey}, K.~R., {et~al.} 2007, \aj, 134, 2398

\bibitem[{{Dalal} {et~al.}(2008){Dalal}, {Dor{\'e}}, {Huterer}, \&
  {Shirokov}}]{dalal08}
{Dalal}, N., {Dor{\'e}}, O., {Huterer}, D., \& {Shirokov}, A. 2008, \prd, 77,
  123514

\bibitem[{{de Grandi} {et~al.}(1999){de Grandi}, {B{\"o}hringer}, {Guzzo},
  {Molendi}, {Chincarini}, {Collins}, {Cruddace}, {Neumann}, {Schindler},
  {Schuecker}, \& {Voges}}]{degrandi99}
{de Grandi}, S., {et~al.} 1999, \apj, 514, 148

\bibitem[{{Duffy} {et~al.}(2008){Duffy}, {Schaye}, {Kay}, \& {Dalla
  Vecchia}}]{duffy08}
{Duffy}, A.~R., {Schaye}, J., {Kay}, S.~T., \& {Dalla Vecchia}, C. 2008,
  \mnras, 390, L64

\bibitem[{{Eddington}(1913)}]{eddington13}
{Eddington}, A.~S. 1913, \mnras, 73, 359

\bibitem[{{Edge} {et~al.}(1994){Edge}, {Boehringer}, {Guzzo}, {Collins},
  {Neumann}, {Chincarini}, {de Grandi}, {Duemmler}, {Ebeling}, {Schindler},
  {Seitter}, {Vettolani}, {Briel}, {Cruddace}, {Gruber}, {Gursky}, {Hartner},
  {MacGillivray}, {Schuecker}, {Shaver}, {Voges}, {Wallin}, {Wolter}, \&
  {Zamorani}}]{edge94}
{Edge}, A.~C., {et~al.} 1994, \aap, 289, L34

\bibitem[{{Enqvist} {et~al.}(2010){Enqvist}, {Hotchkiss}, \&
  {Taanila}}]{enqvist10}
{Enqvist}, K., {Hotchkiss}, S., \& {Taanila}, O. 2010, ArXiv e-prints,
  1012.2732

\bibitem[{{Foley} {et~al.}(2011){Foley}, {Andersson}, {Bazin}, {de Haan},
  {Ruel}, {Ade}, {Aird}, {Armstrong}, {Ashby}, {Bautz}, {Benson}, {Bleem},
  {Bonamente}, {Brodwin}, {Carlstrom}, {Chang}, {Clocchiatti}, {Crawford},
  {Crites}, {Desai}, {Dobbs}, {Dudley}, {Fazio}, {Forman}, {Garmire}, {George},
  {Gladders}, {Gonzalez}, {Halverson}, {High}, {Holder}, {Holzapfel}, {Hoover},
  {Hrubes}, {Jones}, {Joy}, {Keisler}, {Knox}, {Lee}, {Leitch}, {Lueker},
  {Luong-Van}, {Marrone}, {McMahon}, {Mehl}, {Meyer}, {Mohr}, {Montroy},
  {Murray}, {Padin}, {Plagge}, {Pryke}, {Reichardt}, {Rest}, {Ruhl},
  {Saliwanchik}, {Saro}, {Schaffer}, {Shaw}, {Shirokoff}, {Song}, {Spieler},
  {Stalder}, {Stanford}, {Staniszewski}, {Stark}, {Story}, {Stubbs},
  {Vanderlinde}, {Vieira}, {Vikhlinin}, {Williamson}, \& {Zenteno}}]{foley11}
{Foley}, R.~J., {et~al.} 2011, \apj, 731, 86

\bibitem[{{Fowler} {et~al.}(2007){Fowler}, {Niemack}, {Dicker}, {Aboobaker},
  {Ade}, {Battistelli}, {Devlin}, {Fisher}, {Halpern}, {Hargrave}, {Hincks},
  {Kaul}, {Klein}, {Lau}, {Limon}, {Marriage}, {Mauskopf}, {Page}, {Staggs},
  {Swetz}, {Switzer}, {Thornton}, \& {Tucker}}]{fowler07}
{Fowler}, J.~W., {et~al.} 2007, \ao, 46, 3444

\bibitem[{{Haiman} {et~al.}(2001){Haiman}, {Mohr}, \& {Holder}}]{haiman01}
{Haiman}, Z., {Mohr}, J.~J., \& {Holder}, G.~P. 2001, \apj, 553, 545

\bibitem[{{Hall} {et~al.}(2010){Hall}, {Knox}, {Reichardt}, {Ade}, {Aird},
  {Benson}, {Bleem}, {Carlstrom}, {Chang}, {Cho}, {Crawford}, {Crites}, {de
  Haan}, {Dobbs}, {George}, {Halverson}, {Holder}, {Holzapfel}, {Hrubes},
  {Joy}, {Keisler}, {Lee}, {Leitch}, {Lueker}, {McMahon}, {Mehl}, {Meyer},
  {Mohr}, {Montroy}, {Padin}, {Plagge}, {Pryke}, {Ruhl}, {Schaffer}, {Shaw},
  {Shirokoff}, {Spieler}, {Staniszewski}, {Stark}, {Switzer}, {Vanderlinde},
  {Vieira}, {Williamson}, \& {Zahn}}]{hall10}
{Hall}, N.~R., {et~al.} 2010, \apj, 718, 632

\bibitem[{{High} {et~al.}(2010){High}, {Stalder}, {Song}, {Ade}, {Aird},
  {Allam}, {Armstrong}, {Barkhouse}, {Benson}, {Bertin}, {Bhattacharya},
  {Bleem}, {Brodwin}, {Buckley-Geer}, {Carlstrom}, {Challis}, {Chang},
  {Crawford}, {Crites}, {de Haan}, {Desai}, {Dobbs}, {Dudley}, {Foley},
  {George}, {Gladders}, {Halverson}, {Hamuy}, {Hansen}, {Holder}, {Holzapfel},
  {Hrubes}, {Joy}, {Keisler}, {Lee}, {Leitch}, {Lin}, {Lin}, {Loehr}, {Lueker},
  {Marrone}, {McMahon}, {Mehl}, {Meyer}, {Mohr}, {Montroy}, {Morell}, {Ngeow},
  {Padin}, {Plagge}, {Pryke}, {Reichardt}, {Rest}, {Ruel}, {Ruhl}, {Schaffer},
  {Shaw}, {Shirokoff}, {Smith}, {Spieler}, {Staniszewski}, {Stark}, {Stubbs},
  {Tucker}, {Vanderlinde}, {Vieira}, {Williamson}, {Wood-Vasey}, {Yang},
  {Zahn}, \& {Zenteno}}]{high10}
{High}, F.~W., {et~al.} 2010, \apj, 723, 1736

\bibitem[{{High} {et~al.}(2009){High}, {Stubbs}, {Rest}, {Stalder}, \&
  {Challis}}]{high09}
{High}, F.~W., {Stubbs}, C.~W., {Rest}, A., {Stalder}, B., \& {Challis}, P.
  2009, \aj, 138, 110

\bibitem[{{Holder} {et~al.}(2001){Holder}, {Haiman}, \& {Mohr}}]{holder01b}
{Holder}, G., {Haiman}, Z., \& {Mohr}, J.~J. 2001, \apjl, 560, L111

\bibitem[{{Holder} \& {Carlstrom}(2001)}]{holder01a}
{Holder}, G.~P., \& {Carlstrom}, J.~E. 2001, \apj, 558, 515, astro-ph/0105229

\bibitem[{{Hoyle} {et~al.}(2010){Hoyle}, {Jimenez}, \& {Verde}}]{hoyle10}
{Hoyle}, B., {Jimenez}, R., \& {Verde}, L. 2010, ArXiv e-prints, 1009.3884

\bibitem[{{Komatsu} {et~al.}(2011){Komatsu}, {Smith}, {Dunkley}, {Bennett},
  {Gold}, {Hinshaw}, {Jarosik}, {Larson}, {Nolta}, {Page}, {Spergel},
  {Halpern}, {Hill}, {Kogut}, {Limon}, {Meyer}, {Odegard}, {Tucker}, {Weiland},
  {Wollack}, \& {Wright}}]{komatsu11}
{Komatsu}, E., {et~al.} 2011, \apjs, 192, 18

\bibitem[{{Larson} {et~al.}(2011){Larson}, {Dunkley}, {Hinshaw}, {Komatsu},
  {Nolta}, {Bennett}, {Gold}, {Halpern}, {Hill}, {Jarosik}, {Kogut}, {Limon},
  {Meyer}, {Odegard}, {Page}, {Smith}, {Spergel}, {Tucker}, {Weiland},
  {Wollack}, \& {Wright}}]{larson10}
{Larson}, D., {et~al.} 2011, \apjs, 192, 16

\bibitem[{{Lee} {et~al.}(1998){Lee}, {Gildemeister}, {Holmes}, {Lee}, \&
  {Richards}}]{lee98}
{Lee}, S., {Gildemeister}, J.~M., {Holmes}, W., {Lee}, A.~T., \& {Richards},
  P.~L. 1998, \ao, 37, 3391

\bibitem[{{Lima} \& {Hu}(2007)}]{lima07}
{Lima}, M., \& {Hu}, W. 2007, \prd, 76, 123013

\bibitem[{{Mantz} {et~al.}(2010){Mantz}, {Allen}, {Ebeling}, {Rapetti}, \&
  {Drlica-Wagner}}]{mantz10}
{Mantz}, A., {Allen}, S.~W., {Ebeling}, H., {Rapetti}, D., \& {Drlica-Wagner},
  A. 2010, \mnras, 406, 1773

\bibitem[{{Marriage} {et~al.}(2011){Marriage}, {Baptiste Juin}, {Lin},
  {Marsden}, {Nolta}, {Partridge}, {Ade}, {Aguirre}, {Amiri}, {Appel},
  {Barrientos}, {Battistelli}, {Bond}, {Brown}, {Burger}, {Chervenak}, {Das},
  {Devlin}, {Dicker}, {Bertrand Doriese}, {Dunkley}, {D{\"u}nner},
  {Essinger-Hileman}, {Fisher}, {Fowler}, {Hajian}, {Halpern}, {Hasselfield},
  {Hern{\'a}ndez-Monteagudo}, {Hilton}, {Hilton}, {Hincks}, {Hlozek},
  {Huffenberger}, {Handel Hughes}, {Hughes}, {Infante}, {Irwin}, {Kaul},
  {Klein}, {Kosowsky}, {Lau}, {Limon}, {Lupton}, {Martocci}, {Mauskopf},
  {Menanteau}, {Moodley}, {Moseley}, {Netterfield}, {Niemack}, {Page},
  {Parker}, {Quintana}, {Reid}, {Sehgal}, {Sherwin}, {Sievers}, {Spergel},
  {Staggs}, {Swetz}, {Switzer}, {Thornton}, {Trac}, {Tucker}, {Warne},
  {Wilson}, {Wollack}, \& {Zhao}}]{marriage11}
{Marriage}, T.~A., {et~al.} 2011, \apj, 731, 100

\bibitem[{{Matarrese} {et~al.}(2000){Matarrese}, {Verde}, \&
  {Jimenez}}]{matarrese00}
{Matarrese}, S., {Verde}, L., \& {Jimenez}, R. 2000, \apj, 541, 10

\bibitem[{{Melin} {et~al.}(2006){Melin}, {Bartlett}, \&
  {Delabrouille}}]{melin06}
{Melin}, J.-B., {Bartlett}, J.~G., \& {Delabrouille}, J. 2006, \aap, 459, 341

\bibitem[{{Menanteau} {et~al.}(2010){Menanteau}, {Gonz{\'a}lez}, {Juin},
  {Marriage}, {Reese}, {Acquaviva}, {Aguirre}, {Appel}, {Baker}, {Barrientos},
  {Battistelli}, {Bond}, {Das}, {Deshpande}, {Devlin}, {Dicker}, {Dunkley},
  {D{\"u}nner}, {Essinger-Hileman}, {Fowler}, {Hajian}, {Halpern},
  {Hasselfield}, {Hern{\'a}ndez-Monteagudo}, {Hilton}, {Hincks}, {Hlozek},
  {Huffenberger}, {Hughes}, {Infante}, {Irwin}, {Klein}, {Kosowsky}, {Lin},
  {Marsden}, {Moodley}, {Niemack}, {Nolta}, {Page}, {Parker}, {Partridge},
  {Sehgal}, {Sievers}, {Spergel}, {Staggs}, {Swetz}, {Switzer}, {Thornton},
  {Trac}, {Warne}, \& {Wollack}}]{menanteau10b}
{Menanteau}, F., {et~al.} 2010, \apj, 723, 1523

\bibitem[{{Mohr} {et~al.}(2008){Mohr}, {Adams}, {Barkhouse}, {Beldica},
  {Bertin}, {Cai}, {da Costa}, {Darnell}, {Daues}, {Jarvis}, {Gower}, {Lin},
  {Martelli}, {Neilsen}, {Ngeow}, {Ogando}, {Parga}, {Sheldon}, {Tucker},
  {Kuropatkin}, \& {Stoughton}}]{mohr08}
{Mohr}, J.~J., {et~al.} 2008, in Society of Photo-Optical Instrumentation
  Engineers (SPIE) Conference Series, Vol. 7016, Society of Photo-Optical
  Instrumentation Engineers (SPIE) Conference Series

\bibitem[{{Molnar} {et~al.}(2004){Molnar}, {Haiman}, {Birkinshaw}, \&
  {Mushotzky}}]{molnar04}
{Molnar}, S.~M., {Haiman}, Z., {Birkinshaw}, M., \& {Mushotzky}, R.~F. 2004,
  \apj, 601, 22

\bibitem[{{Mortonson} {et~al.}(2011){Mortonson}, {Hu}, \&
  {Huterer}}]{mortonson11}
{Mortonson}, M.~J., {Hu}, W., \& {Huterer}, D. 2011, \prd, 83, 023015

\bibitem[{{Motl} {et~al.}(2005){Motl}, {Hallman}, {Burns}, \&
  {Norman}}]{motl05}
{Motl}, P.~M., {Hallman}, E.~J., {Burns}, J.~O., \& {Norman}, M.~L. 2005,
  \apjl, 623, L63

\bibitem[{{Nagai} {et~al.}(2007){Nagai}, {Kravtsov}, \& {Vikhlinin}}]{nagai07}
{Nagai}, D., {Kravtsov}, A.~V., \& {Vikhlinin}, A. 2007, \apj, 668, 1

\bibitem[{{Ngeow} {et~al.}(2006){Ngeow}, {Mohr}, {Alam}, {Barkhouse},
  {Beldica}, {Cai}, {Daues}, {Plante}, {Annis}, {Lin}, {Tucker}, \&
  {Smith}}]{ngeow06}
{Ngeow}, C., {et~al.} 2006, in Society of Photo-Optical Instrumentation
  Engineers (SPIE) Conference Series, Vol. 6270, Society of Photo-Optical
  Instrumentation Engineers (SPIE) Conference Series

\bibitem[{{Percival} {et~al.}(2010){Percival}, {Reid}, {Eisenstein}, {Bahcall},
  {Budavari}, {Frieman}, {Fukugita}, {Gunn}, {Ivezi{\'c}}, {Knapp}, {Kron},
  {Loveday}, {Lupton}, {McKay}, {Meiksin}, {Nichol}, {Pope}, {Schlegel},
  {Schneider}, {Spergel}, {Stoughton}, {Strauss}, {Szalay}, {Tegmark},
  {Vogeley}, {Weinberg}, {York}, \& {Zehavi}}]{percival10}
{Percival}, W.~J., {et~al.} 2010, \mnras, 401, 2148

\bibitem[{{Pratt} {et~al.}(2009){Pratt}, {Croston}, {Arnaud}, \&
  {B{\"o}hringer}}]{pratt09}
{Pratt}, G.~W., {Croston}, J.~H., {Arnaud}, M., \& {B{\"o}hringer}, H. 2009,
  \aap, 498, 361

\bibitem[{{Riess} {et~al.}(2009){Riess}, {Macri}, {Casertano}, {Sosey},
  {Lampeitl}, {Ferguson}, {Filippenko}, {Jha}, {Li}, {Chornock}, \&
  {Sarkar}}]{riess09}
{Riess}, A.~G., {et~al.} 2009, \apj, 699, 539

\bibitem[{{Rozo} {et~al.}(2010){Rozo}, {Wechsler}, {Rykoff}, {Annis}, {Becker},
  {Evrard}, {Frieman}, {Hansen}, {Hao}, {Johnston}, {Koester}, {McKay},
  {Sheldon}, \& {Weinberg}}]{rozo10}
{Rozo}, E., {et~al.} 2010, \apj, 708, 645

\bibitem[{{Schlegel} {et~al.}(1998){Schlegel}, {Finkbeiner}, \&
  {Davis}}]{schlegel98}
{Schlegel}, D.~J., {Finkbeiner}, D.~P., \& {Davis}, M. 1998, \apj, 500, 525

\bibitem[{{Sehgal} {et~al.}(2010{\natexlab{a}}){Sehgal}, {Bode}, {Das},
  {Hernandez-Monteagudo}, {Huffenberger}, {Lin}, {Ostriker}, \&
  {Trac}}]{sehgal10}
{Sehgal}, N., {Bode}, P., {Das}, S., {Hernandez-Monteagudo}, C.,
  {Huffenberger}, K., {Lin}, Y., {Ostriker}, J.~P., \& {Trac}, H.
  2010{\natexlab{a}}, \apj, 709, 920

\bibitem[{{Sehgal} {et~al.}(2010{\natexlab{b}}){Sehgal}, {Trac}, {Acquaviva},
  {Ade}, {Aguirre}, {Amiri}, {Appel}, {Barrientos}, {Battistelli}, {Bond},
  {Brown}, {Burger}, {Chervenak}, {Das}, {Devlin}, {Dicker}, {Bertrand
  Doriese}, {Dunkley}, {D{\"u}nner}, {Essinger-Hileman}, {Fisher}, {Fowler},
  {Hajian}, {Halpern}, {Hasselfield}, {Hern{\'a}ndez-Monteagudo}, {Hilton},
  {Hilton}, {Hincks}, {Hlozek}, {Holtz}, {Huffenberger}, {Hughes}, {Hughes},
  {Infante}, {Irwin}, {Jones}, {Baptiste Juin}, {Klein}, {Kosowsky}, {Lau},
  {Limon}, {Lin}, {Lupton}, {Marriage}, {Marsden}, {Martocci}, {Mauskopf},
  {Menanteau}, {Moodley}, {Moseley}, {Netterfield}, {Niemack}, {Nolta}, {Page},
  {Parker}, {Partridge}, {Reid}, {Sherwin}, {Sievers}, {Spergel}, {Staggs},
  {Swetz}, {Switzer}, {Thornton}, {Tucker}, {Warne}, {Wollack}, \&
  {Zhao}}]{sehgal10b}
{Sehgal}, N., {et~al.} 2010{\natexlab{b}}, ArXiv e-prints, 1010.1025

\bibitem[{{Shaw} {et~al.}(2008){Shaw}, {Holder}, \& {Bode}}]{shaw08}
{Shaw}, L.~D., {Holder}, G.~P., \& {Bode}, P. 2008, \apj, 686, 206

\bibitem[{{Shaw} {et~al.}(2009){Shaw}, {Zahn}, {Holder}, \&
  {Dor{\'e}}}]{shaw09}
{Shaw}, L.~D., {Zahn}, O., {Holder}, G.~P., \& {Dor{\'e}}, O. 2009, \apj, 702,
  368

\bibitem[{{Shirokoff} {et~al.}(2011){Shirokoff}, {Reichardt}, {Shaw}, {Millea},
  {Ade}, {Aird}, {Benson}, {Bleem}, {Carlstrom}, {Chang}, {Cho}, {Crawford},
  {Crites}, {de Haan}, {Dobbs}, {Dudley}, {George}, {Halverson}, {Holder},
  {HOlzapfel}, {Hrubes}, {Joy}, {Keisler}, {Knox}, {Lee}, {Leitch}, {Lueker},
  {Luong-Van}, {McMahon}, {Mehl}, {Meyer}, {Mohr}, {Montroy}, {Padin},
  {Plagge}, {Pryke}, {Ruhl}, {Schaffer}, {Spieler}, {Staniszewski}, {Stark},
  {Story}, {Vanderlinde}, {Vieira}, {Williamson}, \& {Zahn}}]{shirokoff11}
{Shirokoff}, E., {et~al.} 2011, \apj, in press, arXiv:1012.4788

\bibitem[{{Stanek} {et~al.}(2009){Stanek}, {Rasia}, {Evrard}, {Pearce}, \&
  {Gazzola}}]{stanek09}
{Stanek}, R., {Rasia}, E., {Evrard}, A.~E., {Pearce}, F., \& {Gazzola}, L.
  2009, ArXiv e-prints, 0910.1599

\bibitem[{{Staniszewski} {et~al.}(2009){Staniszewski}, {Ade}, {Aird}, {Benson},
  {Bleem}, {Carlstrom}, {Chang}, {Cho}, {Crawford}, {Crites}, {de Haan},
  {Dobbs}, {Halverson}, {Holder}, {Holzapfel}, {Hrubes}, {Joy}, {Keisler},
  {Lanting}, {Lee}, {Leitch}, {Loehr}, {Lueker}, {McMahon}, {Mehl}, {Meyer},
  {Mohr}, {Montroy}, {Ngeow}, {Padin}, {Plagge}, {Pryke}, {Reichardt}, {Ruhl},
  {Schaffer}, {Shaw}, {Shirokoff}, {Spieler}, {Stalder}, {Stark},
  {Vanderlinde}, {Vieira}, {Zahn}, \& {Zenteno}}]{staniszewski09}
{Staniszewski}, Z., {et~al.} 2009, \apj, 701, 32

\bibitem[{{Struble} \& {Rood}(1999)}]{struble99}
{Struble}, M.~F., \& {Rood}, H.~J. 1999, \apjs, 125, 35

\bibitem[{{Sunyaev} \& {Zel'dovich}(1972)}]{sunyaev72}
{Sunyaev}, R.~A., \& {Zel'dovich}, Y.~B. 1972, Comments on Astrophysics and
  Space Physics, 4, 173

\bibitem[{{Tinker} {et~al.}(2008){Tinker}, {Kravtsov}, {Klypin}, {Abazajian},
  {Warren}, {Yepes}, {Gottl{\"o}ber}, \& {Holz}}]{tinker08}
{Tinker}, J., {Kravtsov}, A.~V., {Klypin}, A., {Abazajian}, K., {Warren}, M.,
  {Yepes}, G., {Gottl{\"o}ber}, S., \& {Holz}, D.~E. 2008, \apj, 688, 709

\bibitem[{{Tucker} {et~al.}(1998){Tucker}, {Blanco}, {Rappoport}, {David},
  {Fabricant}, {Falco}, {Forman}, {Dressler}, \& {Ramella}}]{tucker98}
{Tucker}, W., {et~al.} 1998, \apjl, 496, L5

\bibitem[{{Vanderlinde} {et~al.}(2010){Vanderlinde}, {Crawford}, {de Haan},
  {Dudley}, {Shaw}, {Ade}, {Aird}, {Benson}, {Bleem}, {Brodwin}, {Carlstrom},
  {Chang}, {Crites}, {Desai}, {Dobbs}, {Foley}, {George}, {Gladders}, {Hall},
  {Halverson}, {High}, {Holder}, {Holzapfel}, {Hrubes}, {Joy}, {Keisler},
  {Knox}, {Lee}, {Leitch}, {Loehr}, {Lueker}, {Marrone}, {McMahon}, {Mehl},
  {Meyer}, {Mohr}, {Montroy}, {Ngeow}, {Padin}, {Plagge}, {Pryke}, {Reichardt},
  {Rest}, {Ruel}, {Ruhl}, {Schaffer}, {Shirokoff}, {Song}, {Spieler},
  {Stalder}, {Staniszewski}, {Stark}, {Stubbs}, {van Engelen}, {Vieira},
  {Williamson}, {Yang}, {Zahn}, \& {Zenteno}}]{vanderlinde10}
{Vanderlinde}, K., {et~al.} 2010, \apj, 722, 1180

\bibitem[{{Vikhlinin} {et~al.}(2009{\natexlab{a}}){Vikhlinin}, {Burenin},
  {Ebeling}, {Forman}, {Hornstrup}, {Jones}, {Kravtsov}, {Murray}, {Nagai},
  {Quintana}, \& {Voevodkin}}]{vikhlinin09b}
{Vikhlinin}, A., {et~al.} 2009{\natexlab{a}}, \apj, 692, 1033

\bibitem[{{Vikhlinin} {et~al.}(2009{\natexlab{b}}){Vikhlinin}, {Kravtsov},
  {Burenin}, {Ebeling}, {Forman}, {Hornstrup}, {Jones}, {Murray}, {Nagai},
  {Quintana}, \& {Voevodkin}}]{vikhlinin09}
------. 2009{\natexlab{b}}, \apj, 692, 1060

\bibitem[{Vikhlinin {et~al.}(1998)Vikhlinin, McNamara, Forman, Jones, Quintana,
  \& Hornstrup}]{vikhlinin98}
Vikhlinin, A., McNamara, B., Forman, W., Jones, C., Quintana, H., \& Hornstrup,
  A. 1998, \apj, 502, 558

\bibitem[{{Voges} {et~al.}(1999){Voges}, {Aschenbach}, {Boller},
  {Br{\"a}uninger}, {Briel}, {Burkert}, {Dennerl}, {Englhauser}, {Gruber},
  {Haberl}, {Hartner}, {Hasinger}, {K{\"u}rster}, {Pfeffermann}, {Pietsch},
  {Predehl}, {Rosso}, {Schmitt}, {Tr{\"u}mper}, \& {Zimmermann}}]{voges99}
{Voges}, W., {et~al.} 1999, \aap, 349, 389

\bibitem[{{Wang} \& {Steinhardt}(1998)}]{wang98}
{Wang}, L., \& {Steinhardt}, P.~J. 1998, \apj, 508, 483

\bibitem[{{Wang} {et~al.}(2004){Wang}, {Khoury}, {Haiman}, \& {May}}]{wang04}
{Wang}, S., {Khoury}, J., {Haiman}, Z., \& {May}, M. 2004, \prd, 70, 123008

\end{thebibliography}

\clearpage
\LongTables
\pagestyle{empty}
\begin{landscape}
\begin{center}
\tabletypesize{\small}
\begin{deluxetable}{lrrrlrrrrr}
\tablecaption{Clusters
\label{tab:clusters}}
\tablehead{
\colhead{Object Name} & 
\colhead{R.A.} & 
\colhead{decl.} & 
\colhead{$\xi$} & 
\colhead{Depth} & 
\colhead{$z$} & 
\colhead{$M_{200} \pm \mathrm{stat} \pm \mathrm{syst}$ } &
\colhead{$M_{500} \pm \mathrm{stat} \pm \mathrm{syst}$ } &
\colhead{$F_\mathrm{X}$} & 
\colhead{$L_\mathrm{X}$} \\ 
\colhead{} &
\colhead{} &
\colhead{} &
\colhead{} &
\colhead{} &
\colhead{} &
\colhead{$[10^{14}\,M_\odot\, h^{-1}_{70}]$} & 
\colhead{$[10^{14}\,M_\odot\, h^{-1}_{70}]$} & 
\colhead{$[10^{-13} $ergs cm$^{-2}$ s$^{-1}]$} & 
\colhead{$[10^{44} $ergs s$^{-1}]$} 
}
\startdata
SPT-CL J0040-4407 & 10.202 & -44.131 & $10.1$ & shallow &  0.40 (p) &   $15.9 \pm  2.3 \pm  4.1$ & $ 7.8 \pm  2.0 \pm  2.0$ &  $ 7.2 \pm  2.6$ & $ 3.5 \pm  1.2$ \\ 
SPT-CL J0102-4915\tablenotemark{1}  & 15.728 & -49.257 & $39.5$ & full &  0.78 (p) &   $18.9 \pm  2.9 \pm  3.5$ & $9.9 \pm  2.5 \pm  1.8$ &  $ 6.1 \pm  2.1$ & $11.3 \pm  3.9$ \\ 
SPT-CL J0232-4421\tablenotemark{2}  & 38.070 & -44.351 & $11.4$ & shallow &  0.284 (s) &   $18.8 \pm  2.6 \pm  5.3$ & $8.9 \pm  2.2 \pm  2.4$ &  $30.3 \pm  3.3$ & $ 6.4 \pm  0.7$ \\ 
SPT-CL J0234-5831 & 38.670 & -58.520 & $14.7$ & full &  0.415 (s) &   $12.4 \pm  2.0 \pm  1.7$ & $ 6.2 \pm  1.7 \pm  0.8$ &  $ 11.6 \pm  2.9$ & $ 5.6 \pm  1.4$ \\ 
SPT-CL J0243-4833 & 40.910 & -48.557 & $13.8$ & full &  0.53 (p) &   $10.7 \pm  1.8 \pm  1.4$ & $ 5.5 \pm  1.5 \pm  0.7$ &  $ 9.1 \pm  1.8$ & $ 7.4 \pm  1.5$ \\ 
SPT-CL J0245-5302\tablenotemark{3}  & 41.378 & -53.036 & $19.3$ & full &  0.300 (s) &   $17.0 \pm  2.7 \pm  3.2$ & $ 8.1 \pm  2.1 \pm  1.5$ &  $16.2 \pm  2.1$ & $ 7.2 \pm  0.9$ \\ 
SPT-CL J0254-5856 & 43.563 & -58.949 & $14.3$ & full &  0.438 (s) &   $11.9 \pm  1.9 \pm  1.7$ & $ 6.0 \pm  1.6 \pm  0.8$ &  $ 13.5 \pm  2.8$ & $ 7.2 \pm  1.5$ \\ 
SPT-CL J0304-4401 & 46.064 & -44.030 & $ 8.0$ & shallow &  0.52 (p) &   $12.1 \pm  2.0 \pm  2.6$ & $ 6.2 \pm  1.7 \pm  1.3$ &  $8.9  \pm  1.8$ & $ 7.0 \pm  1.4$ \\ 
SPT-CL J0411-4819 & 62.811 & -48.321 & $14.8$ & full &  0.42 (p) &   $12.4 \pm  2.0 \pm  1.7$ & $ 6.2 \pm  1.7 \pm  0.8$ &  $ 11.9 \pm  3.4$ & $ 6.0 \pm  1.8$ \\ 
SPT-CL J0417-4748 & 64.340 & -47.812 & $13.9$ & full &  0.62 (p) &   $10.0 \pm  1.6 \pm  1.2$ & $ 5.2 \pm  1.4 \pm  0.6$ &  $ 4.8 \pm  0.8$ & $ 5.5 \pm  0.9$ \\ 
SPT-CL J0438-5419\tablenotemark{4}  & 69.569 & -54.321 & $22.3$ & full &  0.45 (p) &   $16.5 \pm  2.6 \pm  2.8$ & $ 8.2 \pm  2.1 \pm  1.4$ &  $ 18.8 \pm  3.0$ & $ 10.7 \pm  1.7$ \\ 
SPT-CL J0549-6204 & 87.326 & -62.083 & $12.6$ & shallow &  0.32 (p) &   $19.9 \pm  2.7 \pm  6.1$ & $9.5 \pm  2.3 \pm  2.8$ &  $25.2 \pm  1.4$ & $ 6.9 \pm  0.4$ \\ 
SPT-CL J0555-6405 & 88.851 & -64.099 & $ 7.1$ & shallow &  0.42 (p) &   $11.3 \pm  2.1 \pm  2.8$ & $ 5.6 \pm  1.6 \pm  1.4$ &  $ 9.5 \pm  0.6$ & $ 4.7 \pm  0.3$ \\ 
SPT-CL J0615-5746 & 93.957 & -57.778 & $11.1$ & shallow &  0.972 (s) &   $13.2 \pm  1.9 \pm  3.5$ & $ 7.1 \pm  1.8 \pm  1.8$ &  $ 4.3 \pm  0.7$ & $12.6 \pm  2.1$ \\ 
SPT-CL J0628-4143\tablenotemark{5}  & 97.201 & -41.720 & $ 8.1$ & shallow &  0.176 (s) &   $14.9 \pm  2.5 \pm  4.0$ & $ 6.9 \pm  1.9 \pm  1.8$ &  $30.0 \pm  2.7$ & $ 2.3 \pm  0.2$ \\ 
SPT-CL J0638-5358\tablenotemark{6}  & 99.693 & -53.974 & $11.1$ & shallow &  0.222 (s) &   $19.1 \pm  2.7 \pm  5.6$ & $8.9 \pm  2.2 \pm  2.5$ &  $44.0 \pm  3.1$ & $ 5.5 \pm  0.4$ \\ 
SPT-CL J0645-5413\tablenotemark{7}  & 101.360 & -54.224 & $10.0$ & shallow &  0.167 (s) &   $18.1 \pm  2.7 \pm  5.4$ & $ 8.3 \pm  2.1 \pm  2.4$ &  $53.7 \pm  4.1$ & $ 3.7 \pm  0.3$ \\ 
SPT-CL J0658-5556\tablenotemark{8}  & 104.625 & -55.949 & $22.0$ & shallow &  0.296 (s) &   $31.2 \pm  3.9 \pm 10.8$ & $14.6 \pm  3.4 \pm  4.9$ &  $44.0 \pm  3.5 $ & $10.1 \pm  0.8$ \\ 
SPT-CL J2023-5535\tablenotemark{9}  & 305.833 & -55.590 & $14.8$ & full &  0.232 (s) &   $14.9 \pm  2.4 \pm  2.7$ & $ 7.0 \pm  1.9 \pm  1.2$ &  $19.2 \pm  3.5$ & $ 2.6 \pm  0.5$ \\ 
SPT-CL J2031-4037\tablenotemark{10}  & 307.960 & -40.619 & $ 9.4$ & shallow &  0.342 (s) &   $15.5 \pm  2.3 \pm  4.3$ & $ 7.5 \pm  1.9 \pm  2.0$ &  $20.6 \pm  4.8$ & $ 6.5 \pm  1.5$ \\ 
SPT-CL J2106-5844 & 316.515 & -58.744 & $22.1$ & full &  1.132 (s) &   $ 9.8 \pm  1.5 \pm  1.4$ & $ 5.4 \pm  1.4 \pm  0.7$ &  $ 2.0 \pm  0.2$ & $13.9 \pm  1.0$ \\ 
SPT-CL J2201-5956\tablenotemark{11}  & 330.462 & -59.944 & $14.5$ & full &  0.098 (s) &   $17.0 \pm  2.8 \pm  3.7$ & $ 7.6 \pm  2.0 \pm  1.6$ &  $125.2 \pm  8.1$ & $ 2.8 \pm  0.2$ \\ 
SPT-CL J2248-4431\tablenotemark{12}  & 342.181 & -44.527 & $20.7$ & shallow &  0.348 (s) &   $29.0 \pm  3.7 \pm  9.6$ & $13.8 \pm  3.2 \pm  4.4$ &  $54.1 \pm  6.1$ & $17.7 \pm  2.0$ \\ 
SPT-CL J2325-4111\tablenotemark{13}  & 351.294 & -41.194 & $ 7.2$ & shallow &  0.37 (p) &   $11.8 \pm  2.1 \pm  3.0$ & $ 5.8 \pm  1.7 \pm  1.4$ &  $15.0 \pm  4.1$ & $5.6 \pm  1.5$ \\ 
SPT-CL J2337-5942 & 354.347 & -59.703 & $16.8$ & full &  0.775 (s) &   $10.2 \pm  1.6 \pm  1.2$ & $ 5.4 \pm  1.4 \pm  0.6$ &  $ 3.1 \pm  0.2$ & $ 8.9 \pm  0.5$ \\ 
SPT-CL J2344-4243 & 356.176 & -42.719 & $12.1$ & shallow &  0.62 (p) &   $16.6 \pm  2.3 \pm  4.4$ & $ 8.5 \pm  2.1 \pm  2.2$ &  $18.3 \pm  5.5$ & $21.1 \pm  6.3$ 
\enddata
\tablenotetext{1}{\scriptsize ACT-CL J0102-4915}
\tablenotetext{2}{\scriptsize RXC J0232.2-4420}
\tablenotetext{3}{\scriptsize ABELL S0295, ACT-CL J0245-5302}
\tablenotetext{4}{\scriptsize ACT-CL J0438-5419}
\tablenotetext{5}{\scriptsize ABELL 3396, RXC J0628.8-4143 }
\tablenotetext{6}{\scriptsize ABELL S0592, RXC J0638.7-5358. ACT-CL J0638-5358}
\tablenotetext{7}{\scriptsize ABELL 3404, RXC J0645.4-5413, ACT-CL J0645-5413 }
\tablenotetext{8}{\scriptsize Bullet, RXC J0658.5-5556, ACT-CL J0658-5557}
\tablenotetext{9}{\scriptsize RXC J2023.4-5535 }
\tablenotetext{10}{\scriptsize RXC J2031.8-4037}
\tablenotetext{11}{\scriptsize ABELL 3827, RXC J2201.9-5956 }
\tablenotetext{12}{\scriptsize ABELL S1063, RXC J2248.7-4431}
\tablenotetext{13}{\scriptsize ABELL S1121}
\tablecomments{$\xi$ is the maximum signal-to-noise obtained over the set of filter scales. $z$ is the redshift where (s) refers to a spectral redshift and (p) refers to a photometric redshift. Table \ref{tab:redshift} shows a detailed breakdown of the redshift observation. The masses $M_{500}(\rho_{crit})$ (where the overdensity is with respect to the critical density rather than the mean density) were calculated by converting from $M_{200}(\rho_{mean})$ assuming an NFW density profile and the mass-concentration relation of \citet{duffy08}. The X-ray Flux and Luminosity are reported for the 0.5-2.0 keV band in the cluster frame. }
\end{deluxetable}
\end{center}
\clearpage
\end{landscape}

\appendix


\section{SZ and Optical/Infrared Images}

\begin{figure*}
  \epsscale{1.15}
 \plotone{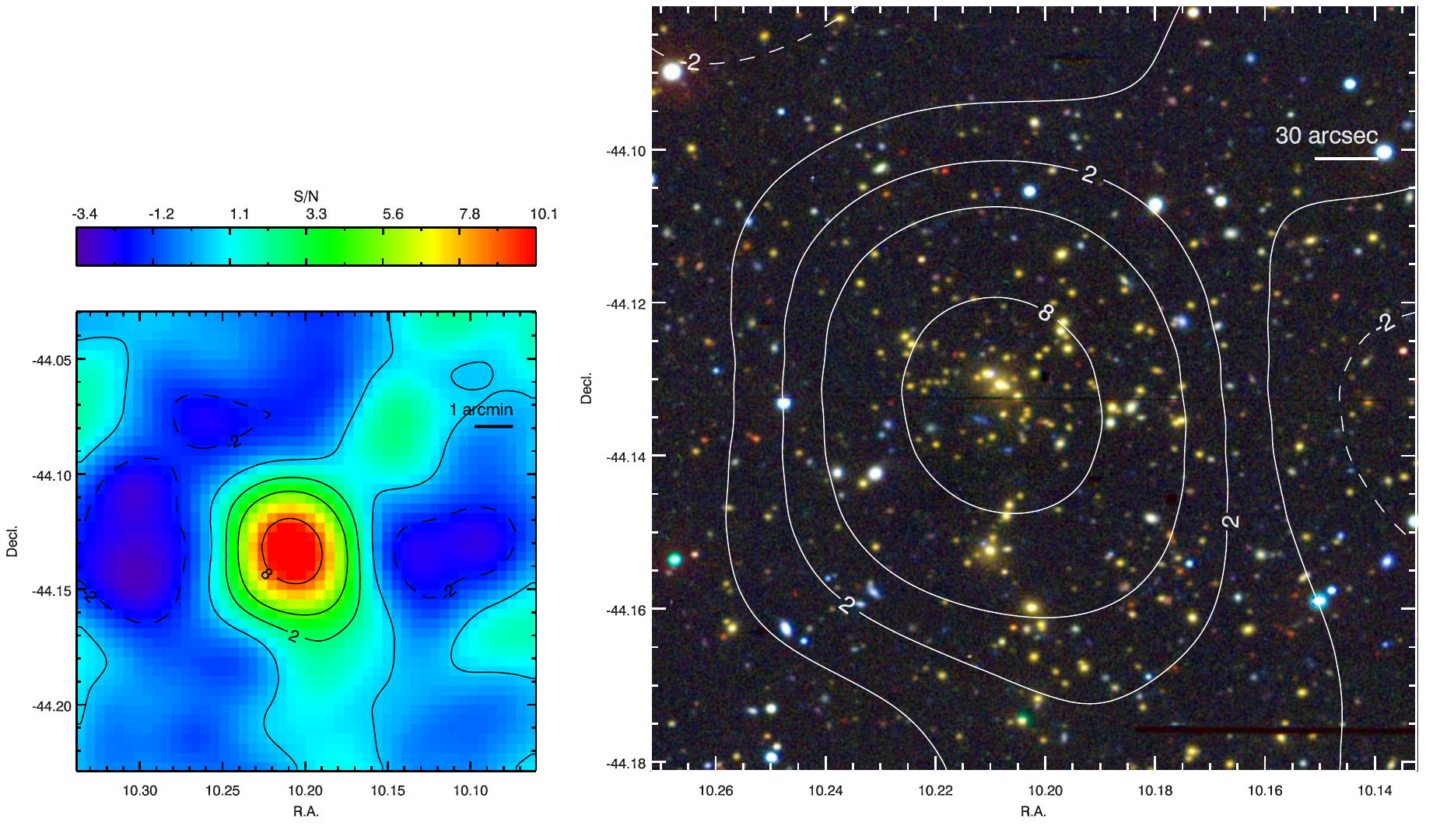}
 \caption{SPT-CL J0040-4407 at $z_{\mathrm{rs}}
   =0.40$. Blanco/MOSAIC-II $irg$ images are shown in the 
   optical/infrared panel.\label{fig:thumb1}}
 \end{figure*}
\begin{figure*}
  \epsscale{1.15}
  \plotone{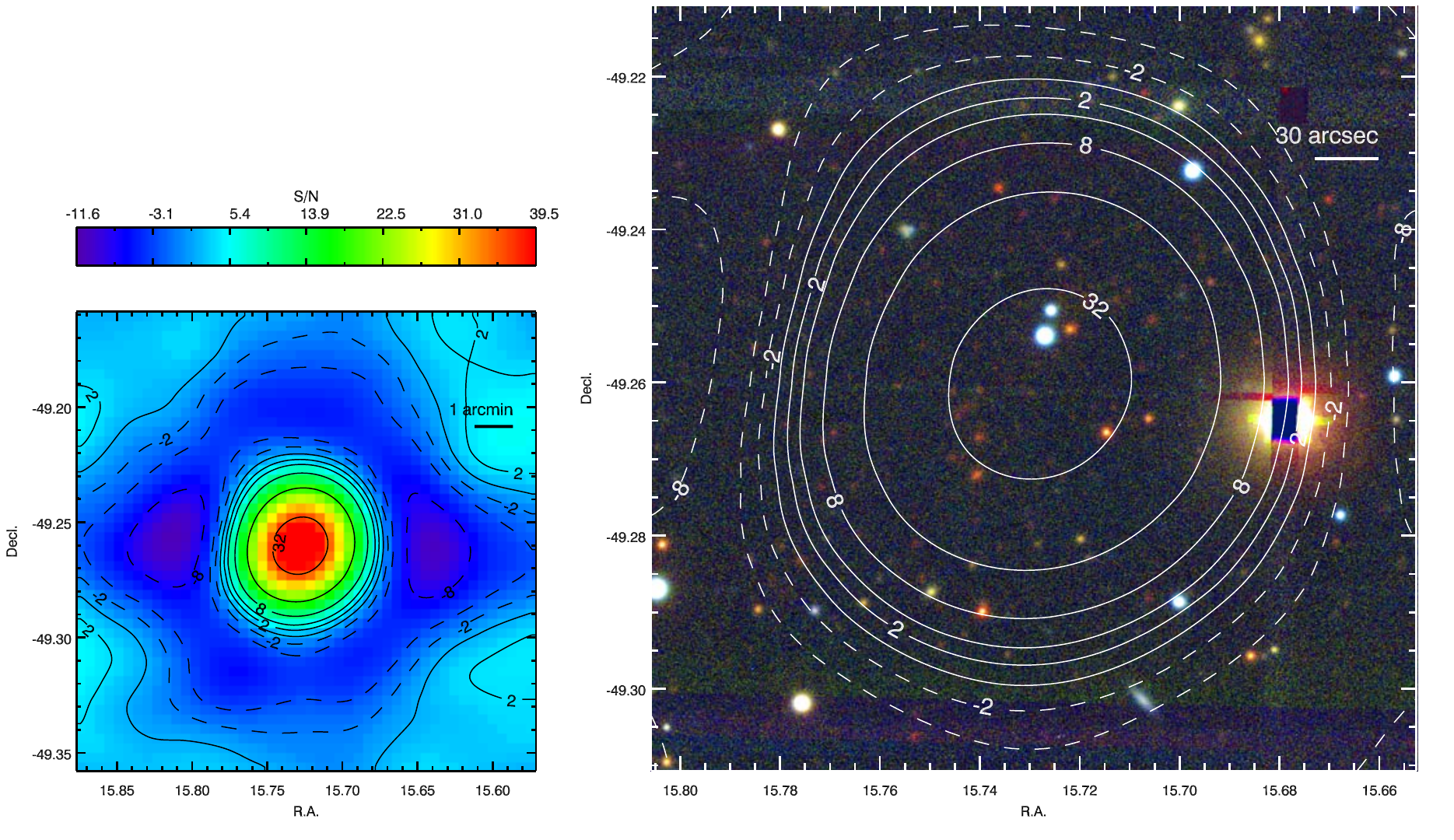}
  \caption{SPT-CL J0102-4915, also known as ACT-CL J0102-4915, at
    $z_{\mathrm{rs}} =0.78$. Blanco/MOSAIC-II $irg$ images are shown
    in the optical/infrared panel.\label{fig:thumb2}}
\end{figure*}
\begin{figure*}
  \epsscale{1.15}
  \plotone{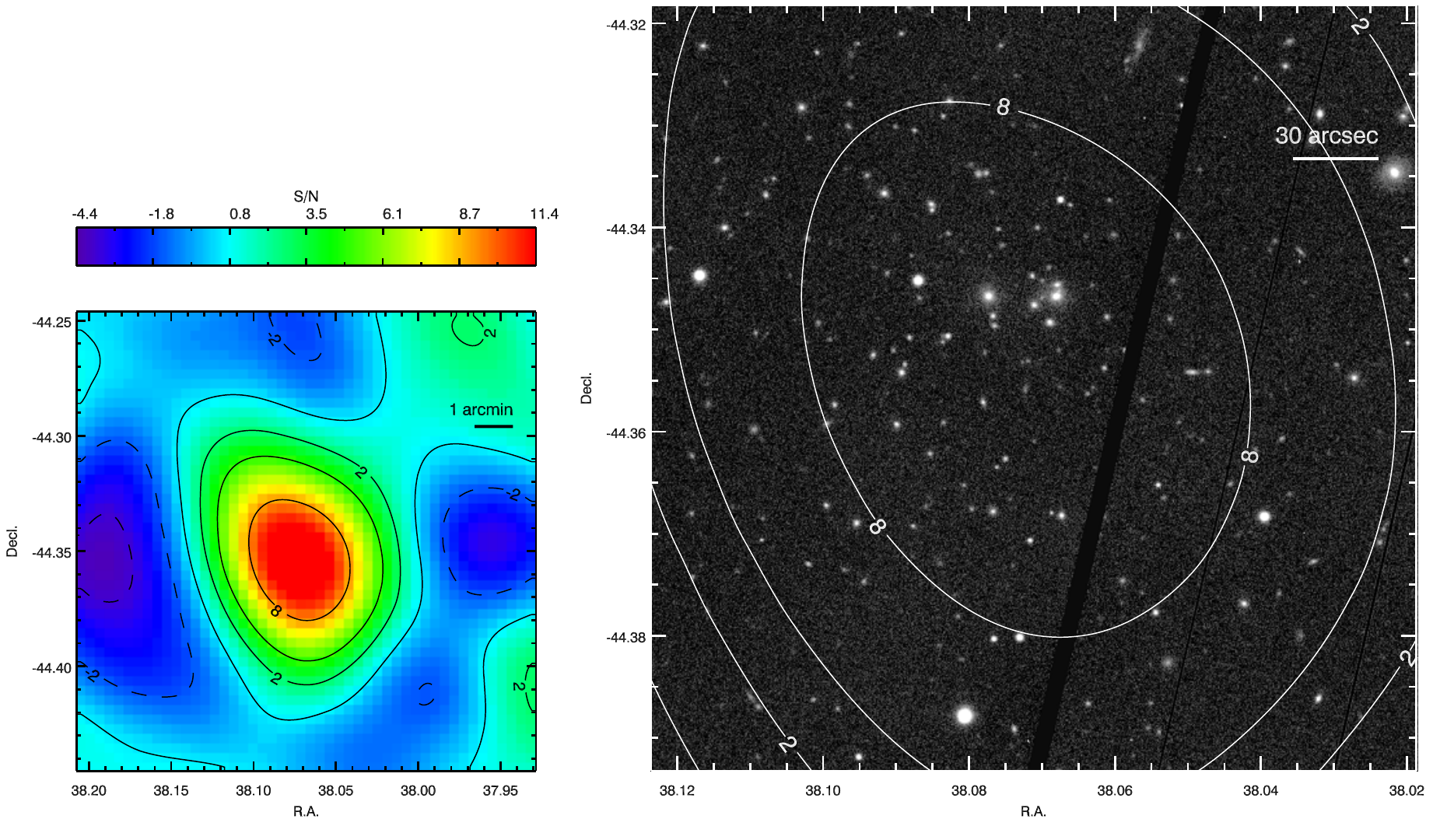}
  \caption{SPT-CL J0232-4421, also known as RXC J0232.2-4420, at
   $z_{\mathrm{spec}} =0.284$. A VLT/FORS2 $R$ image is shown in the
    optical/infrared panel\label{fig:thumb3}}
\end{figure*}
\begin{figure*}
  \epsscale{1.15}
  \plotone{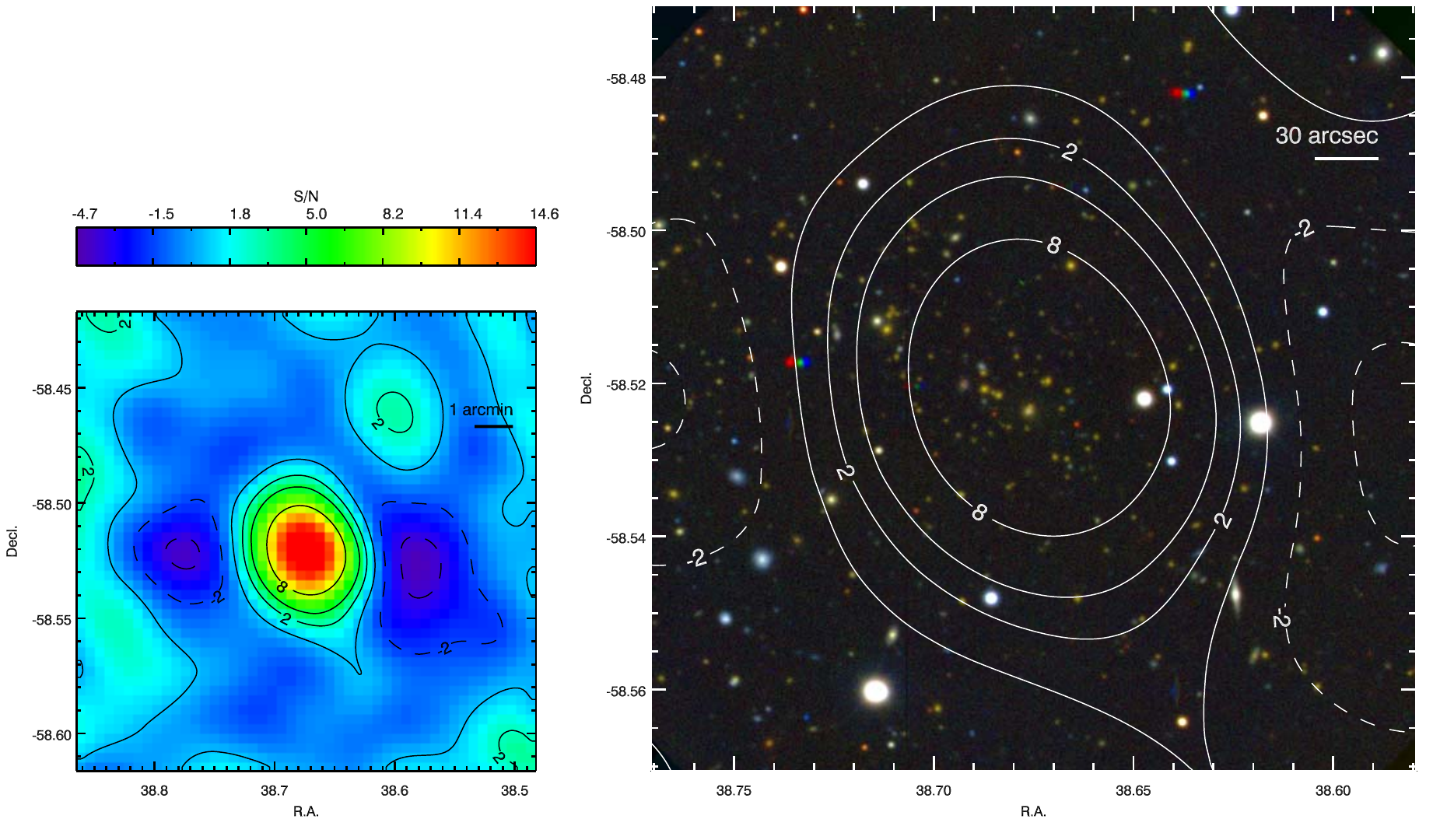}
  \caption{SPT-CL J0234-5831 at $z_{\mathrm{spec}} =0.415$.
    Magellan/LDSS3 $zrg$ images are shown in the 
    optical/infrared panel.\label{fig:thumb4}}
\end{figure*}
\begin{figure*}
  \epsscale{1.15}
  \plotone{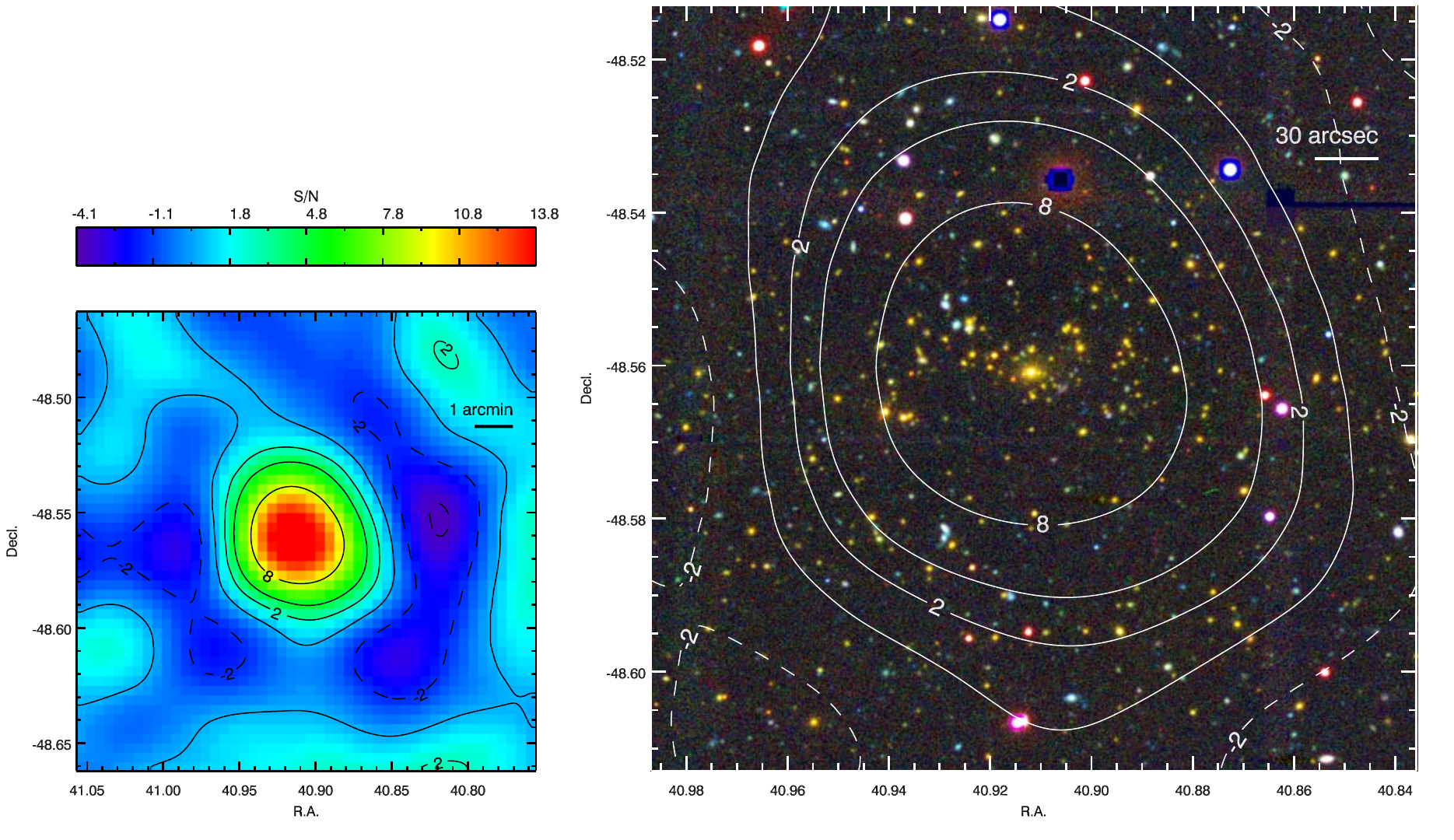}
  \caption{SPT-CL J0243-4833 at $z_{\mathrm{rs}} =0.44$.
    Blanco/MOSAIC-II $irg$ images are shown in the 
    optical/infrared panel.\label{fig:thumb5}}
\end{figure*}
\begin{figure*}
  \epsscale{1.15}
  \plotone{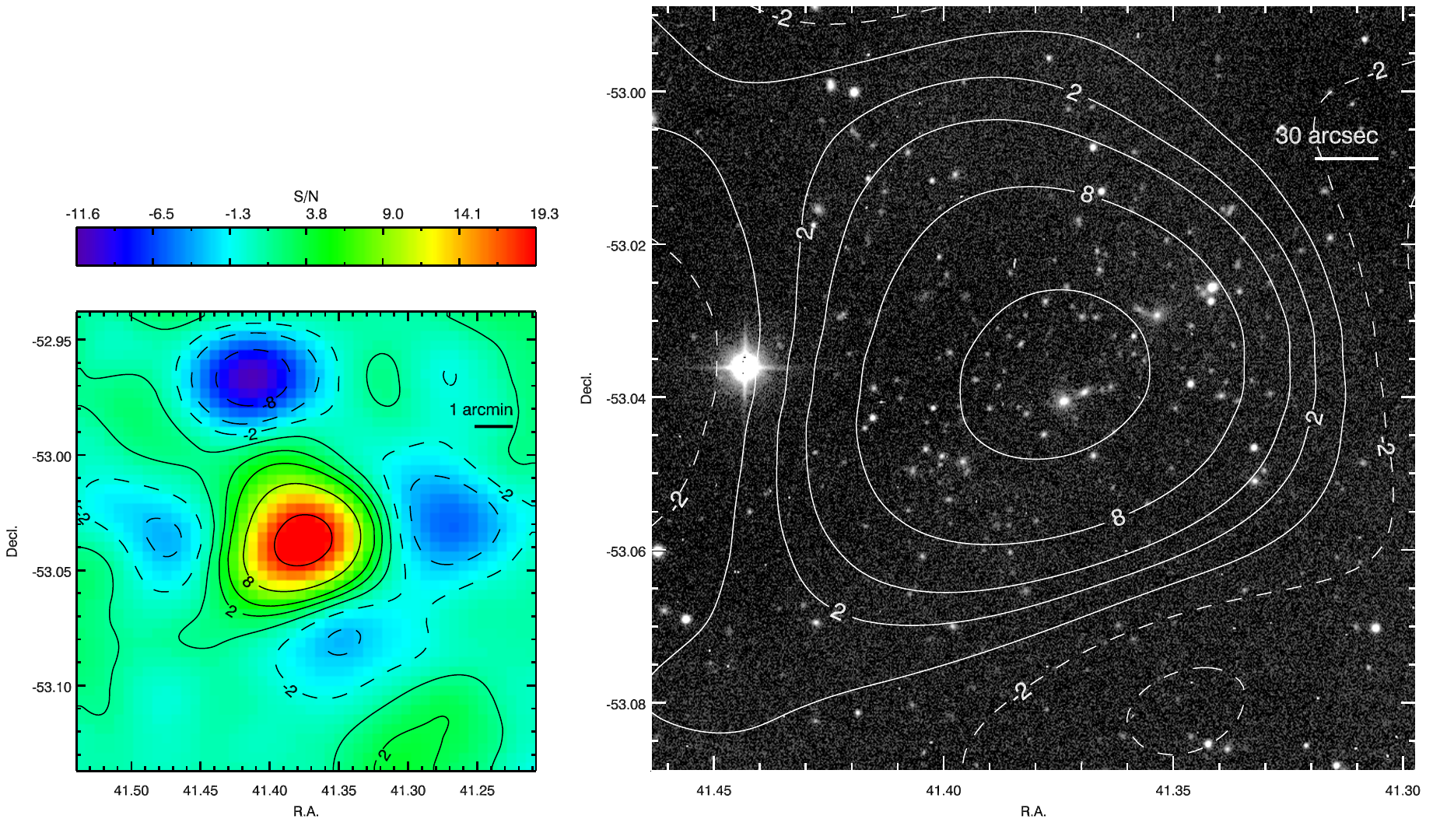}
  \caption{SPT-CL J0245-5302, also known as Abell S0295 and ACT-CL
    J0245-5302, at $z_{\mathrm{spec}}=0.300$. A Swope $R$ image is
    shown in the optical/infrared panel. The  stronger of the two
    point sources discussed in \S \ref{sec:psource} is visible as an $11 \sigma$ negative peak
    approximately five arcminutes north of 
    the cluster. \label{fig:thumb6}} 
\end{figure*}
\begin{figure*}
  \epsscale{1.15}
  \plotone{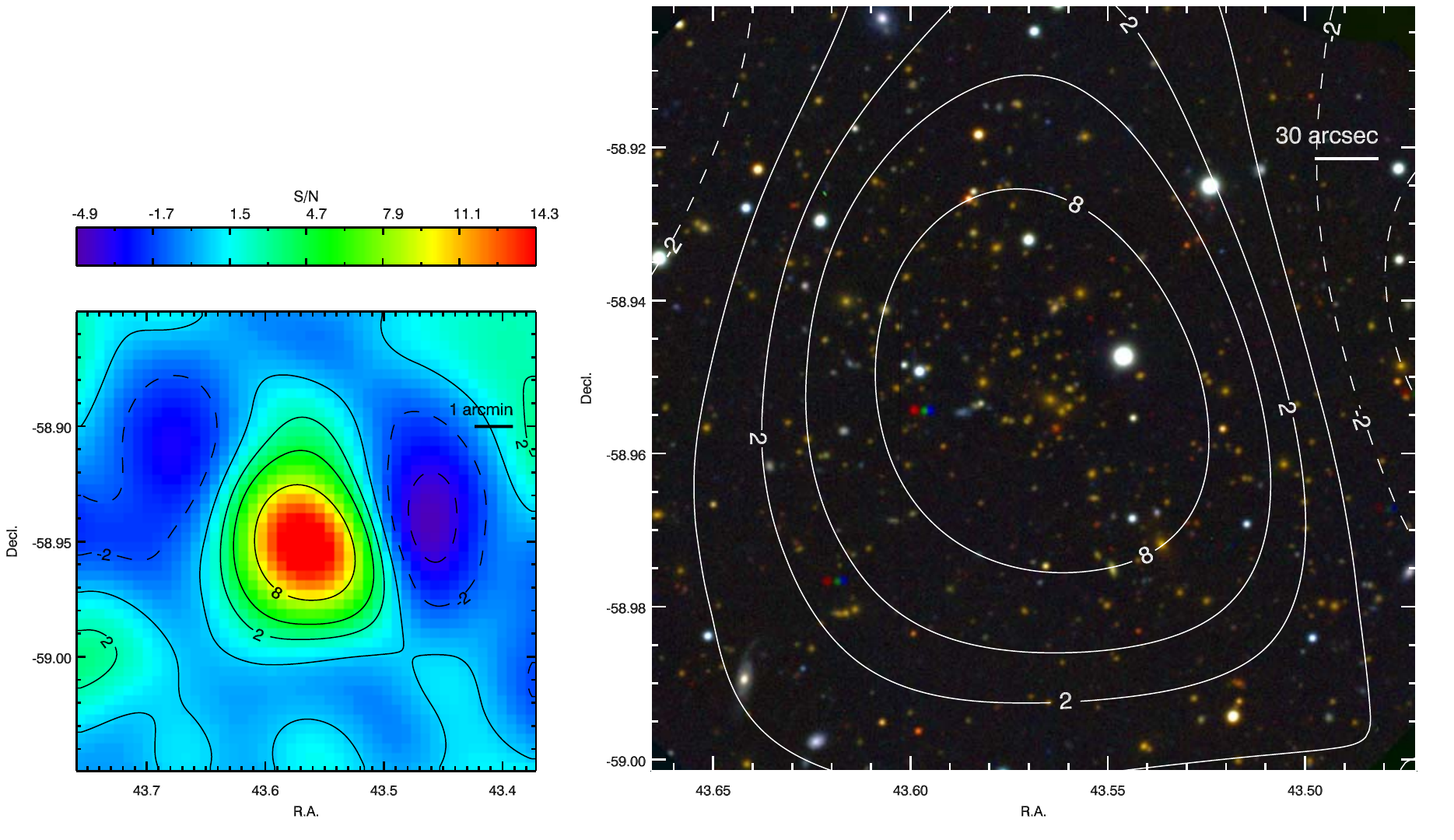}
  \caption{SPT-CL J0254-5856 at $z_{\mathrm{spec}}=0.438$.
    Magellan/LDSS3 $zrg$ images are shown in the 
    optical/infrared panel.\label{fig:thumb7}}
\end{figure*}
\begin{figure*}
  \epsscale{1.15}
  \plotone{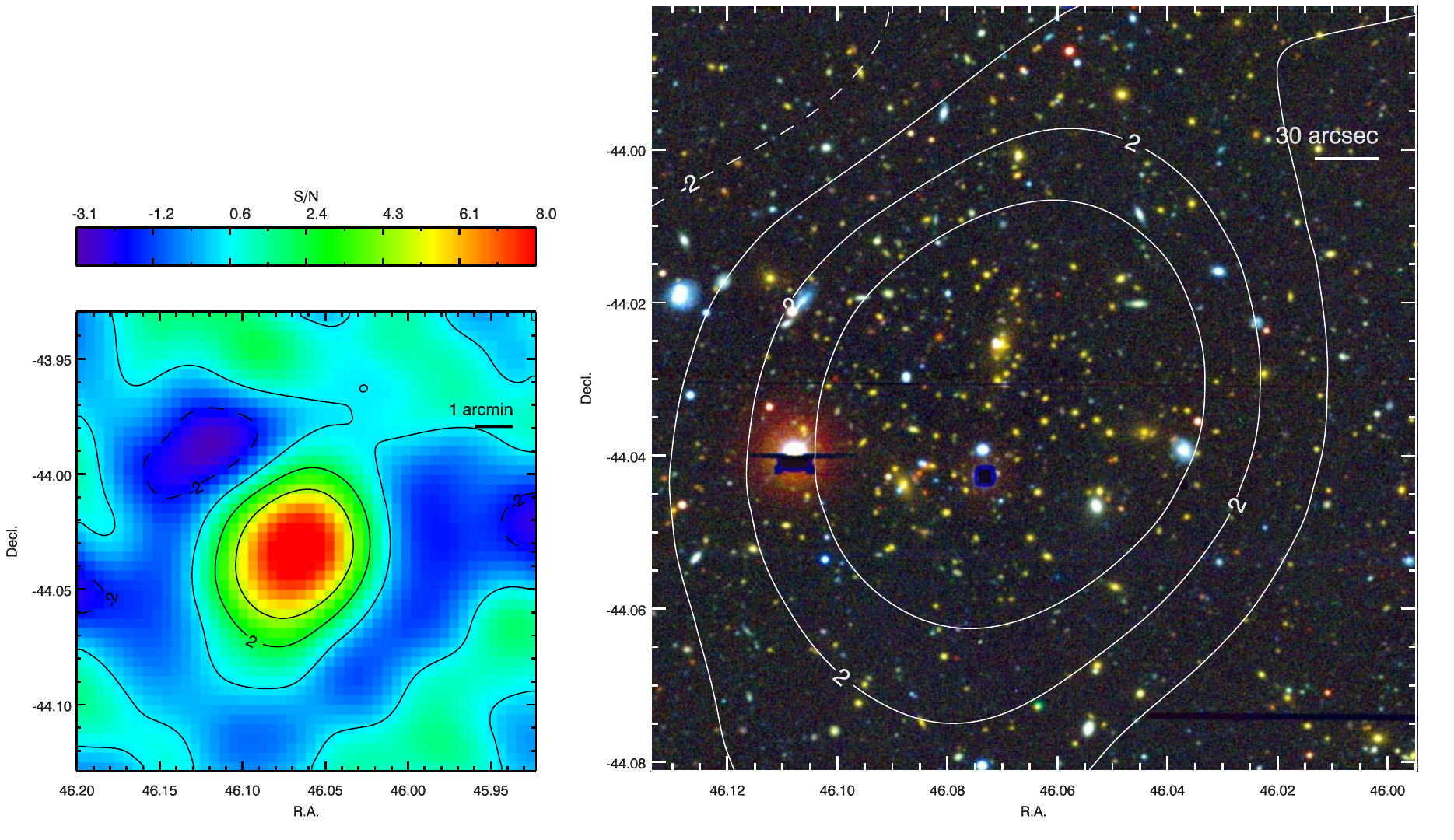}
  \caption{SPT-CL J0304-4401 at $z_{\mathrm{rs}}=0.52$.
    Blanco/MOSAIC-II $irg$ images are shown in the 
    optical/infrared panel.\label{fig:thumb8}}
\end{figure*}
\begin{figure*}
  \epsscale{1.15}
  \plotone{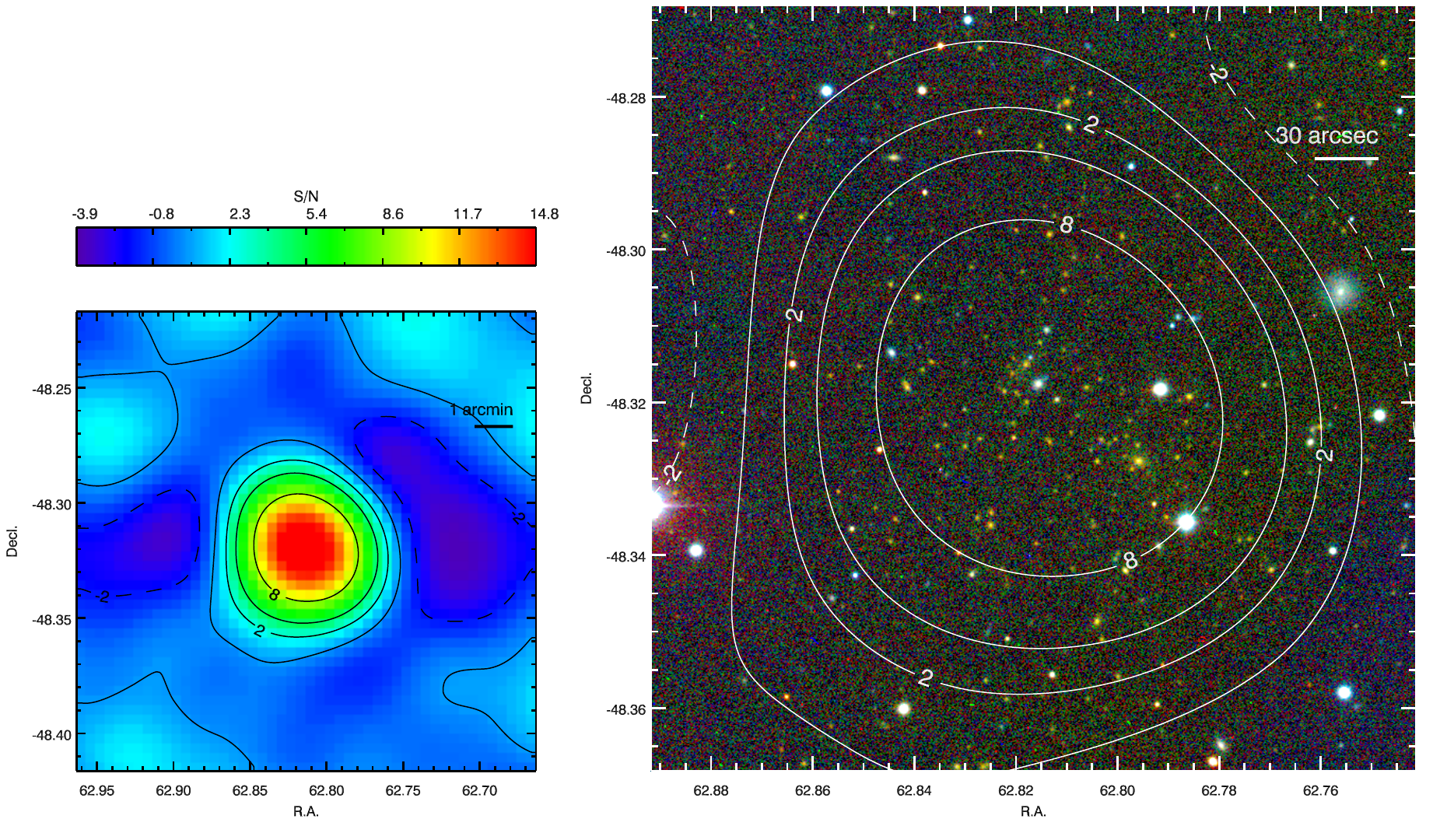}
  \caption{SPT-CL J0411-4819 at $z_{\mathrm{rs}}=0.42$. Swope $IRV$
    images are shown in the optical/infrared
    panel.\label{fig:thumb9}}
\end{figure*}
\begin{figure*}
  \epsscale{1.15}
  \plotone{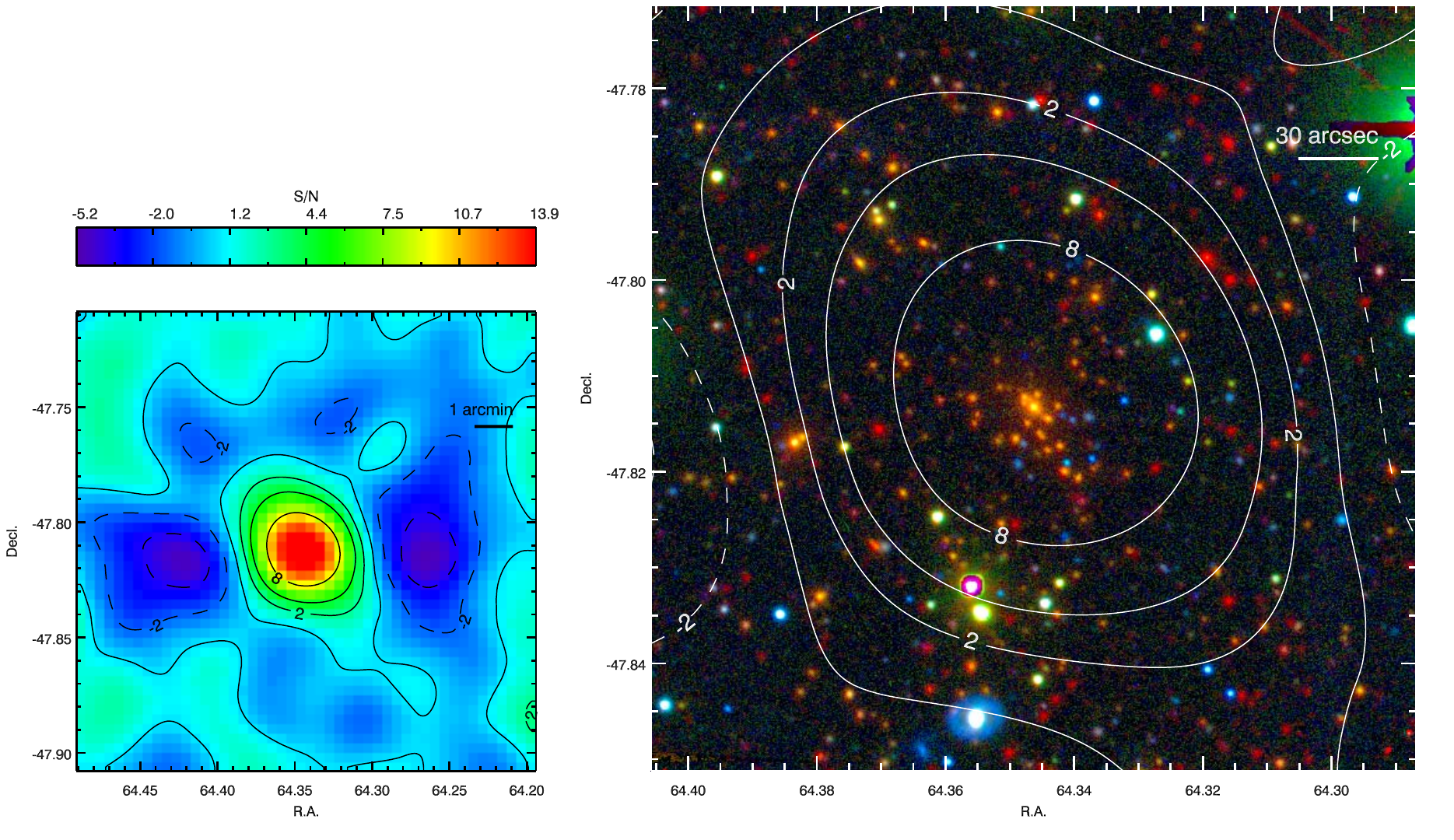}
  \caption{SPT-CL J0417-4748 at $z_{\mathrm{rs}}=0.62$. Spitzer/IRAC
    $[3.6]$ and Blanco/MOSAIC-II $ig$ images are shown in the
    optical/infrared panel.\label{fig:thumb10}}
\end{figure*}
\begin{figure*}
  \epsscale{1.15}
  \plotone{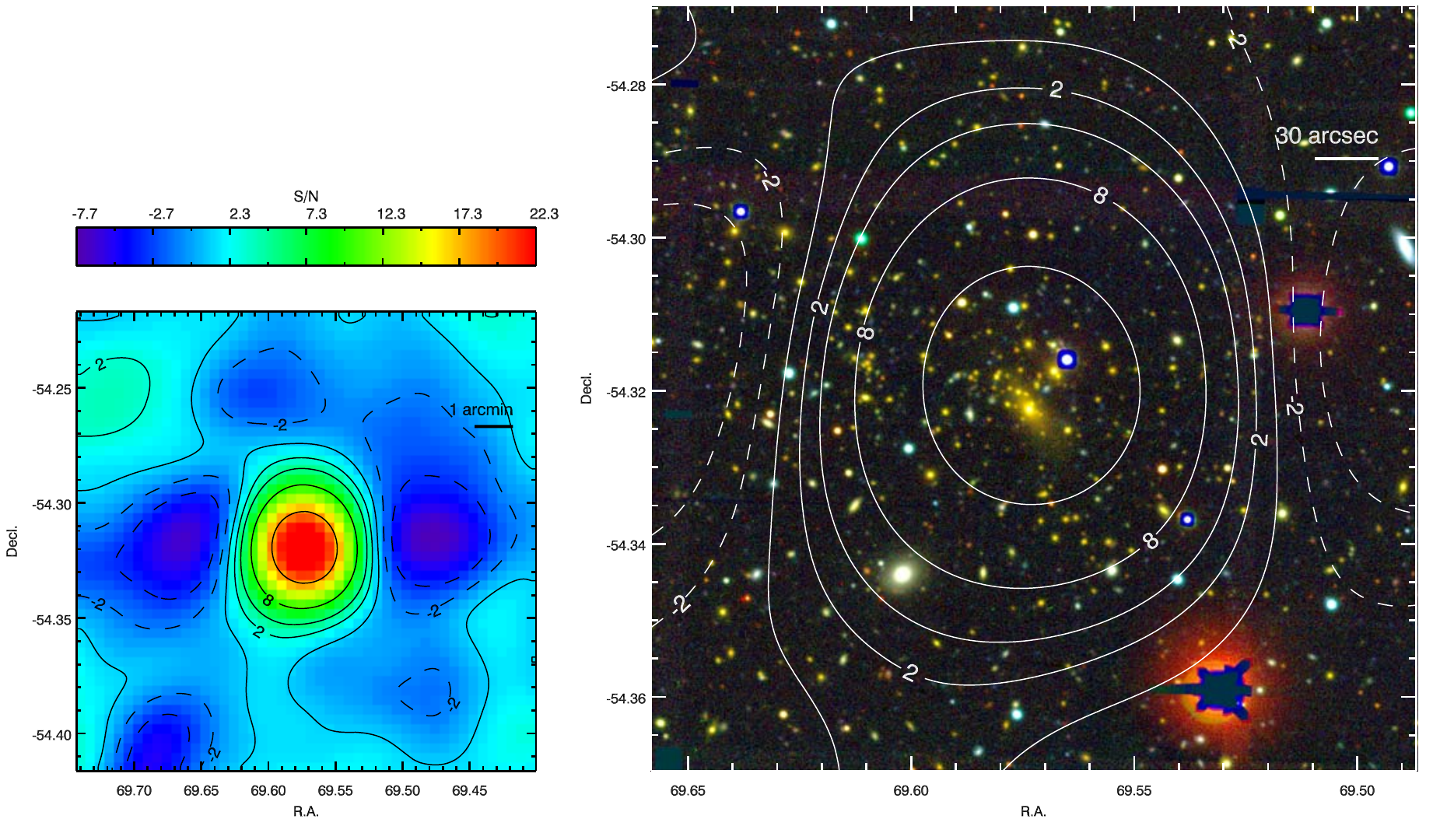}
  \caption{SPT-CL J0438-5419, also known as ACT-CL J0438-5419, at
    $z_{\mathrm{rs}}=0.45$. Blanco/MOSAIC-II $irg$ images are shown in
    the optical/infrared panel.\label{fig:thumb11}}
\end{figure*}
\begin{figure*}
  \epsscale{1.15}
  \plotone{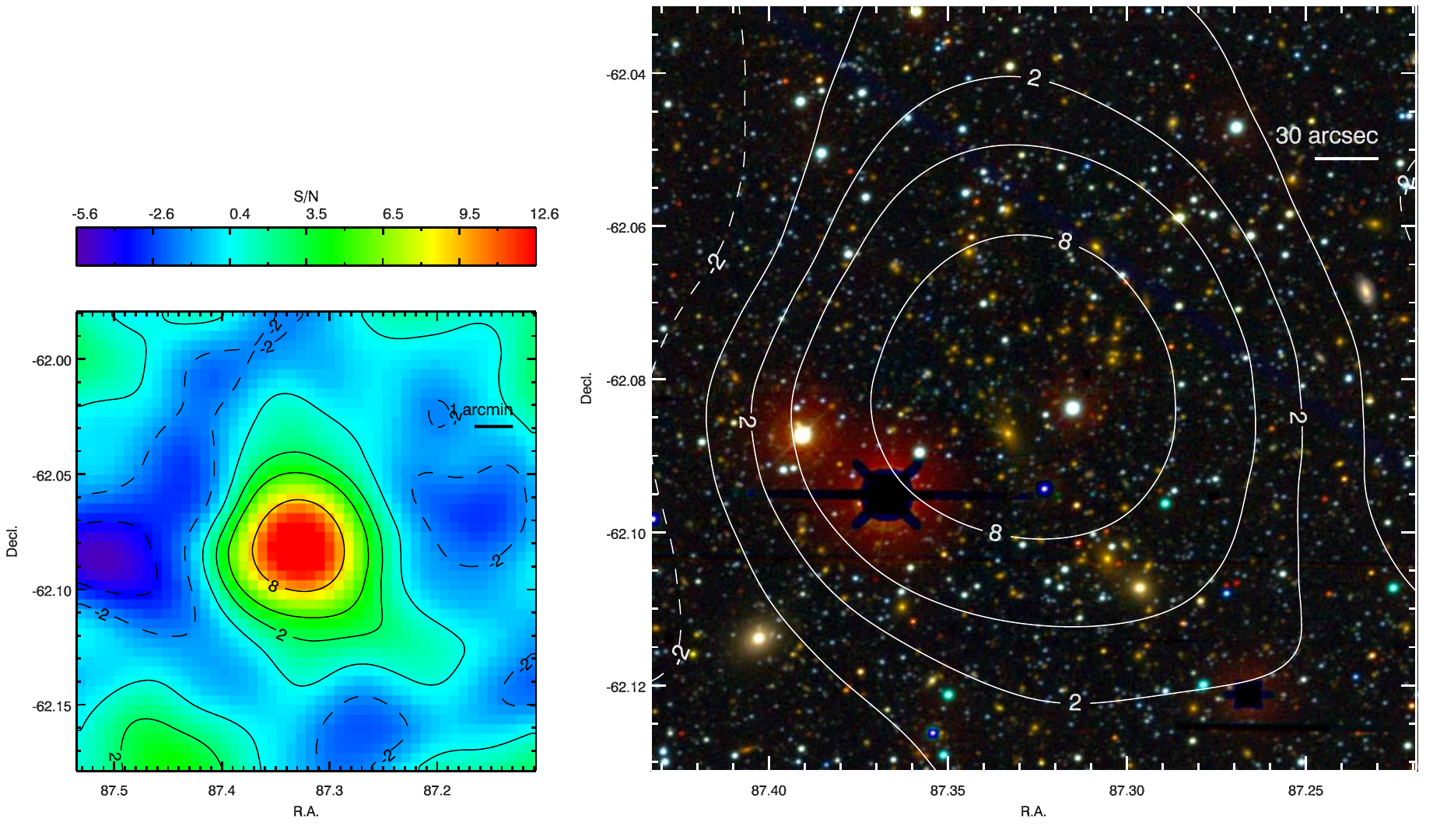}
  \caption{SPT-CL J0549-6204 at $z_{\mathrm{rs}}=0.32$.
    Blanco/MOSAIC-II $irg$ images are shown in the 
    optical/infrared panel.\label{fig:thumb12}}
\end{figure*}
\begin{figure*}
  \epsscale{1.15}
  \plotone{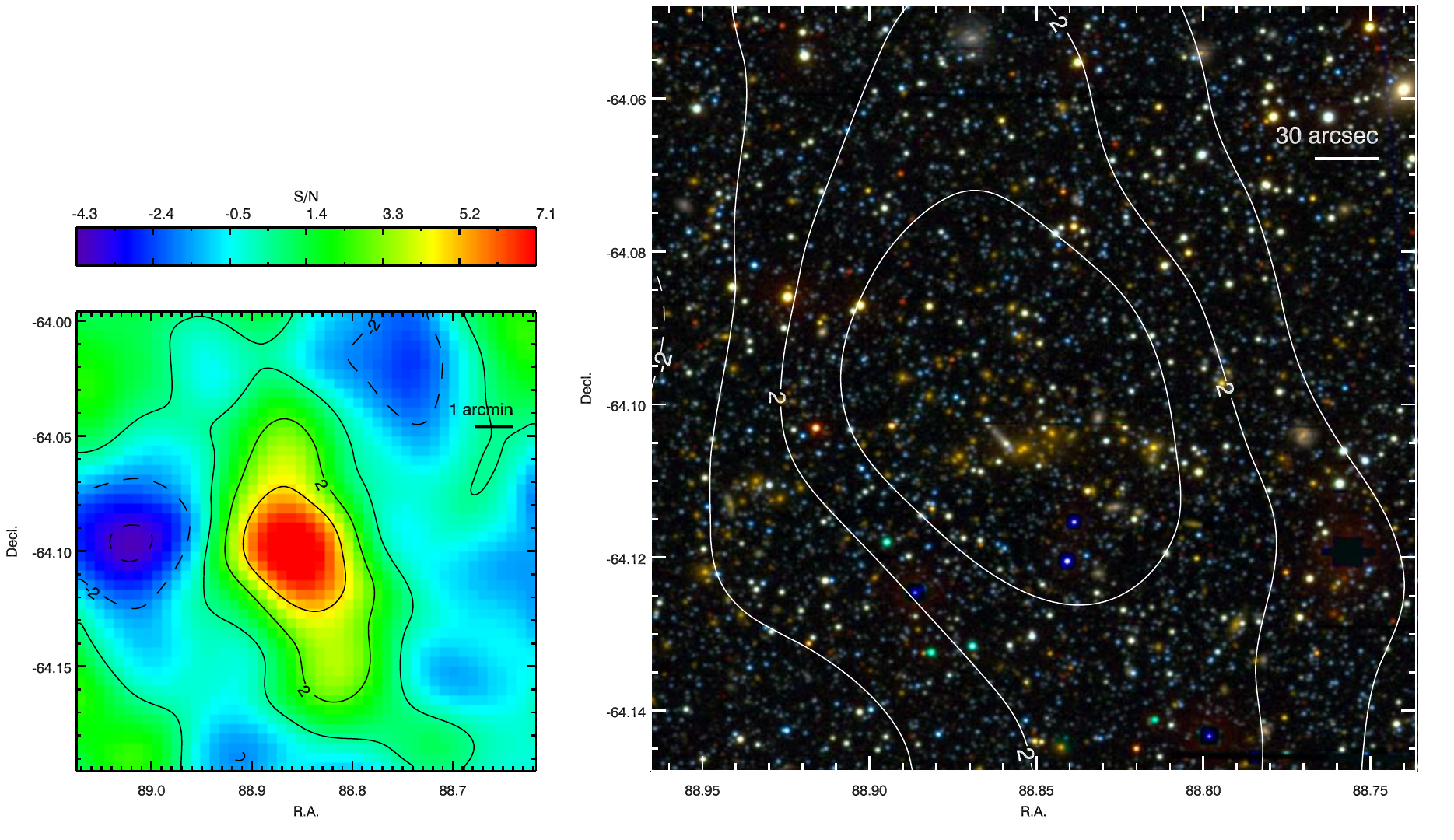}
  \caption{SPT-CL J0555-6405 at $z_{\mathrm{rs}}=0.42$.
    Blanco/MOSAIC-II $irg$ images are shown in the 
    optical/infrared panel.\label{fig:thumb13}}
\end{figure*}
\begin{figure*}
  \epsscale{1.15}
  \plotone{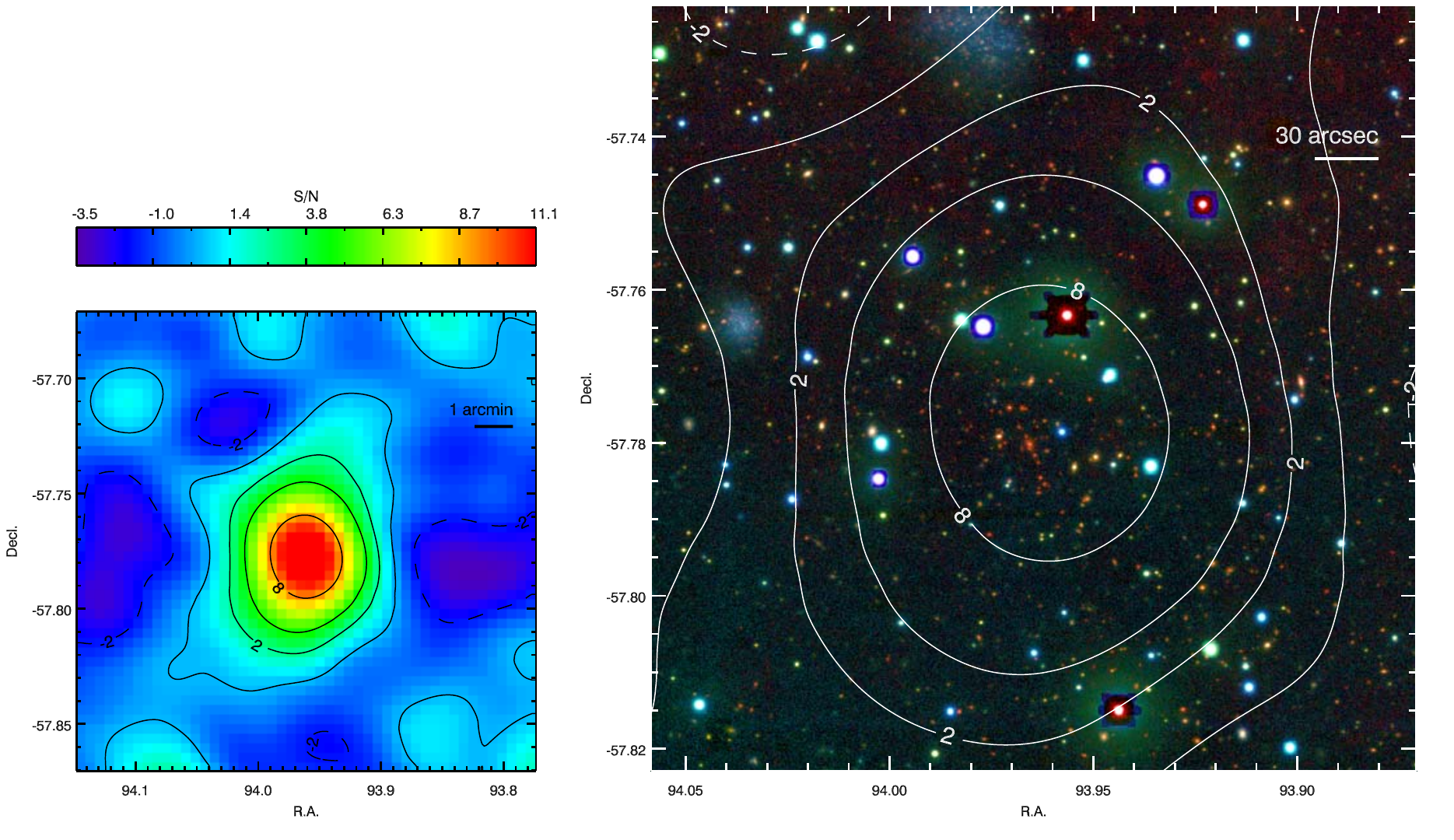}
  \caption{SPT-CL J0615-5746 at $z_{\mathrm{spec}}=0.972$. Blanco/NEWFIRM
    $K_s$ and Blanco/MOSAIC-II $rg$ images are shown in the 
    optical/infrared panel.\label{fig:thumb14}}
\end{figure*}
\begin{figure*}
  \epsscale{1.15}
  \plotone{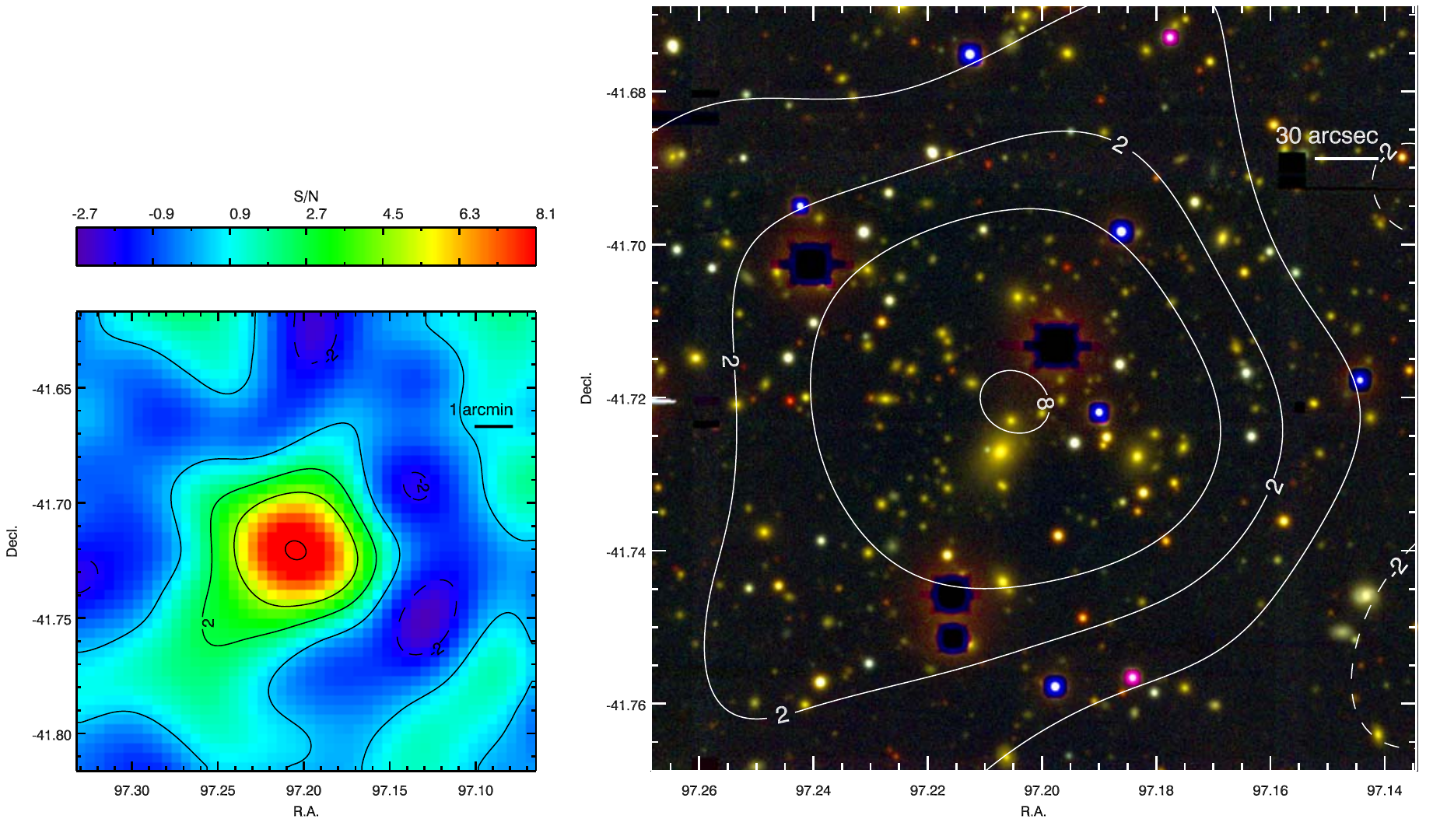}
  \caption{SPT-CL J0628-4143, also known as Abell 3396 and
    RXCJ0628.8-4143, at $z_{\mathrm{spec}}=0.176$. Blanco/MOSAIC-II
    $irg$ images are shown in the optical/infrared
    panel.\label{fig:thumb15}}
\end{figure*}
\begin{figure*}
  \epsscale{1.15}
  \plotone{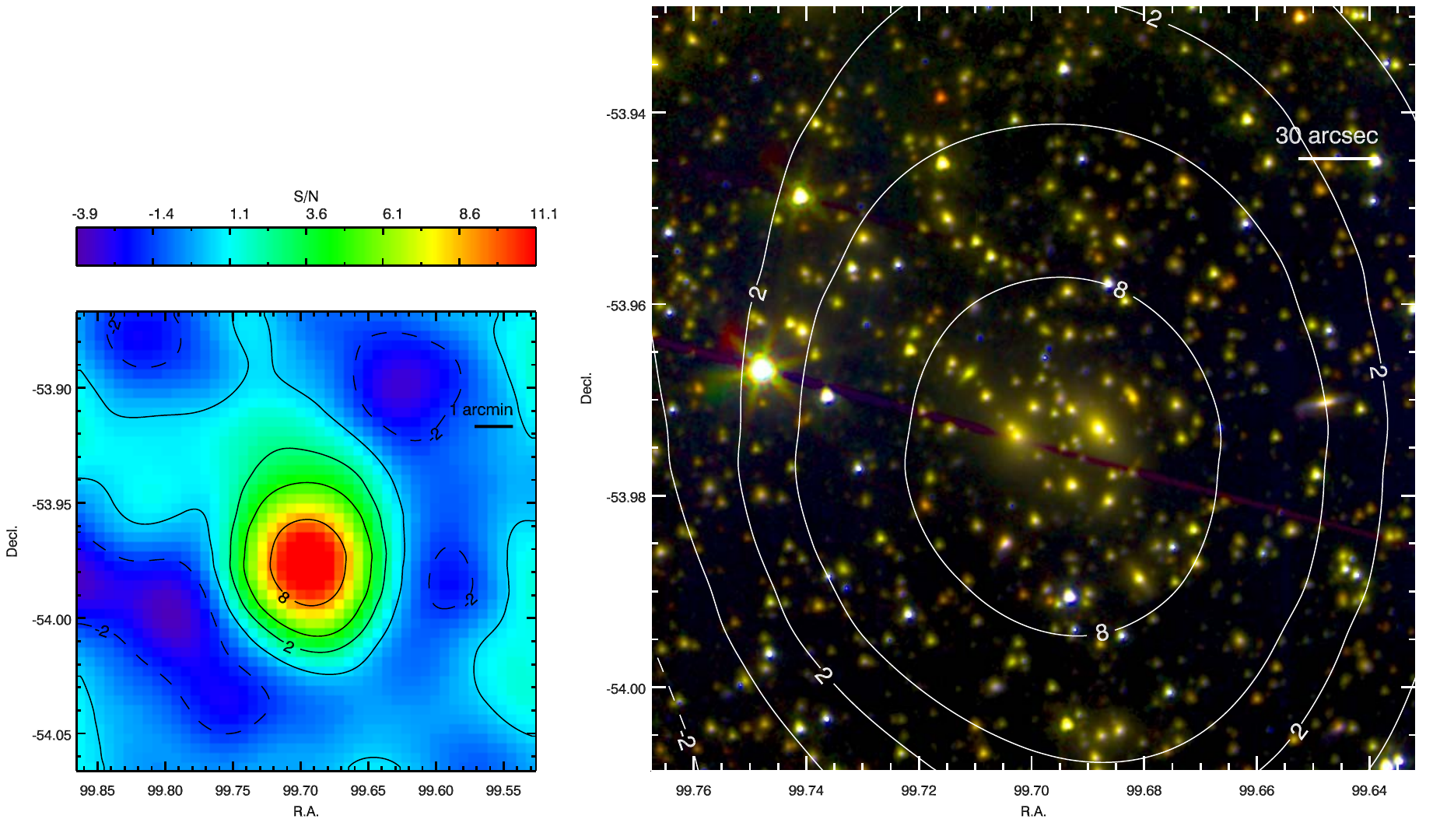}
  \caption{SPT-CL J0638-5358, also known as ABELL S0592,
    RXCJ0638.7-5358, and ACT-CL J0638-5358, at
    $z_{\mathrm{spec}}=0.222$. Spitzer/IRAC $[4.5][3.6]$ and
    Gemini/GMOS $r$ images are shown in the optical/infrared
    panel.\label{fig:thumb16}}
\end{figure*}
\begin{figure*}
  \epsscale{1.15}
  \plotone{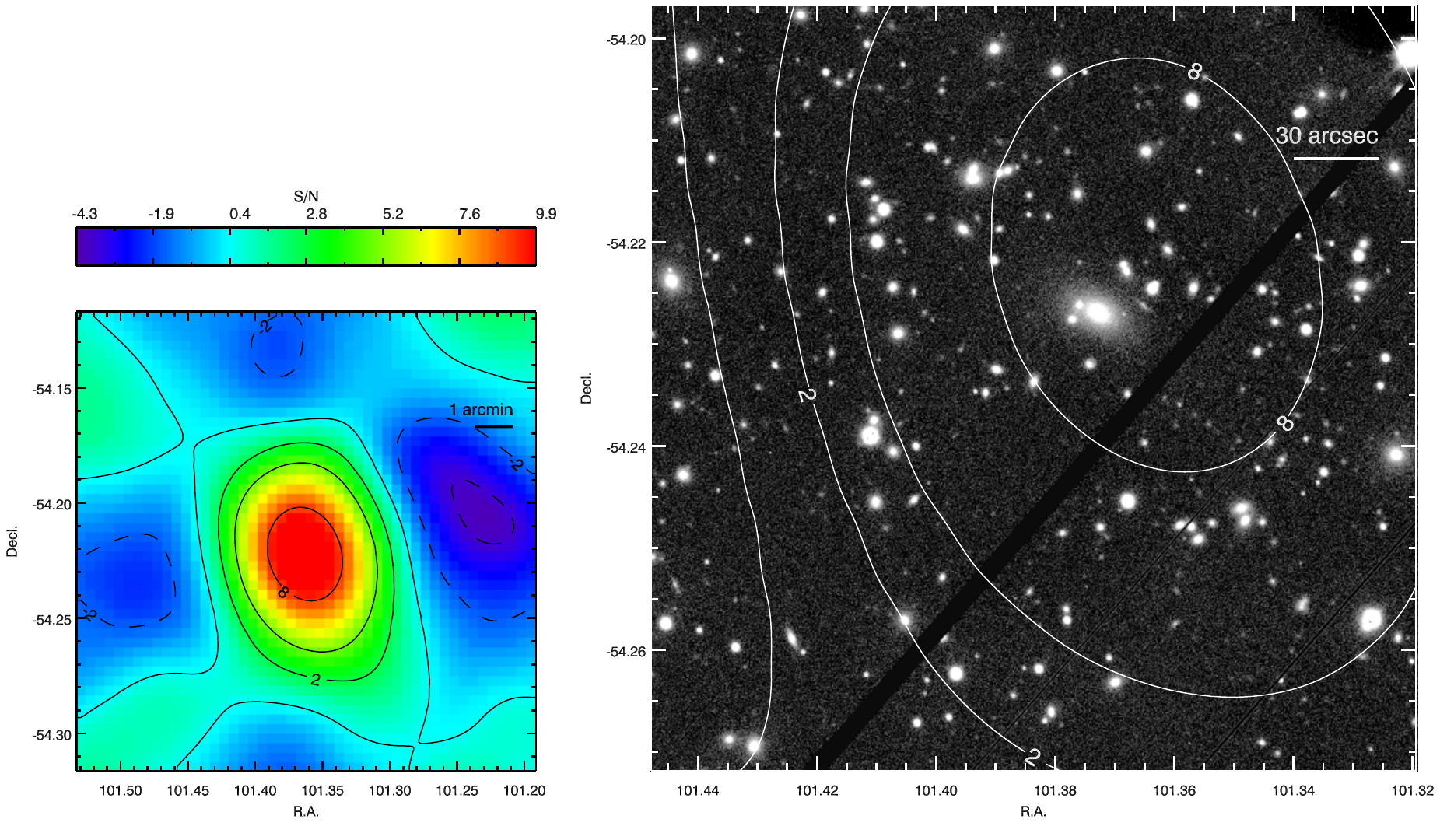}
  \caption{SPT-CL J0645-5413, also known as ABELL 3404,
    RXCJ0645.4-5413, and ACT-CL J0645-5413, at
    $z_{\mathrm{spec}}=0.167$. A VLT/FORS2 $R$ image is shown in the
    optical/infrared panel.\label{fig:thumb17}}
\end{figure*}
\begin{figure*}
  \epsscale{1.15}
  \plotone{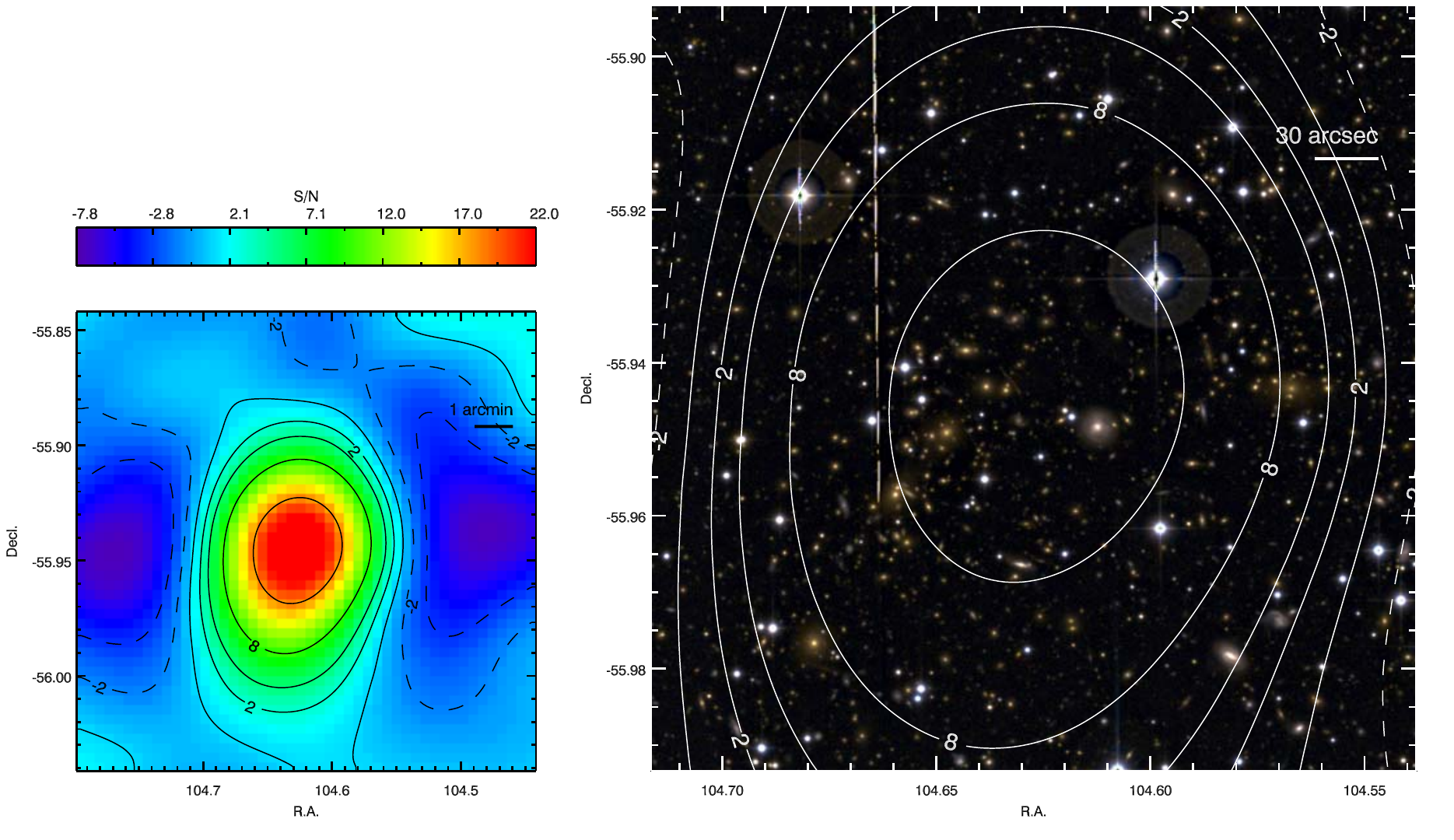}
  \caption{SPT-CL J0658-5556, also known as the Bullet Cluster,
    RXCJ0658.5-5556, and ACT-CL J0658-5557, at
    $z_{\mathrm{spec}}=0.296$. MPG-ESO/WFI $R$ and $V$ images are
    shown in the optical/infrared panel.\label{fig:thumb18}}
\end{figure*}
\begin{figure*}
  \epsscale{1.15}
  \plotone{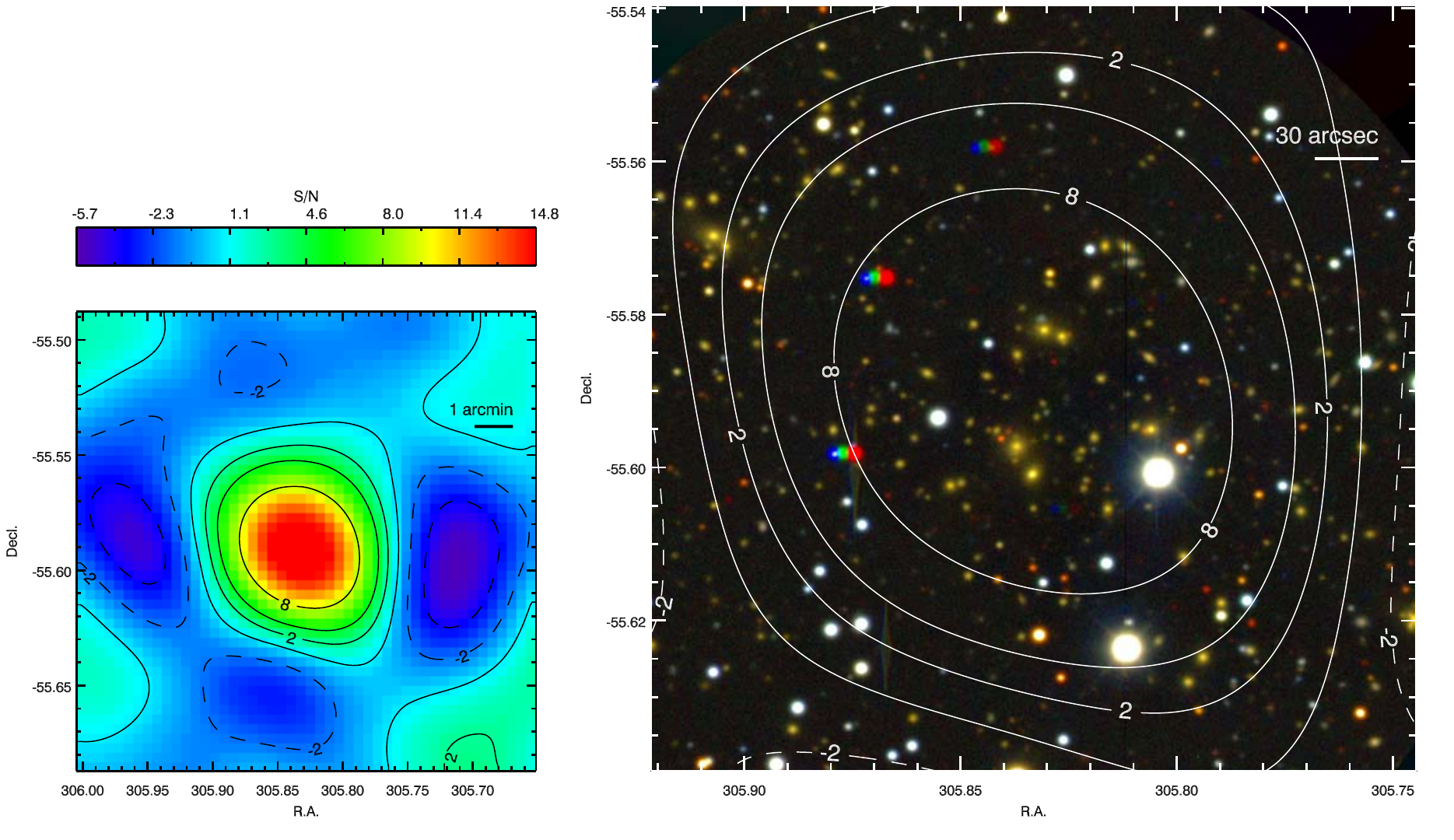}
  \caption{SPT-CL J2023-5535, also known as RXCJ2023.4-5535, at
    $z_{\mathrm{spec}}=0.232$. Magellan/LDSS3 $irg$ images are shown
    in the optical/infrared panel.\label{fig:thumb19}}
\end{figure*}
\begin{figure*}
  \epsscale{1.15}
  \plotone{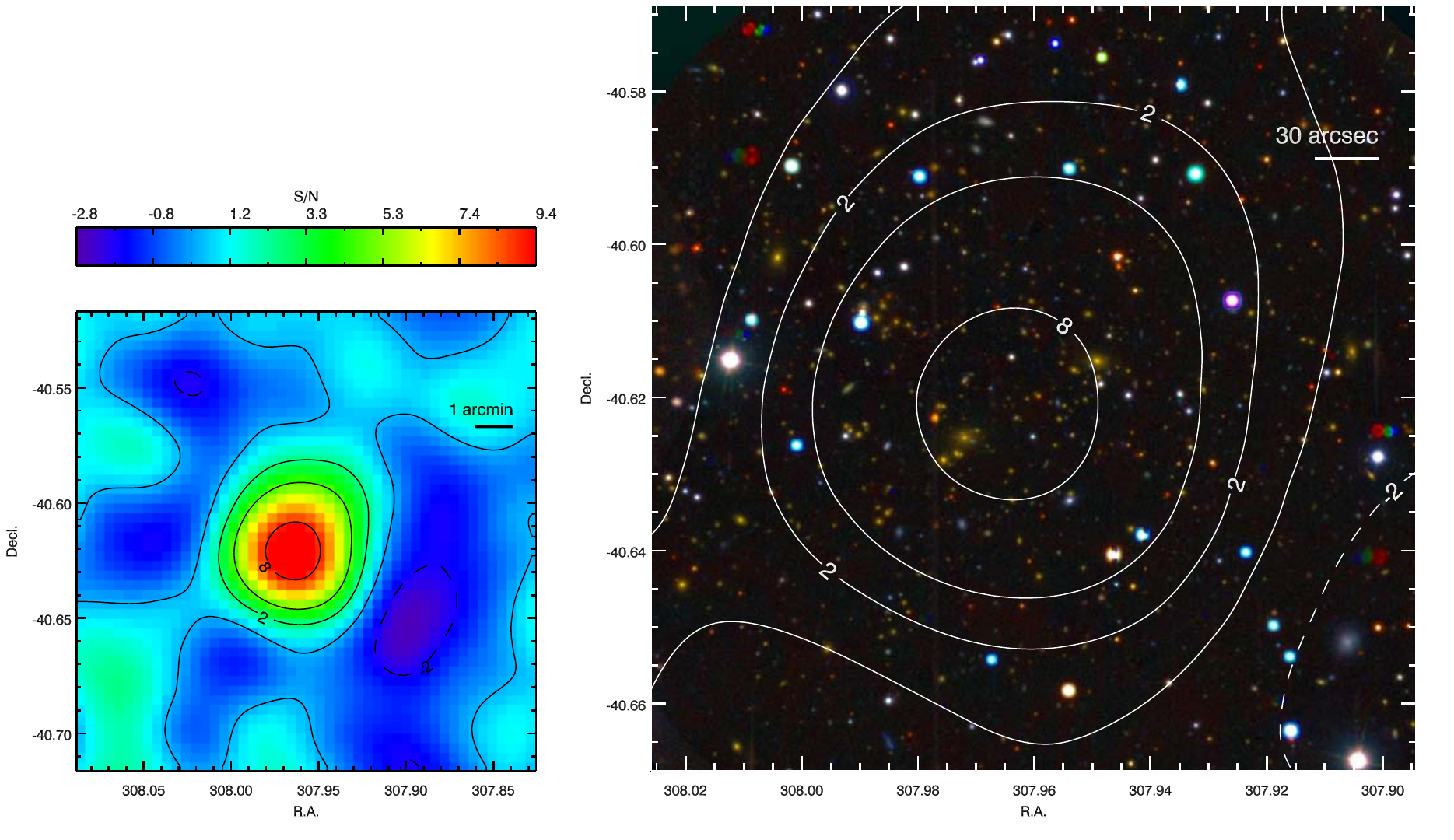}
  \caption{SPT-CL J2031-4037, also known as RXC J2031.8-4037, at
    $z_{\mathrm{spec}}=0.342$. Magellan/LDSS3 $irg$ images are shown
    in the optical/infrared panel.\label{fig:thumb20}}
\end{figure*}
\begin{figure*}
  \epsscale{1.15}
  \plotone{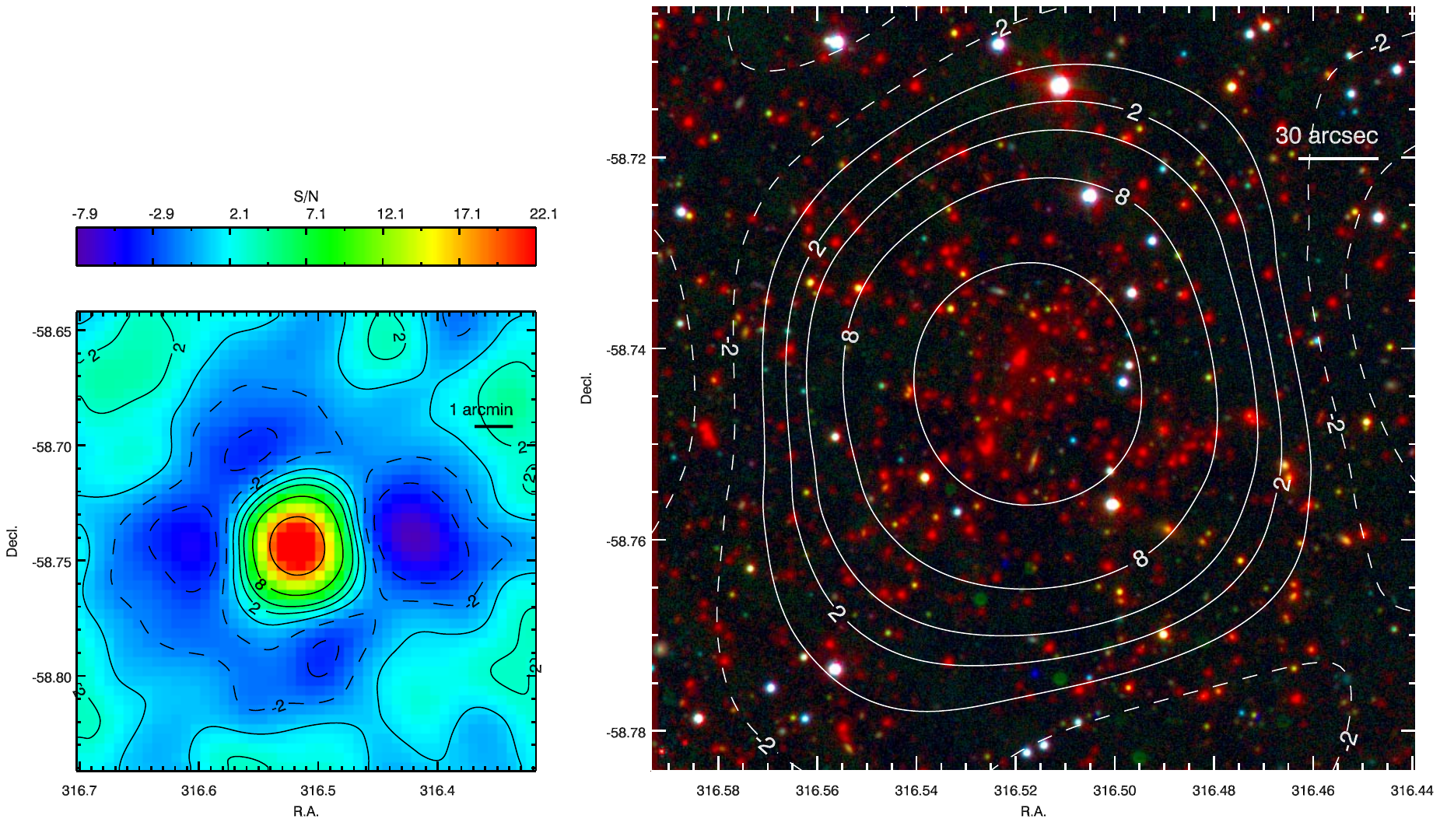}
  \caption{SPT-CL J2106-5844 at $z_{\mathrm{spec}}=1.133$.
    Spitzer/IRAC $[3.6]$ and Magellan/LDSS3 $ig$ images are shown in
    the optical/infrared panel.\label{fig:thumb21}}
\end{figure*}
\begin{figure*}
  \epsscale{1.15}
  \plotone{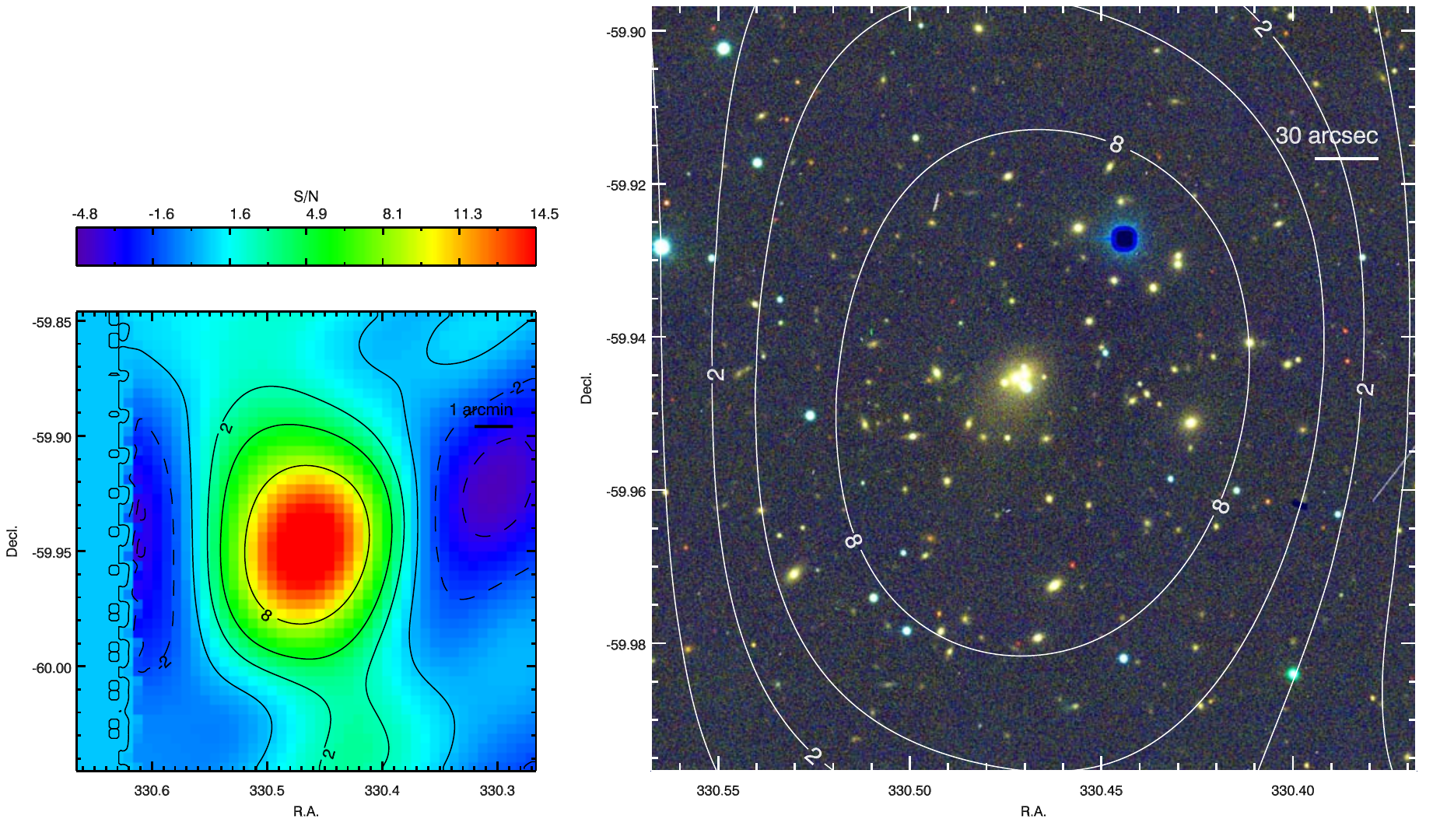}
  \caption{SPT-CL J2201-5956, also known as Abell 3827 and
    RXCJ2201.9-5956, at $z_{\mathrm{spec}}=0.098$. IMACS f/2 $irg$
    images are shown in the optical/infrared panel.  This
    detection is at the eastern edge of the survey
    field.\label{fig:thumb22}}
\end{figure*}
\begin{figure*}
  \epsscale{1.15}
  \plotone{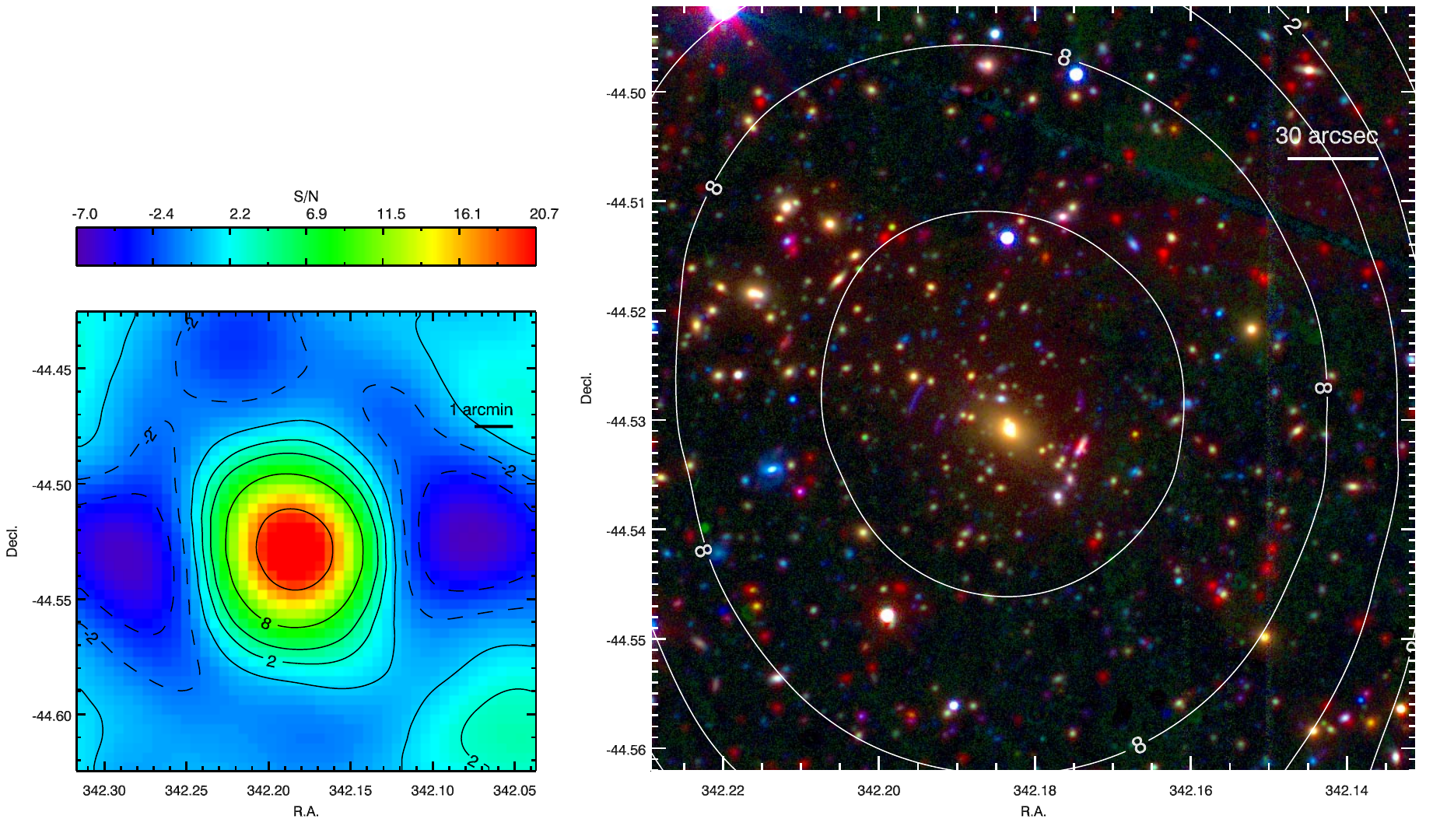}
 \caption{SPT-CL J2248-4431, also known as Abell S1063 and
    RXCJ2248.7-4431, at $z_{\mathrm{spec}}=0.348$.  
    Spitzer/IRAC $[3.6]$ and Magellan/LDSS3 $ig$ 
    images are shown in
    the optical/infrared panel.\label{fig:thumb23}}
\end{figure*}
\begin{figure*}
  \epsscale{1.15}
  \plotone{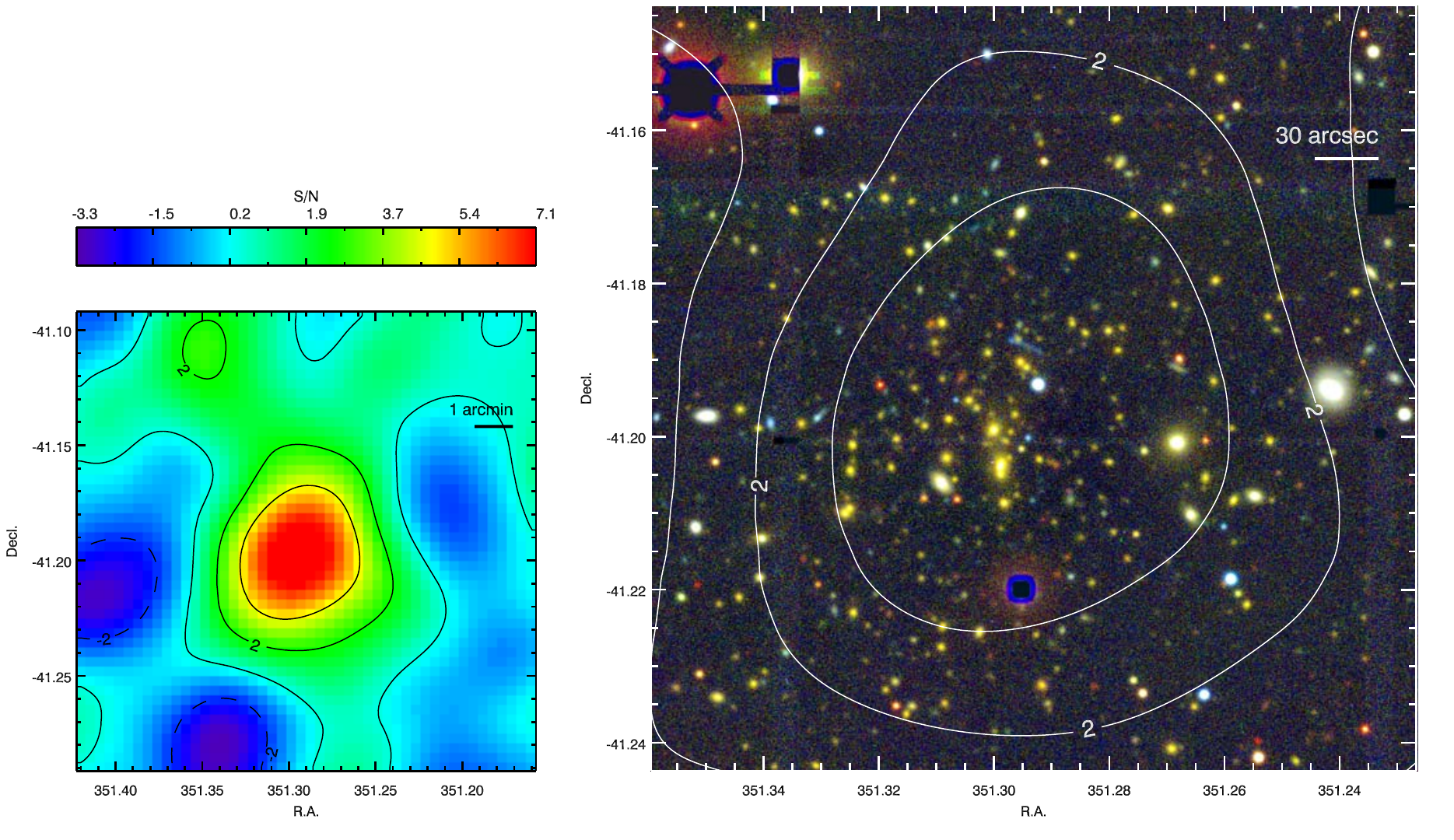}
  \caption{SPT-CL J2325-4111, also known as Abell S1121, at
    $z_{\mathrm{rs}}=0.37$. Blanco/MOSAIC-II $irg$ images are shown in
    the optical/infrared panel.\label{fig:thumb24}}
\end{figure*}
\begin{figure*}
  \epsscale{1.15}
  \plotone{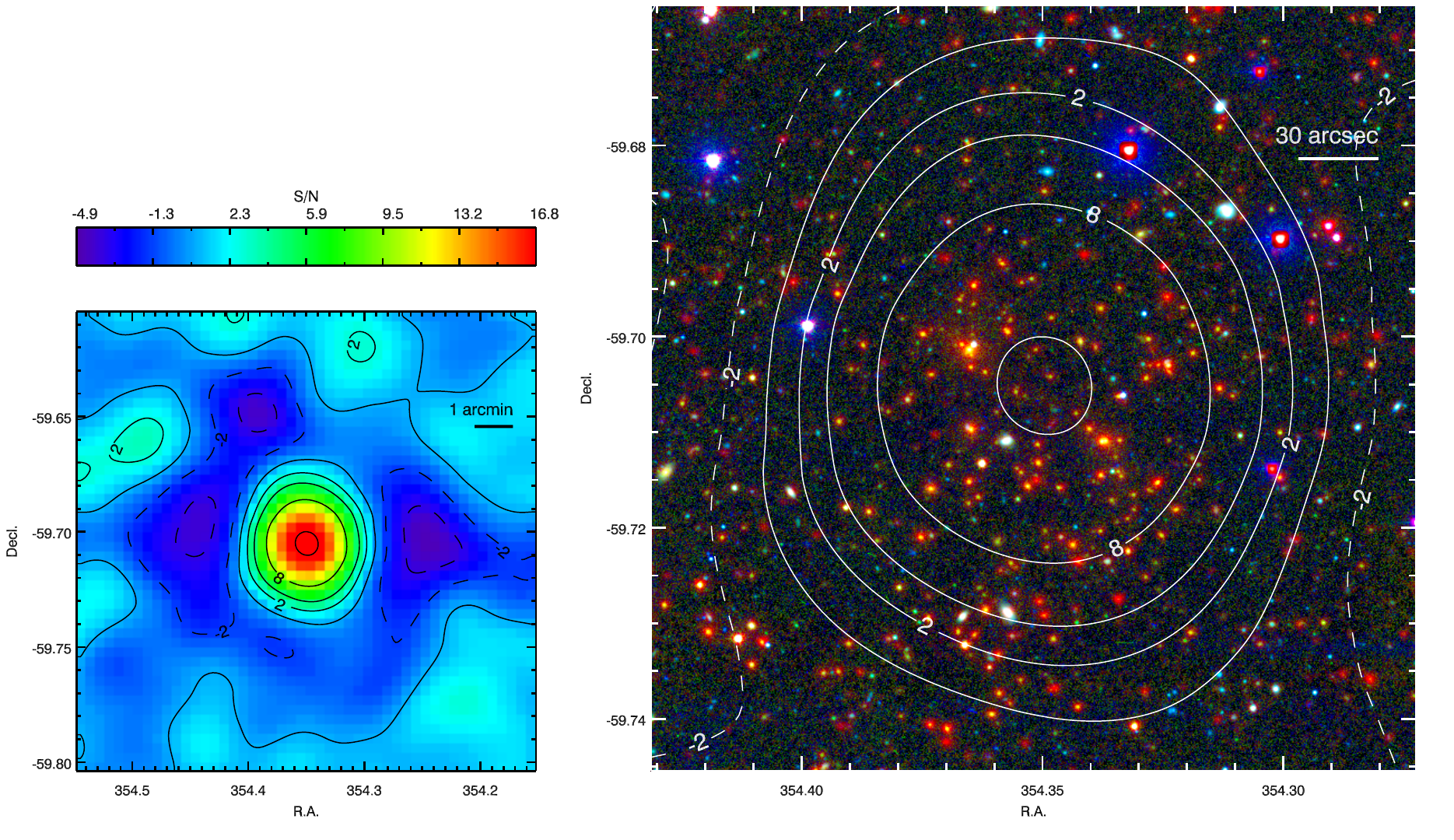}
  \caption{SPT-CL J2337-5942 at $z_{\mathrm{spec}}=0.775$.
    Spitzer/IRAC $[3.6]$ and Magellan/IMACS f/2 $ig$ images are shown
    in the optical/infrared panel.\label{fig:thumb25}}
\end{figure*}
\begin{figure*}
  \epsscale{1.15}
  \plotone{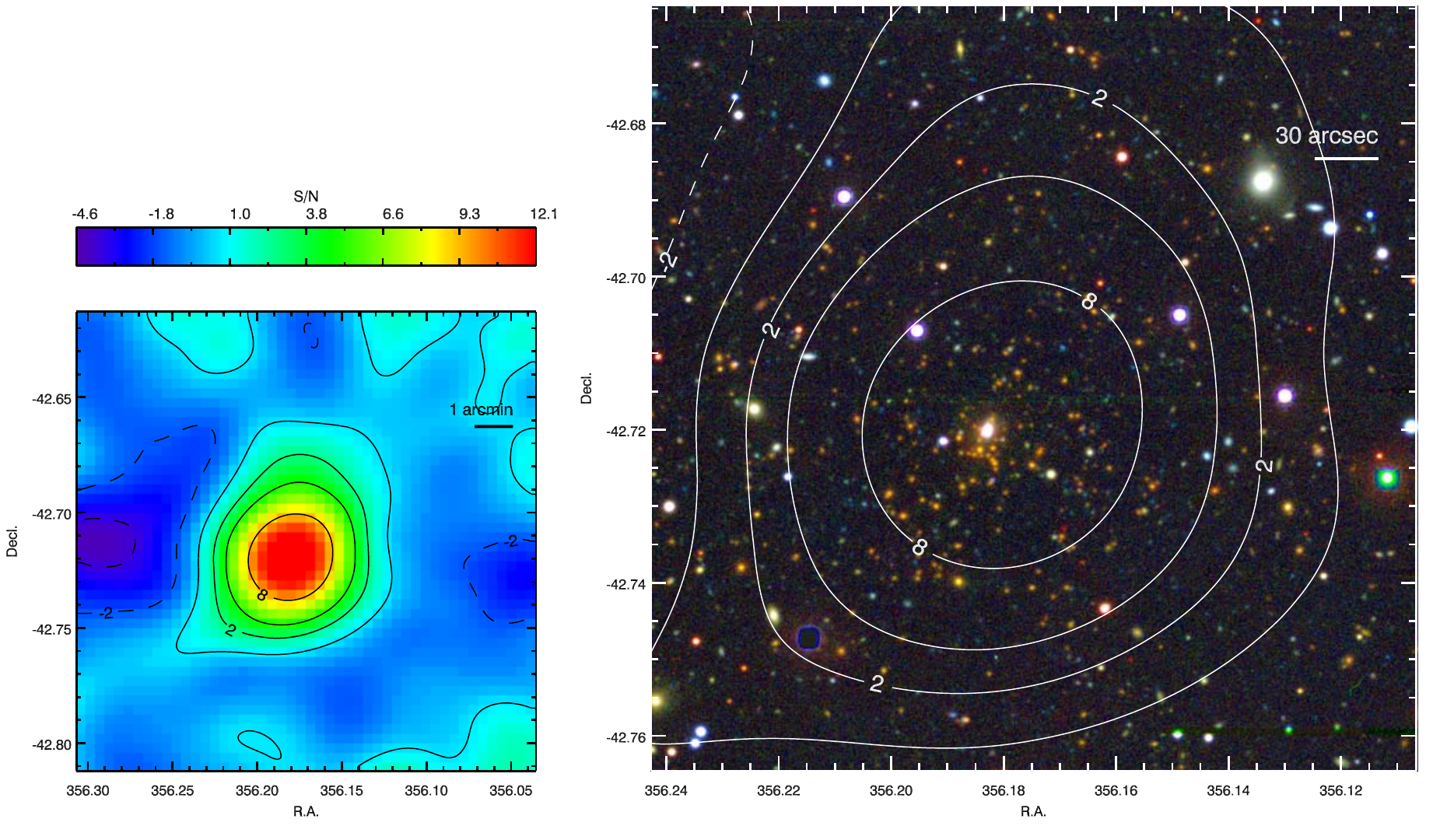}
  \caption{SPT-CL J2344-4243 at $z_{\mathrm{rs}}=0.62$.
    Blanco/MOSAIC-II $irg$ images are shown in the 
    optical/infrared panel.\label{fig:thumb26}}
\end{figure*}

Figures \ref{fig:thumb1}--\ref{fig:thumb26} show SZ detection
significance maps ({\it left panels}) and optical and infrared images
({\it right panels}) of the clusters. In all images, north is up, east
is left.
The SZ-only insets subtend 12 arcminutes on a side. The mapping
between color and SZ significance $\xi$ is different in all SZ
thumbnails, spanning the full range of SZ pixel values in the region
of sky shown. The peak value in each thumbnail is equal to the quoted
SZ detection significance in Table \ref{tab:clusters}. Contours denote
significance values of $(-8,-4,-2,0,2,4,8,16,32)$ in all thumbnails.  Contours are
dashed where $\xi$ is negative, and solid where $\xi$ is positive.
The negative lobes around some of the most significantly detected
clusters in the SZ images are due to the filtering of the time-ordered
data and the maps 

The optical/infrared images have the same contours as their
corresponding SZ thumbnail overlaid. False-color composites are
presented for clusters where multiband imaging is available, either
from our own observations or from public archives. Otherwise,
black-and-white images are shown.

\clearpage

\clearpage

\clearpage

\clearpage

\clearpage

\clearpage

\clearpage

\clearpage

\clearpage

\clearpage

\clearpage

\clearpage

\clearpage

\end{document}